\documentclass[trackchanges, twocolumn]{aastex701}

\usepackage{graphicx, subfig, caption, subcaption}
\usepackage{orcidlink}
\usepackage{academicons}
\usepackage{hhline}
\usepackage{adjustbox}
\usepackage{float}
\usepackage{booktabs}
\usepackage{amsmath}
\usepackage{siunitx}
\usepackage{subcaption}

\definecolor{orcidlogocol}{HTML}{A6CE39}
\newcommand {\jwst}{\textit{JWST}}
\newcommand {\hst}{\textit{HST}}
\newcommand {\spitzer}{\textit{Spitzer}}
\newcommand {\gaia}{\textit{Gaia}}
\newcommand {\herschel}{\textit{Herschel}}

\newcommand {\msun}{M$_\odot$}

\begin{document}

\title{The Stellar Population of NGC 346 in the Small Magellanic Cloud with JWST}

\author[orcid=0000-0002-0577-1950]{Jeroen Jaspers}
\affiliation{Department of Physics, Maynooth University, Maynooth, Co. Kildare, Ireland}
\affiliation{Astronomy \& Astrophysics Section, School of Cosmic Physics, Dublin Institute for Advanced Studies, 31 Fitzwilliam Place, Dublin D02 XF86, Ireland}
\email[]{jeroen.jaspers@mu.ie}  

\author[orcid=0000-0001-6872-2358]{Patrick J. Kavanagh} 
\affiliation{Department of Physics, Maynooth University, Maynooth, Co. Kildare, Ireland}
\email[]{}

\author[orcid=0000-0001-7906-3829]{Guido De Marchi} 
\affiliation{European Space Research and Technology Centre, Keplerlaan 1, 2200 AG Noordwijk, Netherlands}
\email[]{}

\author[orcid=0000-0002-7512-1662]{Connor Nally} 
\affiliation{Institute for Astronomy, University of Edinburgh, Blackford Hill, Edinburgh, EH9 3HJ, UK}
\email[]{}

\author[orcid=0000-0003-4870-5547]{Olivia C. Jones} 
\affiliation{UK Astronomy Technology Centre, Royal Observatory, Blackford Hill, Edinburgh, EH9 3HJ, UK}
\email[]{}

\author[orcid=0000-0002-2667-1676]{Nolan Habel} 
\affiliation{Jet Propulsion Laboratory, California Institute of Technology, 4800 Oak Grove Dr., Pasadena, CA 91109, USA}
\email[]{}

\author[orcid=0000-0002-6091-7924]{Peter Zeidler} 
\affiliation{AURA for the European Space Agency, ESA Office, STScI, 3700 San Martin Drive, Baltimore, MD 21218, USA}
\email[]{}

\author[orcid=0000-0002-0522-3743]{Margaret Meixner} 
\affiliation{Jet Propulsion Laboratory, California Institute of Technology, 4800 Oak Grove Dr., Pasadena, CA 91109, USA}
\email[]{}

\author[orcid=0000-0003-2954-7643]{Elena Sabbi} 
\affiliation{Space Telescope Science Institute, 3700 San Martin Drive, Baltimore, MD 21218, USA}
\email[]{}

\author[orcid=0000-0002-2954-8622]{Alec S. Hirschauer}
\affiliation{Department of Physics \& Engineering Physics, Morgan State University, 1700 East Cold Spring Lane, Baltimore, MD 21251, USA}
\email[]{}

\author[orcid=0000-0002-1892-2180]{Kattia Biazzo}
\affiliation{INAF, Astronomical Observatory of Rome, via Frascati 33, Monteporzio Catone I-00078, Italy}
\email[]{}

\author[orcid=0000-0003-4023-8657]{Laura Lenki\'{c}}
\affiliation{IPAC, California Institute of Technology, 1200 East California Boulevard, Pasadena, CA 91125, USA}
\affiliation{Jet Propulsion Laboratory, California Institute of Technology, 4800 Oak Grove Dr., Pasadena, CA 91109, USA}
\email[]{}

\author[orcid=0000-0001-6576-6339]{Omnarayani Nayak}
\affiliation{United States Naval Observatory, 3450 Massachusetts Avenue NW, Washington DC 20392, USA}
\email[]{}

\author[orcid=0000-0002-9573-3199]{Massimo Roberto}
\affiliation{Space Telescope Science Institute, 3700 San Martin Drive, Baltimore, MD 21218, USA}
\affiliation{Department of Physics and Astronomy, Johns Hopkins University, 3400 North Charles Street, Baltimore, MD 21218, USA}
\email[]{}

\author[orcid=0000-0001-5742-2261]{Ciaran Rogers}
\affiliation{Space Telescope Science Institute, 3700 San Martin Drive, Baltimore, MD 21218, USA}
\email[]{}

\author[orcid=0000-0001-9855-8261]{Beth A. Sargent}
\affiliation{Space Telescope Science Institute, 3700 San Martin Drive, Baltimore, MD 21218, USA}
\affiliation{Department of Physics and Astronomy, Johns Hopkins University, 3400 North Charles Street, Baltimore, MD 21218, USA}
\email[]{}

\begin{abstract}

NGC~346 is a massive star-forming region located at a distance of $\sim$62 kpc, in the Small Magellanic Cloud (SMC). Due to its low metallicity (Z $\sim$1/5 Z$_{\odot}$), it is an ideal environment to study star formation and stellar population analogues to those at Cosmic Noon.
In this work, we produce a combined \jwst\ NIRCam and MIRI photometric catalogue of NGC~346.
We characterise different stellar populations in the region: the upper main sequence (UMS), red giant branch (RGB), and red clump (RC), as well as pre-main sequence (pre-MS) stars and young stellar objects (YSOs).
We performed point-spread function (PSF) weighted photometry in 11 wavelength bands across NIRCam and MIRI and utilised multiple colour-magnitude-diagrams to identify the various stellar populations in the field. 
Our final photometric catalogue of NGC~346 comprises 249,519 unique sources, including 2,024 UMS stars, 2,755 RGB stars and 742 RC stars. In addition, we identified 6,274 candidate pre-MS stars, 7,350 candidate YSOs and 23,819 IR-excess sources.
Combining these three categories, we characterised 1,583 strong and 3,761 likely pre-MS/YSO candidates.
By utilising the F115W-F200W vs F115W-F187N colour-colour diagram, we found 239 non-spurious sources with Pa$\alpha$ excess, indicating accretion and star formation in NGC~346.
Using \jwst\ NIRCam and MIRI observations, we produced the deepest catalogue to date of the star formation region NGC~346 in the near- and mid-IR range (1-21$~\mu$m).
Our catalogue and characterisations of the young and old populations will provide the basis for detailed follow-up studies.

\end{abstract}


\section{Introduction}
Studies of star-forming regions in low-metallicity galaxies provide key insights into star formation and stellar evolution in the early Universe.
In such early galaxies, spatially resolving stellar populations is challenging, given the distances to these low-metallicity galaxies.
However, nearby analogues provide important laboratories to understand the physical mechanisms governing star formation and evolution at low metallicity.
Recent studies with the {\it James Webb Space Telescope} (\jwst) of galaxies at a redshift of z~$\sim$~2~-~3 showed that the low-mass end of these galaxies (log($M_\star/M_\odot$)~$\approx$~7.2~-~9.7) have metallicities spanning z~$\sim$~0.04~-~0.5 $Z_\odot$ \citep{Raptis2025}. This is the epoch of peak star formation in the Universe, known as ``Cosmic Noon'' \citep[2~$\lesssim$~z~$\lesssim$~3; ][]{Madau2014}.  
The Small Magellanic Cloud (SMC) is of particular importance given its metallicity of $\sim$1/5~Z$_\odot$ \citep{Russell1992, Peimbert2000, Choudhury2020}, and its total stellar mass \citep[log($M_\star/M_\odot$)~=~8.7; ][]{Rubele2018} are representative of these low-mass galaxies at ``Cosmic Noon''.
In addition, at a distance of 62~($\pm$~1)~kpc \citep{deGrijs2015}, the SMC is near enough for modern observatories to resolve the stellar populations to sub-solar masses.
\newline
\indent NGC~346 is the brightest star formation region in the SMC \citep{deGrijs2015}. It has a complex star formation history (SFH) and contains multiple stellar populations \citep[e.g.,][]{Cignoni2011}, including low-mass young stellar objects \citep[YSOs; ][]{Simon2007} and more massive and evolved stars.
The 30 O-type stars with masses 35-100 M$_\odot$ \citep{Massey1989, Evans2006, Dufton2019}, the largest grouping in the SMC, power NGC~346's \ion{H}{2} region, dominating its radiative and mechanical feedback.
There is a range of stellar and interstellar environments within NGC~346 \citep{Hony2015}. Within its interstellar medium (ISM), a wide range of substructures is revealed in polycyclic aromatic hydrocarbon (PAH) emission (8 $\mathrm{\mu}$m), warm dust (24 $\mathrm{\mu}$m), and molecular gas \citep[CO J=2-1; ][]{Rubio2000, Contursi2000, Hony2015}.  \newline
\indent The region possesses a complex distribution of hierarchically-linked star clusters of varying complexities and ages, distributed across the field \citep{Sabbi2008, Hennekemper2008, Gouliermis2014}.
The youngest populations have estimated ages from 0.01--0.05 Myr \citep{Rubio2018} to 1--3 Myr \citep{Bouret2003, Sabbi2007, DeMarchi2011, Dufton2019}. \citet{Sabbi2007} identified sixteen sub-clusters within the larger NGC~346 structure.
A separate, intermediate-age star cluster, BS~90 \citep{Bica1995}, is located in the north and was found to have an age of 4.3~($\pm$~0.1)~Gyr \citep{Sabbi2007}. 
Past studies of NGC~346 in both the optical and infrared (IR) revealed young stellar populations, including thousands of pre-main-sequence (pre-MS) stars discovered by \citet{Nota2006} using Hubble Space Telescope (\hst) imaging.
These pre-MS stars have masses as low as 0.6--3 M$_\odot$ \citep{Sabbi2007, Hennekemper2008, DeMarchi2011} and are scattered throughout the dusty filaments of NGC~346.
IR surveys of NGC~346 with \spitzer\ and \herschel\ discovered $\sim$100 YSO candidates at early stages in their evolution \citep{Simon2007, Bolatto2007, Gordon2011, Meixner2013, Rubio2018}.
These YSOs have typical masses of 8 M$_\odot$ with some as low as 1.5 M$_\odot$, and are estimated to have formed in the last $\sim$1 Myr \citep{Simon2007, Sewilo2013, Seale2014}.
Using these sources, \citet{Simon2007} estimated a star formation rate of $>3.2\times10^{-3}$ M$_\odot$ yr$\mathrm{^{-1}}$ for NGC~346. \newline
Using \hst\ proper motions \citep{Sabbi2022} and \textit{VLT/MUSE} radial velocities \citep{Zeidler2022}, it was shown that the stars in NGC~346 move along a wide spiral.
The clusters of pre-MS stars and YSOs appear primarily located within the areas where the coherent motion field shows significant changes, suggesting that turbulence is driving star formation in NGC~346.\newline
\indent Thanks to the point-source sensitivity and high spatial resolution of the \jwst, we can for the first time detect and spatially resolve subsolar-mass pre-MS stars and YSOs in the near- and mid-IR (1-28 $\mu$m) in nearby galaxies. 
Initial results from Near Infrared Camera \citep[NIRCam;][]{Rieke2005, Rieke2023}, reported in \citet{Jones2023} and combined NIRCam and Mid-Infrared Instrument \citep[MIRI;][]{Rieke2015, Wright2023} observations reported in \citet{Habel2024} identified sub-solar mass YSOs in NGC~346.
Using aperture photometry and NIRCam colour-magnitude diagrams (CMDs) and the F115W-F187N vs F200W-F444W colour-colour-diagram (CCD), \citet{Jones2023} found 136 YSOs without Pa$\mathrm{\alpha}$, 216 YSOs with Pa$\mathrm{\alpha}$ and 179 pre-MS stars.
\citet{Habel2024} performed PSF-weighted photometry on the NIRCam filters (except F200W) and followed a similar, albeit stricter approach than \citet{Jones2023}, characterising 19 YSOs without Pa$\mathrm{\alpha}$, 163 YSOs with Pa$\mathrm{\alpha}$ and 14 pre-MS stars.
In addition, \citet{Habel2024} performed aperture photometry on the MIRI filters and utilised various MIRI CMDs to select sources with an IR-excess in the mid-IR. 
Using this selection and performing spectral energy distribution (SED) fitting, they found 23 MIRI YSO candidates with masses ranging between 0.95 and 4.15~\msun.
This work builds upon these previous studies by implementing updated data processing and source extraction for both NIRCam and MIRI data, creating a photometric catalogue including all characterised pre-MS stars and YSOs, in contrast to \citet{Jones2023, Habel2024} who focused on a fraction of their characterised pre-MS stars and YSOs to identify targets for spectroscopic follow-up observations. \newline
\indent We perform PSF-weighted photometry on all available filters and classify detected sources, providing the most comprehensive source catalogue for NGC~346 in the near- and mid-IR to date.
Our paper is arranged as follows: in Sect.~\ref{sec:observations} we describe our observations and data reduction procedures, in Sect.~\ref{sec:stellar_and_diffuse} we give an overview of the stellar and diffuse emission in the field, in Sect.~\ref{sec:photometric_analysis} we detail our photometric analysis, and in Sect.~\ref{sec:classification} we explain the classification of different populations.
We give our results and discussion in Sect.~\ref{sec:results}, before we summarise our findings in Sect.~\ref{sec:conclusion}.

\section{Observations and data processing} \label{sec:observations}
The \jwst\ observations used in this work were obtained as part of guaranteed time (GTO) programme \#1227 (PI: M. Meixner). These observations are described in detail in \citet{Jones2023} and \citet{Habel2024}.
For convenience, the instrument configurations are summarised in Tables \ref{tab:NIRCam_obs_param} and \ref{tab:MIRI_obs_param} in Appendix~\ref{sec:appendix}.
The NIRCam mosaic was constructed from three observations with the A and B modules with NIRCam as the primary instrument, and a fourth deeper observation by NIRCam taken as a parallel to the program's MIRI imaging \citep[see Fig.~1 in][]{Habel2024}.
The variation in exposure time around the NIRCam mosaic means that the limiting magnitude of sources in parts of the mosaic are higher than in others, requiring tailored source detection, photometry, and background treatment that are region specific.
Given the updates to the \jwst\ Calibration Pipeline and reference files since \citet{Habel2024}, we reprocessed our data so that these calibrations are applied.
Again, the data reduction and analysis procedures are described in \citet{Jones2023} and \citet{Habel2024}, but we give a brief overview here.

\subsection{NIRCam} \label{subsec:NIRCam_processing}
The uncalibrated NIRCam dithers were processed using \jwst\ pipeline version 1.12.3 and Calibration Reference Data System (CRDS) version 11.17.14 and context `jwst\_1135.pmap'.
Default parameters were used for both the \texttt{Detector1Pipeline} and \texttt{Image2Pipeline} stages.
The calibrated dithers were then corrected for 1/f noise \citep{Schlawin2020} by using the \texttt{image1overf.py} tool from \citet{Willot2022}. 
We aligned the resulting dithers relative to each other using the \jwst/Hubble Alignment Tool \citep[JHAT][]{Rest2023} and then to \textit{Gaia} Data Release 3\footnote{\url{https://www.cosmos.esa.int/web/gaia/dr3}} \citep[DR3,][]{Gaia2021} using JHAT.
In the case of the long wavelength channels, we used counterparts to our sources from the \gaia\ DR3 aligned \hst\ catalogue of \citet{DeMarchi2011}. As described in \citet{Habel2024}, this was not possible for the short wavelength channels as several of the individual dithers at the extremities of the mosaic did not overlap with the \hst\ catalogue. \citet{Habel2024} aligned the short wavelength channels to \gaia\ DR3 by bootstrapping the \gaia\ DR3 alignment to sources detected in the F277W mosaic. We made use of the resulting source coordinates to align our reprocessed images. We combined the resulting images to create the NIRCam mosaics using \texttt{Image3Pipeline} with the \texttt{tweakreg} step disabled, given the images were already aligned. Examples of our NIRCam mosaics are shown in Fig.~\ref{fig:4_filter_images}, top row.


\subsection{MIRI} \label{subsec:MIRI_processing}
For our MIRI reprocessing, we used \jwst\ pipeline version 1.13.4 with CRDS version 11.17.14 and context `jwst\_1185.pmap' to create calibrated MIRI images and the mosaics. We aligned the resulting images to \gaia\ DR3 using the \texttt{tweakreg} step in \texttt{Image3Pipeline} and combined them into mosaics. Given the lower spatial resolution and pixel scale of MIRI \citep{Wright2023}, image alignment using \texttt{tweakreg} is sufficient. Examples of our MIRI mosaics are shown in Fig.~\ref{fig:4_filter_images}, bottom row.

\section{Stellar and diffuse emission} \label{sec:stellar_and_diffuse}
\begin{deluxetable}{lrrl}
\tablewidth{0pt}
\tablecaption{Information for the NIRCam and MIRI filters used in this work. \footnote{The table is adapted from \url{https://jwst-docs.stsci.edu/\#gsc.tab=0}} \label{tab:filters}}
\tablehead{
\colhead{Filter} & \colhead{Pivot} & \colhead{Bandwidth} & \colhead{Use} \\
\colhead{ } & \colhead{$\lambda$ (\micron)} & \colhead{$\Delta \lambda$\ (\micron)} &\colhead{}
}
\startdata
F115W & 1.154 & 0.225 & Stellar continuum \\
F187N & 1.874 & 0.024 & Pa$\alpha$ \\
F200W & 1.988 & 0.461 & Continuum (stellar/dust) \\
F277W & 2.776 & 0.672 & Continuum (stellar/dust) \\
F335M & 3.362 & 0.348 & PAHs, $\mathrm{CH_4}$ \\
F444W & 4.402 & 1.024 & Dust ($\sim$650~K) continuum \\
F770W & 7.639 & 1.95 & PAHs \\
F1000W & 9.953 & 1.80 & Silicates \\
F1130W & 11.309 & 0.73 & PAHs \\
F1500W & 15.064 & 2.92 & Dust ($\sim$200~K) continuum \\
F2100W & 20.795 & 4.58 & Dust ($\sim$140~K) continuum \\
\enddata
\end{deluxetable}

\begin{figure*}
    \centering
    \includegraphics[width=0.9\linewidth]{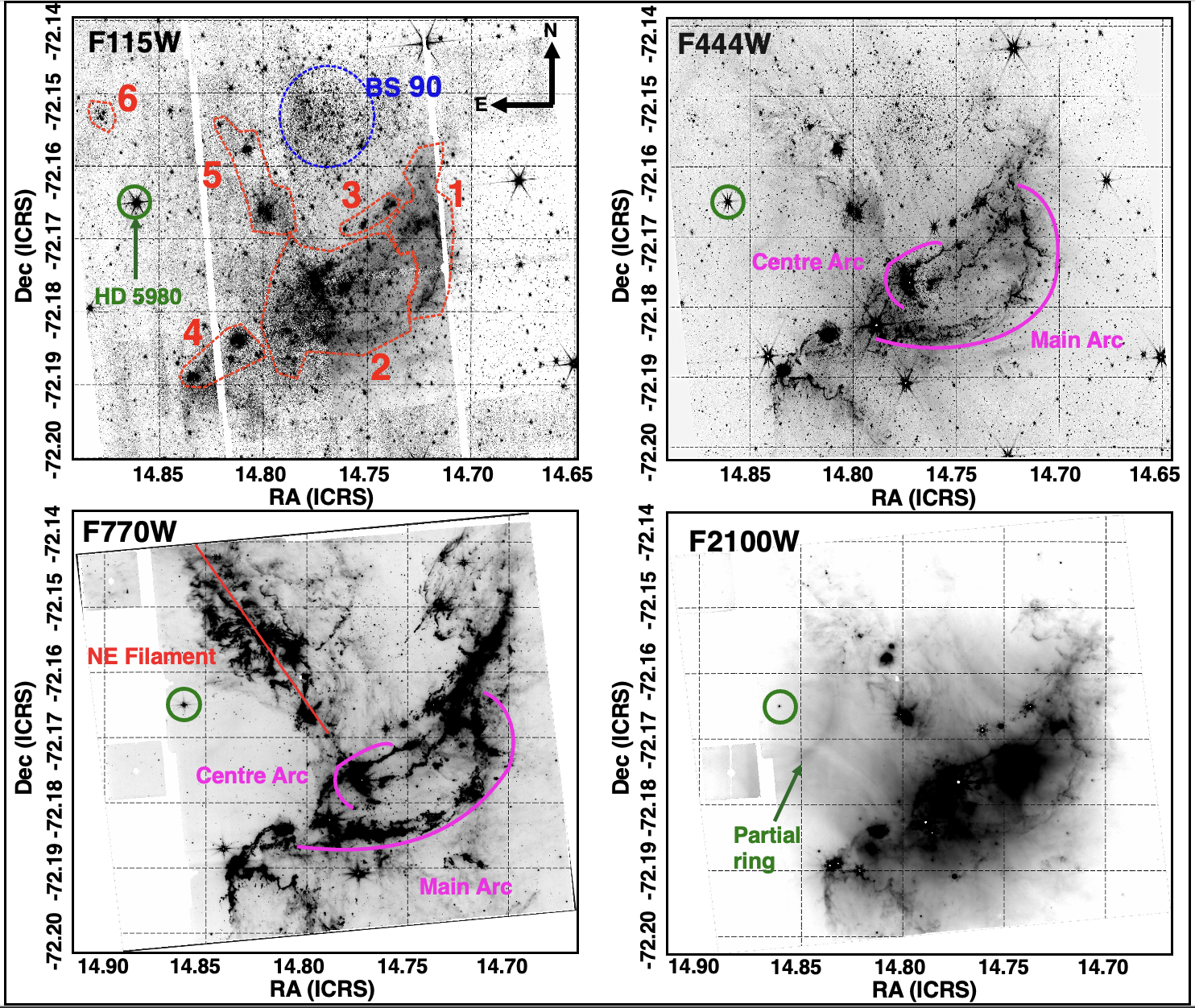}
    \caption{Examples of our NGC~346 mosaics with significant structures and regions indicated.
    {\it Top left}: NIRCam F115W filter, highlighting the stellar population. The blue circle indicates the location of the intermediate-age star cluster BS~90 \citep{Bica1995}. Our region definitions (see Sect. \ref{sec:stellar_and_diffuse}), are shown by the six red polygons. These contain all sixteen of the sub-clusters identified by \citet{Sabbi2007} (see section \ref{subsec:stellar_content} for details).
    {\it Top right}: NIRCam F444W filter. The filamentary structures of NGC~346 are evident. The two most prominent, the Centre Arc and Main Arc, are indicated in magenta. The diffuse emission structures trace warm dust (T$\sim$650~K).
    {\it Bottom left}: MIRI F770W filter. There are significant diffuse structures, indicating the location of PAHs. The Main arc, Centre Arc, and northeast filament are shown.
    {\it Bottom right}: MIRI F2100W filter. The emission is much more extended. The emission traces warm dust (T$\sim$140~K), which is brightest in the arcs but also suffuses into lower density regions to the northeast and southwest. The partial ring centred on the Wolf-Rayet star HD 5980 (circled in green in each panel) is indicated with a green arrow.}
    \label{fig:4_filter_images}
\end{figure*}

Four of our \jwst\ mosaics are shown in grey-scale in Fig. \ref{fig:4_filter_images}, each highlighting a different aspect of the complex field of NGC~346. Table \ref{tab:filters} lists the primary origins of emission in each filter. In the following, we describe the stellar content and diffuse structures traced by our observations.

\subsection{Stellar content}\label{subsec:stellar_content}
The NIRCam F115W filter best traces the stellar content in the NGC~346 region, shown in Fig.~\ref{fig:4_filter_images}, top left.
The image is dominated by stars.
The intermediate-age star cluster BS~90 \citep{Bica1995} is located in the north of the image and indicated in blue in Fig.~\ref{fig:4_filter_images}.
The 16 sub-clusters identified by \citet{Sabbi2007}, named Sc-1 to Sc-16, are located throughout the image.
Rather than show each individually, we define six polygonal regions through NGC~346 (shown in red in Fig.~\ref{fig:4_filter_images}, top left), which contain these 16 clusters.
Many of the sub-clusters are embedded in dust and nebulosity, indicating that they are sites of recent or possibly ongoing star formation.
\citet{Sabbi2007} found the ages of sub-clusters Sc-1 to Sc-15, located in regions 2--5, to be $\sim$3--4$(\pm$1) Myr.
Sub-cluster Sc-16, located in region 6, is located to the northeast of the main body of NGC~346.
This sub-cluster was found to be older than the others with an age of $\sim$15$(\pm$2.5) Myr \citep{Sabbi2007}.
The brightest and most massive stars ($\sim$50--60 $\mathrm{M_\odot}$) are located within the centre of region 2.
The sub-clusters in region 5 are located in the northeast (NE) filament (see Fig.~\ref{fig:4_filter_images}, bottom left), and predominantly comprise low-mass pre-MS stars \citep{Sabbi2007}.
These are co-located with a dense clump of molecular CO gas as observed by \citet{Rubio2000}.

\subsection{Warm dust}\label{subsec:warm_dust}
The NIRCam F444W filter, shown in Fig. \ref{fig:4_filter_images} - top right, is dominated by the emission from warm dust (T$\sim$650~K), showing the complex filamentary structure of NGC~346, especially the Main Arc \citep{DeMarchi2011b, Jones2023}.
The BS~90 cluster stars are still visible in the north, while the NGC~346 sub-clusters are less apparent since they are embedded in the diffuse emission.
The Main and Centre Arcs \citep{Jones2023} of NGC~346 are clearly visible, indicated in magenta in Fig \ref{fig:4_filter_images} - top right). \newline
\indent The MIRI F2100W filter, shown in Fig \ref{fig:4_filter_images} - bottom right, also traces the distribution of warm dust (T$\sim$140~K).
The intermediate-age cluster BS~90 is not detected at this wavelength.
Emission from the younger sub-clusters is evident along the Main and Centre Arcs.
The diffuse emission is also brightest in these regions, given the higher density and radiation field from the most massive stars.
Above and below the arcs, there are extended faint filamentary structures to the northeast and south-west.
Their morphology suggests they are likely the result of dust being entrained in the powerful cluster winds emanating from the massive stellar population \citep[e.g. ][]{Rogers2013}.

In the F2100W image, a faint, partial ring is located to the east of the NE filament, stretching from (-72.174; 14.85) to (-72.168; 14.83). This is centred on the Wolf-Rayet star HD~5980 and appears to delineate the wind interaction region between the cluster and HD~5980 outflows.

\subsection{PAHs} \label{subsec:PAHs}
The MIRI F770W filter is sensitive to 7.7\micron\ PAH emission, which in this case is likely produced by heating of the PAHs by the UV radiation field of the massive stars.
The image is shown in Fig~\ref{fig:4_filter_images}, bottom left.
Both the Main and Centre Arcs are bright in PAH emission.
Another very prominent feature is the NE filament, perpendicular to the main body of NGC~346.
This structure extends further than in the \hst\ data \citep{Sabbi2007}.

\section{Photometric analysis}\label{sec:photometric_analysis}
\subsection{Source detection and photometry }\label{subsec:detection_photometry}
We used the \texttt{STARBUGII} tool \citep[version 0.7.5, ][]{Starbug, Nally2024} for source detection and photometry of the NIRCam and MIRI images.
Developed specifically for performing PSF photometry in crowded fields with complex diffuse emission, \texttt{STARBUGII} supports both aperture and PSF photometry and is capable of band-matching individual filter catalogues into a unified multi-wavelength catalogue.

\subsubsection{NIRCam photometry} \label{subsubsection_NIRCam_photometry}
We performed the source detection with \texttt{STARBUGII} on the NIRCam mosaics, setting the parameter \texttt{SIGSRC} to 4.0 or 5.0 for the medium and wide filters, and to 4.0 or 10.0 for the F187N filter. 
Since each NIRCam mosaic consists of areas with different exposure times \citep[Fig 1 in][]{Habel2024}, we used different values for this \texttt{SIGSRC} parameter based on the ``WHT'' extension of the mosaic FITS files.
This is a 2-D weight image giving the relative weight of the output pixels (see Fig. \ref{fig:weight_image}).
Pixels with a deeper exposure are given a higher weight than pixels with a shallower exposure.
\texttt{SIGSRC} was adapted to specific regions of the mosaic accordingly (see Table \ref{tab:source_detection_params}).

\texttt{STARBUGII} calculates the geometry of each point source and assigns \textit{sharpness}, \textit{roundness}, and \textit{smoothness} values.
By setting upper and lower limits on each of these values, we remove sources such as cosmic rays and resolved background galaxies.
All \texttt{STARBUGII} values used in this work can be found in Table \ref{tab:source_detection_params}.

Before starting the PSF weighted photometry, we initially ran aperture photometry on the detected sources.
For all the NIRCam filters, we applied \texttt{STARBUGII}'s default aperture radius of 1.5 pixels with an inner sky annulus of 3.0 pixels and an outer sky annulus of 4.5 pixels, with aperture correction factors from the pipeline reference file \texttt{jwst\_nircam\_apcorr\_0004.fits}.
We note that we used the aperture photometry only to retrieve the centroids of the stars as a starting point for PSF-weighted photometry and to set a broad cut-off value to ensure that the sources with the poorest signal-to-noise were removed.
We excluded any source with an uncertainty in magnitude, larger than 2 in every individual filter.
Including these sources would cause this PSF-weighted photometry processing to take an excessive amount of time for little gain.

The resulting catalogues for both the areas with a deeper and shallower exposure are then combined into a single catalogue using \texttt{TOPCAT} \citep[version 4.8-8, ][]{TOPCAT}.
The matching radius per filter is shown as \textit{MATCH\_THRESH} in Table \ref{tab:source_detection_params}.
The resulting aperture catalogue is used as input for PSF photometry, which was performed on the individual dithers.
We used the background tool in \texttt{STARBUGII} with default parameters to model the diffuse emission present in the image, by masking the detected sources and creating an image of only the nebulous emission.
This representation of the nebulous emission was subtracted before PSFs were fitted.
This method is described in detail in \citet{Nally2024}.

Using \texttt{STARBUGII}, we generated a $5\arcsec$ radius PSF from \texttt{WEBBPSF} \citep[version 1.2.1, ][]{Perrin2014} for each detector subarray in NIRCam.
These PSFs were fitted on the centroid positions given by the aperture photometry catalogue, while allowing to fit a new centroid position within $0\farcs1$.
If the routine cannot fit a new centroid position, it is force-fitted on the initial position.
This creates individual dither PSF-weighted catalogues that are then cross-matched using \texttt{STARBUGII} generic matching, resulting in a PSF-weighted catalogue for each filter.
We cleaned these catalogues by excluding sources with an uncertainty in the PSF-weighted magnitude larger than 0.70, which was chosen to retain fainter sources.
In Section \ref{subsubsec:multi-filter}, we discuss the implications of choosing this magnitude uncertainty cut-off. 

\subsubsection{MIRI photometry} \label{subsubsection_MIRI_photometry}
We followed a similar procedure for the MIRI photometry as for NIRCam, with some minor changes.
The MIRI mosaics do not have areas with different exposure times, so only a single value for \texttt{SIGSRC} was required.
We ran source detection twice per filter, changing the value for the ``\texttt{RICKER\_R}'' parameter between the two runs.
This parameter sets the radius for the Ricker wavelet used in the convolution routine, making source detection more sensitive in regions with more extended diffuse emission.
The values used for ``\texttt{RICKER\_R}'', the parameters for applying geometric limits on the \textit{sharpness}, \textit{roundness} and \textit{smoothness} of the sources and the values for the aperture photometry parameters are given in Table \ref{tab:source_detection_params}.
The two aperture catalogues are matched together using \texttt{TOPCAT} \citep{TOPCAT}.
The matching radius used per filter are shown in Table \ref{tab:source_detection_params} as \texttt{MATCH\_THRESH}.

\begin{figure*}\centering
\subfloat[Contours based on the NIRCam shallow exposure area `backgrounds' (white boxes).]{\label{fig:density_map_1}\includegraphics[width=.48\linewidth]{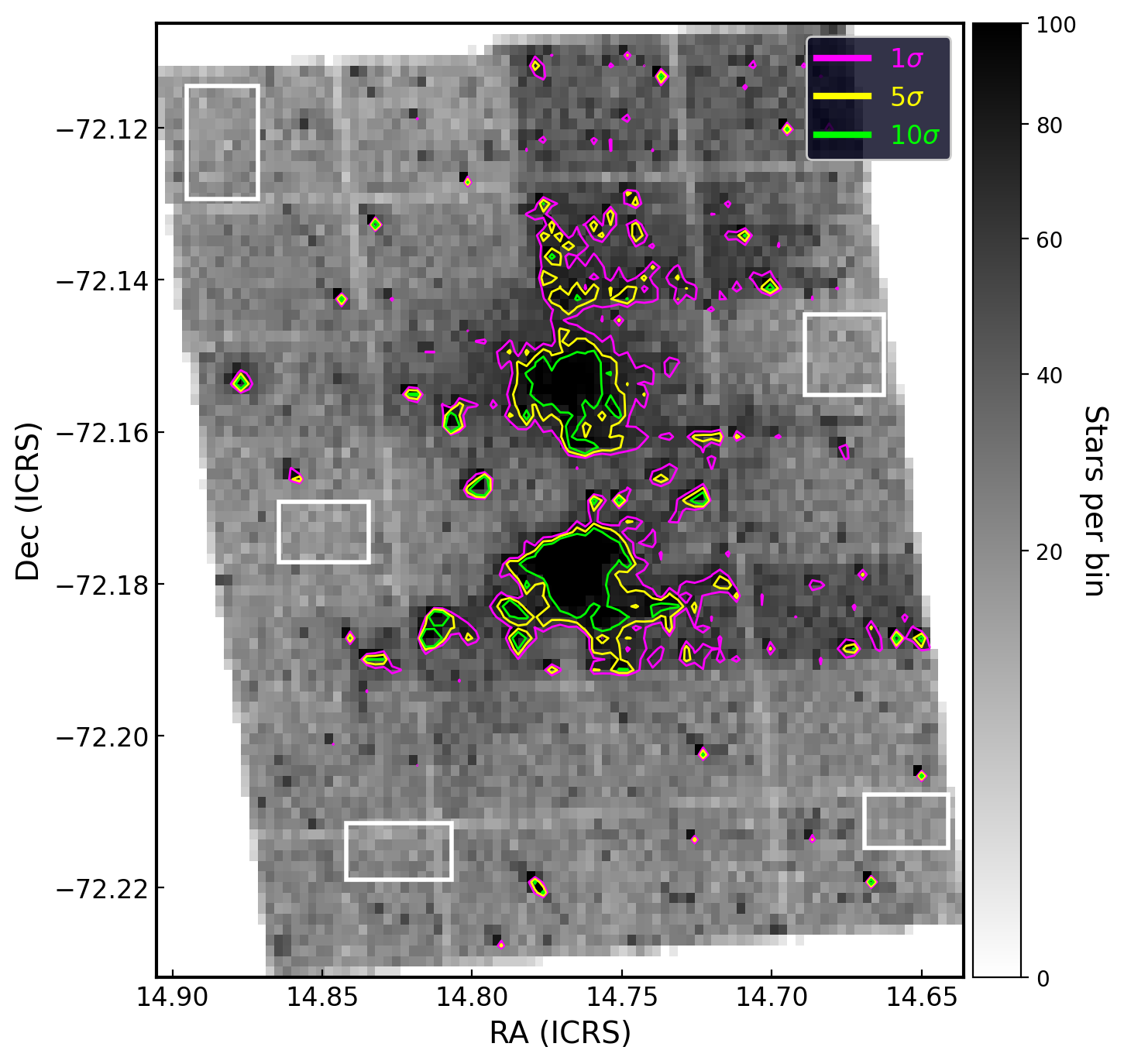}} \hfill
\subfloat[Contours based on the NIRCam deep exposure area `backgrounds' (white boxes).]{\label{fig:density_map_2}\includegraphics[width=.48\linewidth]{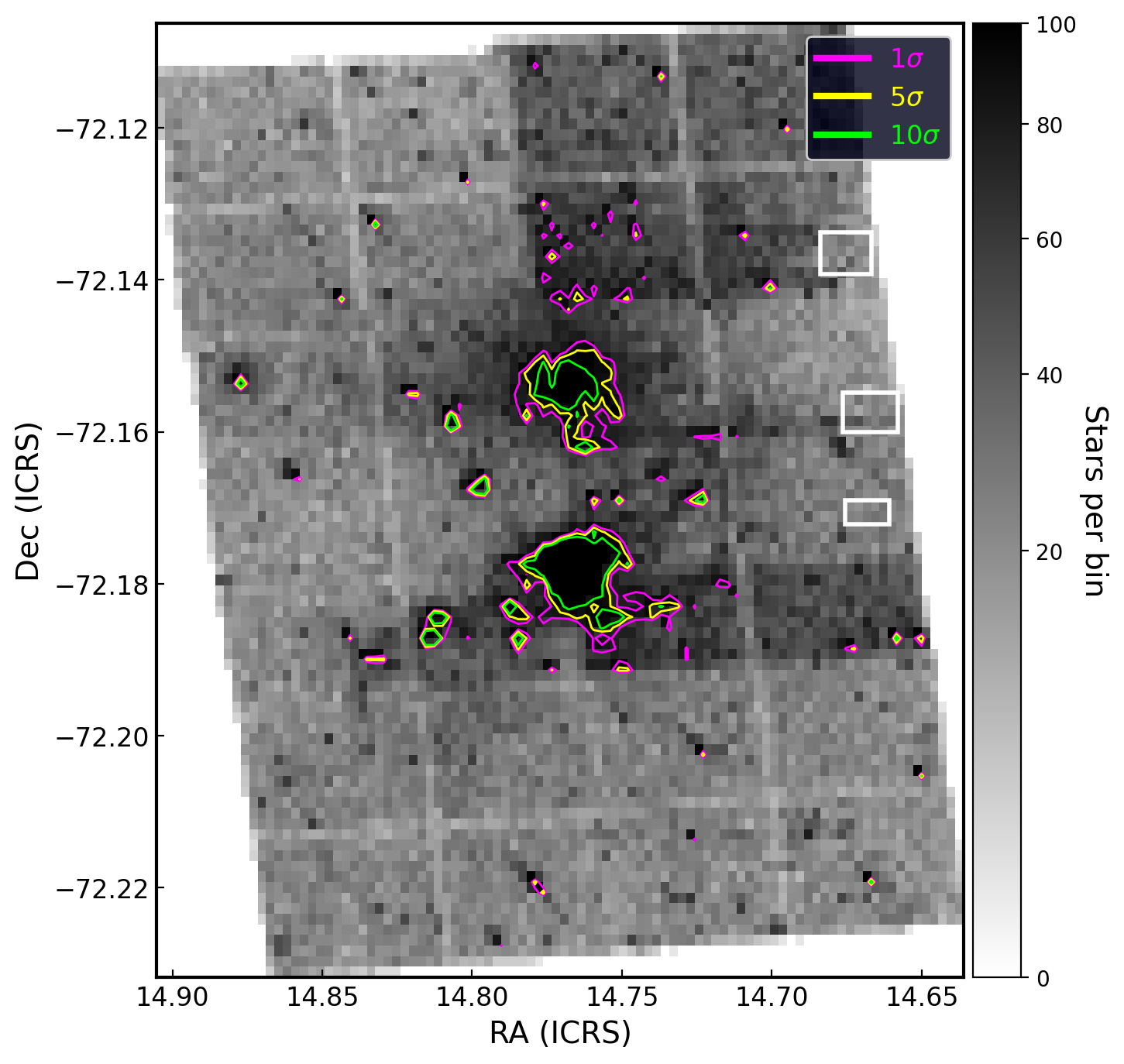}} \par 
\subfloat[Contours based on the MIRI `backgrounds' (white boxes).]{\label{fig:density_map_3}\includegraphics[width=.48\linewidth]{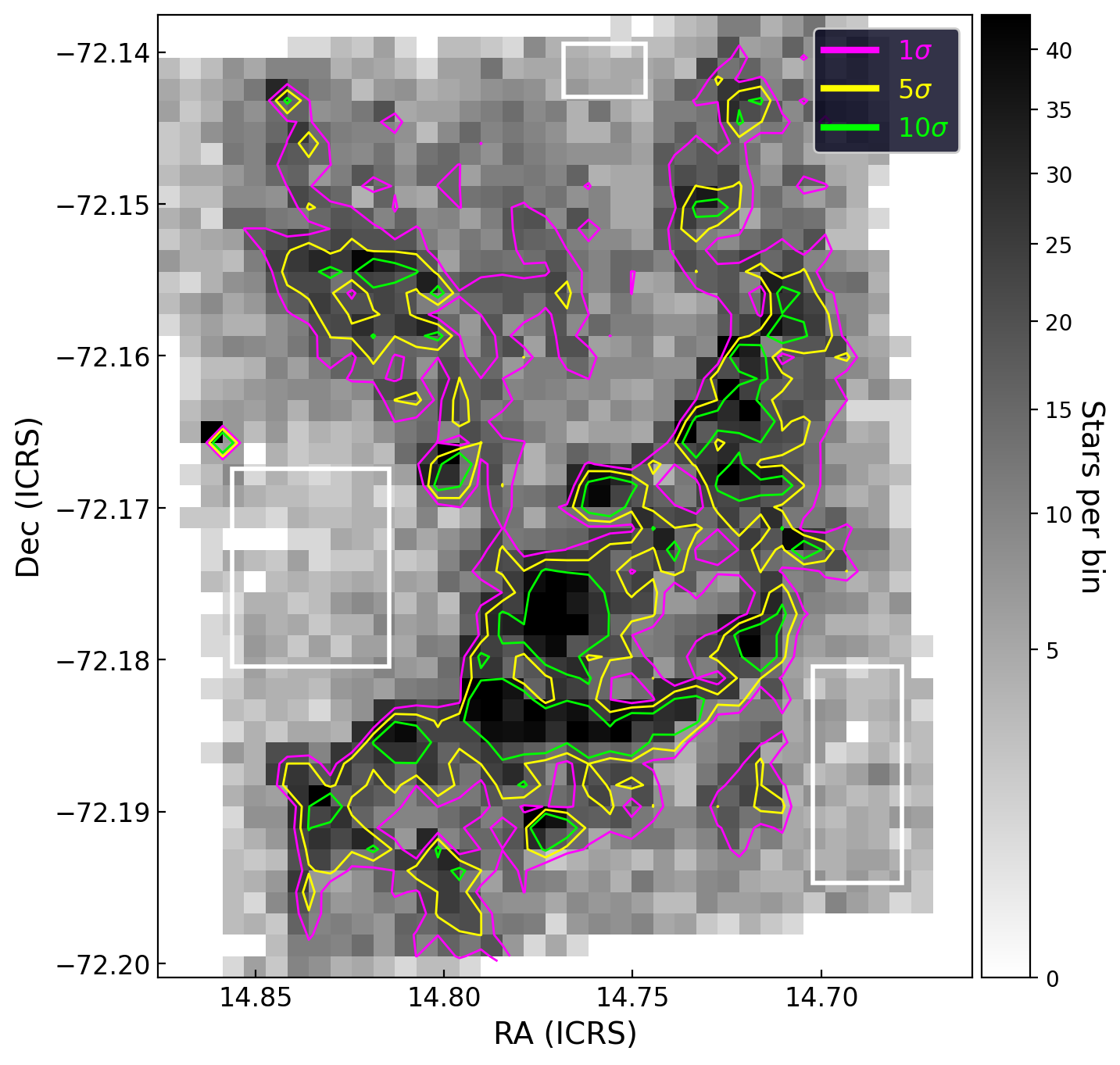}}
\caption{Source density maps of NIRCam and MIRI.
    The magenta, yellow, and green contours are respectively 1$\mathrm{\sigma}$, 5$\mathrm{\sigma}$ and 10$\mathrm{\sigma}$ above the mean of the field population density, derived from the `backgrounds'.}
\label{fig:density_map}
\end{figure*}

\begin{figure*}\centering
\subfloat[ Luminosity functions of the full catalogue.]{\label{fig:completeness_whole}\includegraphics[width=.5\linewidth]{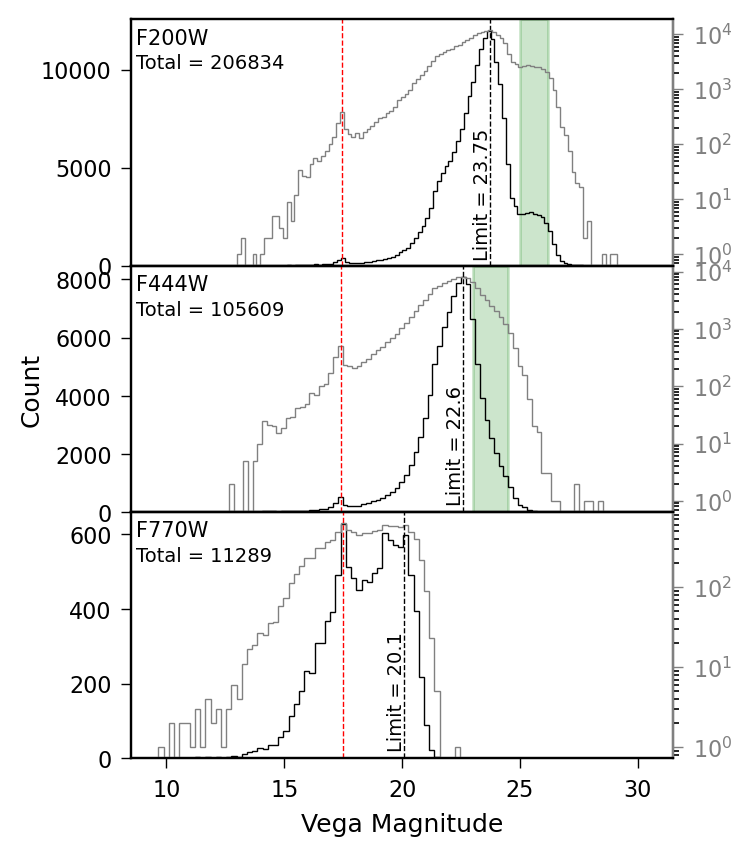}}\hfill
\subfloat[Luminosity functions of the NGC~346 regions combined (see Fig.~\ref{fig:4_filter_images} - top left panel).]{\label{fig:completeness_NGC346}\includegraphics[width=.5\linewidth]{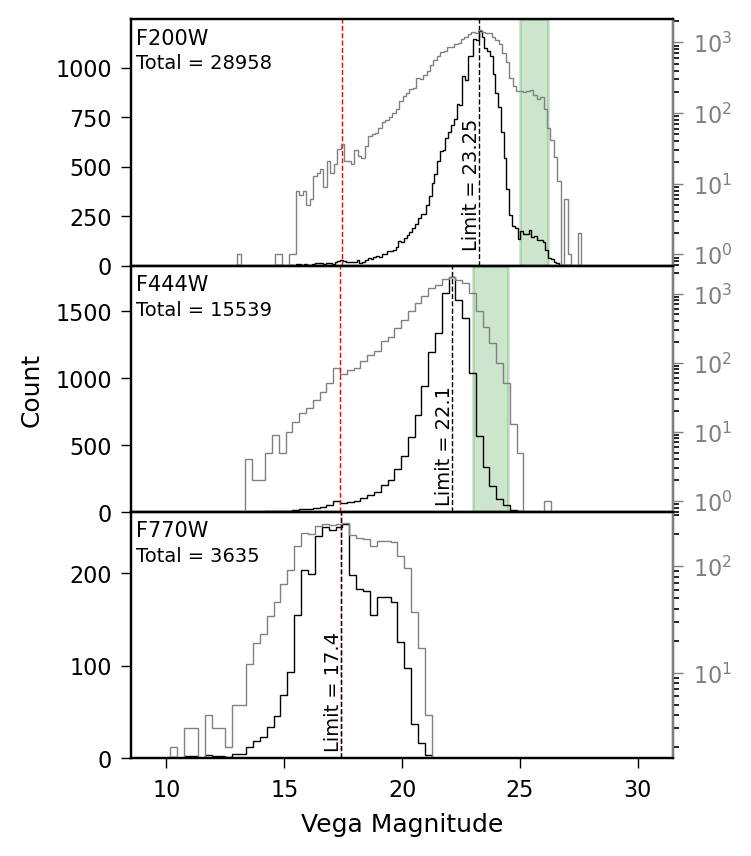}} \par 
\caption{Luminosity functions of sources detected in the full catalogue (panel a) and the NGC~346 regions combined (panel b) in 3 filters: F200W (top), F444W (middle) and F770W (bottom).
The completeness limit of each filter is shown with a dashed black line.
The location of the red clump (around 17.45 Vega mag) in these filters is shown with a red dashed line.
The distribution of the sources is shown in both linear and logarithmic scales, in black and gray, respectively.
The F200W and F444W luminosity functions also show a green area, indicating the slower drop-off in the number of sources at those magnitudes due to the areas with deeper exposure in the NIRCam mosaics.}
\label{fig:completeness}
\end{figure*}

Before performing the PSF photometry for MIRI, we masked all the sources located in the Lyot coronograph field, as sources around the Lyot stop led to spurious detections. 
The PSF photometry for MIRI then followed the same steps as for NIRCam, adding one additional cleaning step.
The PSF weighted catalogue was cleaned of sources with an uncertainty on the magnitude larger than 0.70, chosen to retain fainter sources.
In addition, we cleaned sources containing any flag given during the \texttt{STARBUGII} PSF photometry, for example, sources that have an asymmetric flux distribution between the exposures of a filter (see \citealt{Nally2024} for more details).

\subsubsection{Multi-filter detections} \label{subsubsec:multi-filter}
The PSF-weighted catalogues of NIRCam and MIRI were cross-matched into a single, multi-wavelength catalogue using \texttt{STARBUGII} band-matching.
The matching radii used for the cross-matching are identical to the values for \texttt{MATCH\_THRESH} given in Table \ref{tab:source_detection_params}.
The multi-wavelength source catalogue incorporates detections that satisfied a cross-matching criterion across multiple photometric bands, as well as single-band detections.
To ensure a robust sample for physical property characterisation, we excluded all single-band detections.
This exclusion criterion was specifically applied because the primary objective of this investigation is to derive physical parameters for sources with confirmed photometric data across the spectrum. 
This requirement simultaneously rejects the very faintest sources and mitigates a majority of spurious detections (e.g., those found in the PSF wings of bright objects) since their inconsistent centroids across different filters prevent successful cross-matching.
The final multi-wavelength photometric catalogue contains 249,519 sources, an increase of $\sim$23\% from \citet{Habel2024}. This increase can be attributed to performing the source detection on the mosaics instead of the individual dithers, which allows for the detection of dimmer sources.
Additionally, setting the cut-off in magnitude uncertainty to 0.70 instead of using a signal-to-noise of 10 results in more sources.
Finally, the quality of our data products has slightly improved due to the updates of the \jwst\ Calibration Pipeline and reference files.

\subsubsection{Implications of magnitude uncertainty cut-off}\label{subsubsec:consequences}
We chose to apply a photometric uncertainty cut-off of 0.70 mag for the PSF photometry in all filters.
This was chosen to retain fainter sources, which leaves the opportunity to attempt to classify sub-solar mass YSOs and pre-MS stars.
The best of these candidates could potentially be targets for deeper follow-up studies.
For comparison, when choosing a more stringent cut-off in the photometric uncertainty of 0.30 mag, the final catalogue would consist of 184,565 sources, which is $\sim$74\% of the full catalogue presented in this work.
Applying this stricter photometric uncertainty cut-off in the NIRCam medium and wide filters would result in keeping $\sim$59-69\% of the sources, while for F187N only $\sim$41.4\% of the sources remain.
The number of sources kept in the MIRI bands with a 0.30 mag uncertainty cut-off would range between $\sim$92-99\%.
This shows that utilising an uncertainty cut-off of 0.70 mag in PSF photometry is already strict for MIRI, as opposed to NIRCam.
The large difference in the number of sources with the different cut-offs in the NIRCam bands can partly be attributed to the sources that received a quality flag other than `0' (clean detection) during the PSF photometry step of STARBUGII \citep{Starbug}. These sources account for slightly less than half of the $\sim$~31~–~42\% difference between a cut-off of 0.30 or 0.70 in the NIRCam medium and wide filters. In the F187N NIRCam filter, these sources only account for $\sim$4\% of the difference, due to the larger uncertainties of the sources in the low exposure areas of the F187N mosaic.


\subsection{Contamination} \label{subsec_contamination}
\subsubsection{Foreground stars}\label{subsubsec:foreground}
To remove foreground stars, we used the \textit{Gaia} DR3 catalogue.
We first cleaned the \gaia\ catalogue by removing sources with poor astrometry and possible non-single objects, as explained by \citet{Fabricius2021} and \citet{Nally2024}.
We matched these sources with our NGC~346 photometric catalogue, using a matching radius of $0\farcs 2$, resulting in 541 matches.
For each of these, we created SEDs with the combined photometric data of \gaia\ and \jwst\ to identify sources that are mismatched between the two catalogues.
If the flux of the source in the `$\mathrm{G_{rp}}$' \gaia\ filter was at least five times higher than the flux of the source in the F115W and F200W NIRCam filters, it was considered a mismatch.
For sources without a detection in F115W (e.g. due to saturation), we only used the F200W flux to determine a mismatch.
To determine if the sources were actual foreground sources, we utilised the distance data based on the low-resolution BP/RP spectra available in the \gaia\ catalogue, called ``distance\_gspphot''.
We chose to limit the distance of possible foreground sources to 5~kpc, as distances for stars at high galactic altitude is accurate up to roughly 5~kpc \citep[see Fig. 7 - bottom left panel in][]{Bailer-Jones2021}.
326 foreground sources remained after excluding those with mismatches and ensuring a distance within 5~kpc, which were all removed from our catalogue.
We note that the highest magnitude of the cleaned \gaia\ sources was $\sim$20 in Vega mag.
We therefore cannot exclude that a number of faint foreground contaminants may still be present in our catalogue.
In addition, foreground sources with distances more than 5~kpc can also be present in our catalogue.




\subsubsection{Contaminating sources} \label{subsubsec:bkg_srcs}
Our photometric catalogue of NGC~346 is also contaminated with SMC field stars and background objects.
These cannot easily be separated from the NGC~346 sources.
Instead, we estimated the contamination due to these sources using `background' regions in the field of views (FOVs).
We created density maps of all sources detected in NIRCam and MIRI (Fig. \ref{fig:density_map}). 
We used the contours of the stellar density maps as guides for the regions we selected to represent the SMC field stars and background objects, choosing regions with the lowest stellar densities and furthest away from NGC~346.
To estimate the total number of contaminating sources per filter, we calculated the average number of sources per arcmin$^2$ in these regions and multiplied this by the full size of the mosaic.
The NIRCam mosaic consists of areas with different exposure times with the NGC~346 cluster covered by both.
The mosaicking of these areas created a too complex `field' of different exposure times to separate it into just shallow and deep (see Fig. \ref{fig:weight_image} in Appendix A).
Therefore, we had to choose whether to use the `background' of the shallower or deeper exposure areas for this work.
The number of contaminating sources per filter based on both are shown in Appendix A in Tables \ref{tab:background_NIRCam_2} and \ref{tab:background_NIRCam_1}, respectively.
The difference in the percentage of contaminating sources between the shallower and deeper `background' areas ranges from 7.8 - 22.5\% for the four NIRCam wideband filters.
This difference increases to 37.7\% for the F335M filter and to 93.4\% for the F187N filter.
The 93.4\% difference for the F187N filter comes from the fact that based on the deep exposure backgrounds, more sources would be expected in the whole field than actually detected in this filter.
In this work, we chose to calculate the contamination based on the deeper exposure `background' areas which is the most conservative and means we overestimate the number of contaminating sources. \newline
\indent We chose three regions outside the stellar density contours in the MIRI mosaic (see Fig. \ref{fig:density_map_3}) to represent the SMC field stars and background objects in the MIRI filters.
The number of contaminating sources per MIRI filter is shown in the appendix in Table \ref{tab:background_MIRI}.

\subsection{Completeness}
To assess the completeness of our catalogue, we employed luminosity functions, which are shown in Fig. \ref{fig:completeness} for the F200W, F444W, and F770W filters.
For each wavelength band, we used Knuth's Rule \citep{Knuth2006} to select the optimised bin widths.
We estimated the completeness limit per filter by locating the turnover of the luminosity functions at the faint end of our histograms within 0.05 mag accuracy.
In this work, we set this turnover in the luminosity function as the completeness limit, which is shown by the black dashed lines in Fig. \ref{fig:completeness} and also listed in Table \ref{tab:completeness}.
We note that our completeness limits for the NIRCam filters reach 0.35~-~1.05 mag and for the MIRI filters reach 1.20~-~1.95 mag lower than in \citet{Habel2024}.
There are three reasons for this.
First, we performed the source detection on the mosaics instead of the individual dithers, allowing us to detect fainter sources.
Second, we implemented a photometric uncertainty cut-off of 0.70 mag and not an S/N of 10. Finally, we used an updated version of the \jwst\ pipeline, creating higher-quality data products.


\begin{deluxetable*}{lrrc}
\tablewidth{0pt}
\tablecaption{NIRCam and MIRI full catalogue source count and completeness limit within 0.05 magnitude. \label{tab:completeness}}
\tablehead{
\colhead{Filter} & \colhead{Source} & \colhead{\% more} & \colhead{Completeness} \\
\colhead{ } & \colhead{Count} & \colhead{sources than} & \colhead{limit} \\
\colhead{ } & \colhead{ } & \colhead{\citet{Habel2024}} & \colhead{[VegaMag]}
}
\startdata
F115W & 201,533 & 53.3 & 24.80 \\
F187N & 27,974 & 106.3 & 20.65 \\
F200W & 206,834 & 24.9 & 23.75 \\
F277W & 173,918 & 98.5 & 23.55 \\
F335M & 111,091 & 109.7 & 22.60 \\
F444W & 105,609 & 204.3 & 22.60 \\
F770W & 11,289 & 256.9 & 20.10 \\
F1000W & 7,637 & 299.0 & 18.60 \\
F1130W & 6,312 & 439.5 & 16.95 \\
F1500W & 3,585 & 492.6 & 16.40 \\
F2100W & 993 & 280.5 & 13.50 \\
\enddata
\end{deluxetable*}

In the luminosity functions (see Fig. \ref{fig:completeness}) of the wavelengths between 1.15 and 10 $\mu$m, we detected evidence of the red clump (RC), which appears as a secondary peak at the bright end of the distributions at $\sim$17.45 mag (Fig. \ref{fig:completeness}).
In the F115W filter, the RC appeared at $\sim$18.45 mag.
We observed that the RC contributes a significant portion of the sources in the F770W photometry.
This results in a second peak separate from the peak where the completeness fall-off begins.
Since this `RC' peak is higher than the completeness fall-off peak, it could be argued that the `RC' peak should be used for the completeness limit, as fewer sources have been detected at higher magnitudes.
Therefore, using the peak in the luminosity function as our completeness limit is an approximate way to estimate the completeness, especially for the MIRI filters.

The difference in integration time across the NIRCam mosaic results in a slower fall-off on the fainter side of the luminosity function (Fig. \ref{fig:completeness}).
For the F444W filter, this is visible in the slower fall-off between approximately $\sim$23.0~-~24.4 mag and the faster fall-off after $\sim$24.4 mag, while in the F200W luminosity function a plateau is visible between approximately $\sim$25~-~26.2 mag (Fig. \ref{fig:completeness}, green area) after the first fall-off from 23.8 mag.
A greater number of fainter sources is detected in the areas on the mosaic with the deeper integration, resulting in these features in the NIRCam luminosity functions.

Fig. \ref{fig:completeness_NGC346} shows the luminosity functions of the NGC~346 polygon regions combined (see Fig. \ref{fig:4_filter_images} - top left) for the F200W, F444W and F770W filters.
These regions contain the dense gas and dust as shown in Fig \ref{fig:4_filter_images}, resulting in confusion between bright dust emission and faint sources, which leads to less faint sources detected, lowering the completeness limits for each filter.
Especially in the MIRI filters, the completeness limits differ strongly, because the majority of the faintest MIRI sources were detected on the outskirts of the main body of NGC~346.

\section{Source classification}\label{sec:classification}
\subsection{Extinction}\label{subsec:extinction}
Determining the extinction in the region is important for this work as it is used to set boundaries for source classification (see Sect. \ref{subsec:characterisation}).
The isochrones we used depend on the extinction.
In addition, the different populations can have different amounts of extinction, as described in detail below.

\subsubsection{Extinction law} \label{subsubsec:reddening}
\begin{figure}
    \centering
    \includegraphics[width=1.0\linewidth]{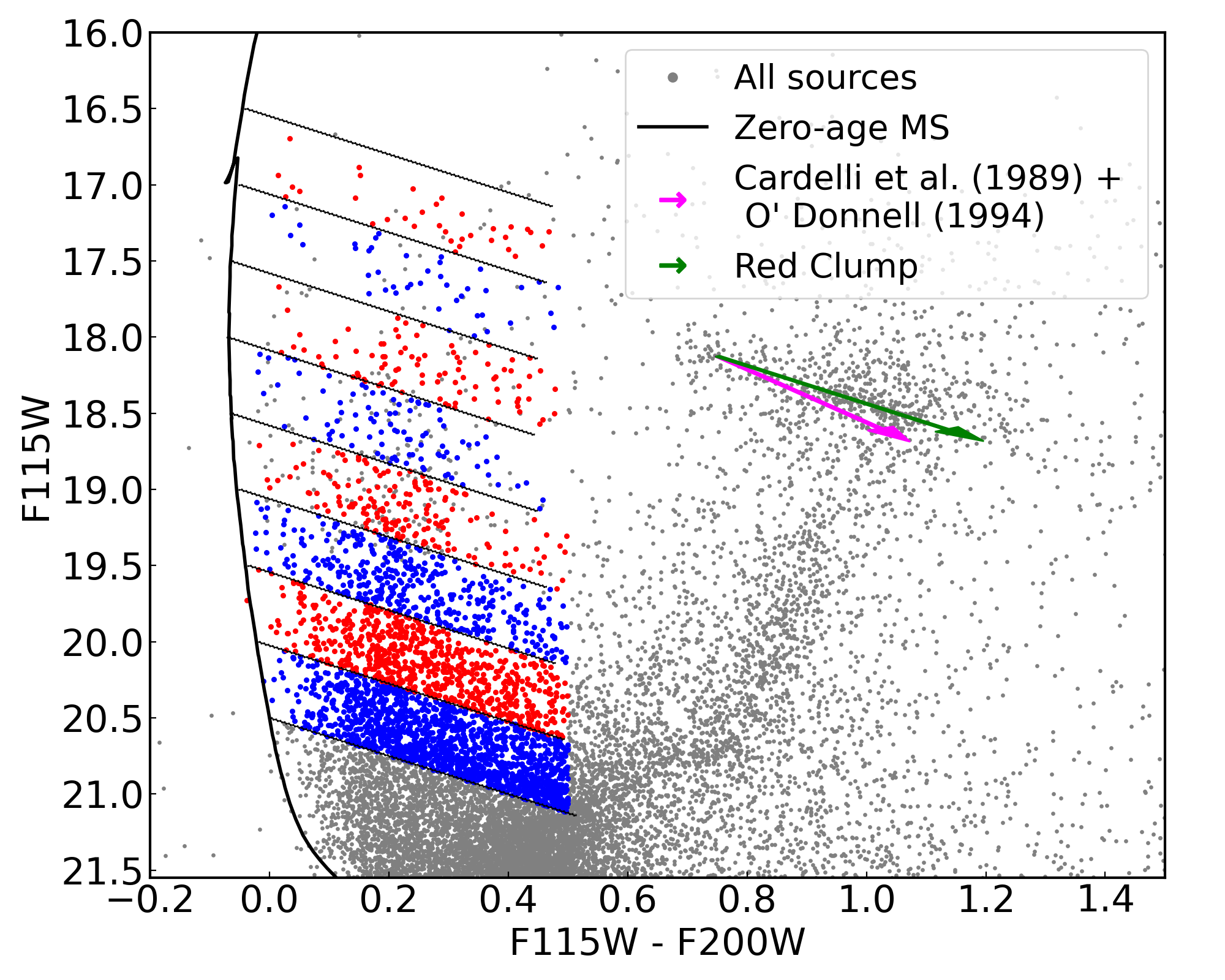}
    \caption{Observed F115W vs F115W-F200W CMD, focused on the upper part of the CMD.
    Two reddening vectors are shown with an $\mathrm{A_V}$ = 1.5 for illustration purposes, following the slope of the extinction curves of \citet{Cardelli1989, O'Donnell1994} for an $\mathrm{R_V}$=3.1 (magenta) and the slope of the RC (green).
    As described in Sect. \ref{subsubsec:extinction_populations}, a zero-age MS isochrone with $\mathrm{A_V}$ = 0, metallicity Z = 0.004 and distance modulus (m - M) = 18.96~($\pm$~0.035) is over-plotted in black.
    The slanted black lines show the reddening vectors for $\mathrm{A_V}$ = 2.0. They are drawn every 0.5 mag for stars brighter than F115W = 20.5 and show similar UMS stars that can have been displaced by reddening as blue and red dots. }
    \label{fig:extinction_arrows}
\end{figure}

Fig. \ref{fig:extinction_arrows} shows the F115W-F200W vs F115W CMD, zoomed in on the upper part with a magnitude of F115W $\mathrm{\leq}$ 21.5.
Two different reddening vectors are shown on top of the RC.
The pink reddening vector was created adopting the extinction curves from \citet{Cardelli1989, O'Donnell1994}, with $R_V$ = 3.1, whereas the green reddening vector follows the elongation of the RC.
During the RGB phase of low-mass stars, helium is deposited onto their degenerate core.
When a core mass of $\mathrm{M_{core}}$ = 0.48 $\mathrm{M_\odot}$ is reached \citep{Bildsten2012}, the electron degeneracy is broken in a helium flash \citep{Habing2003} and helium-burning ignites in the core.
Because the initial characteristics of these helium-burning cores are uniform, the stars are located in a tight knot on the CMD, creating the RC.
As the F115W-F200W vs F115W CMD in Fig. \ref{fig:extinction_arrows} shows, the RC has a slight elongation.
By using the direction of this elongation of the RC, it is possible to derive the extinction law for the region, as \citet {DeMarchi2014} showed for the 30 Doradus Nebula in the LMC.
The difference in reddening between the two reddening vectors for the adopted $\mathrm{A_V}$ = 0.9 for the older population (see Section \ref{subsubsec:extinction_populations}) is $<$ 0.07 in the F115W-F200W colour.
This is well within the magnitude uncertainty cut-off of 0.70 mag.
In addition, there is another possible reason for the elongation of the RC, which is the shape of the SMC.
The SMC extends considerably along the line of sight \citep{Yanchulova2021, Murray2024}, so RC stars at different distances could result in a slight elongation.
These two effects, acting together, can influence the elongation of the RC in the CMD.
Because of this, we chose to adopt the extinction curve of \citet{Cardelli1989, O'Donnell1994} with $R_V$ = 3.1.
These are also incorporated in the PARSEC isochrone tracks version 1.2S \citep{Bressan2012, Tang2014, Chen2014, Chen2015} and COLIBRI S\_37 isochrone tracks \citep{Marigo2013, Rosenfield2016, Pastorelli2019, Pastorelli2020} used in this work.

\begin{figure*}[h!]\centering
\subfloat[Three isochrones with an age of 3.0 Gyr, metallicity Z = 0.004 and distance modulus (m - M) = 18.96~($\pm$~0.035) are shown in each panel.
The isochrones have different values for $\mathrm{A_V}$: 0.7 (green), 0.9 (magenta) and 1.1 (yellow).]{\label{fig:isochrones_RGB}\includegraphics[width=0.99\linewidth]{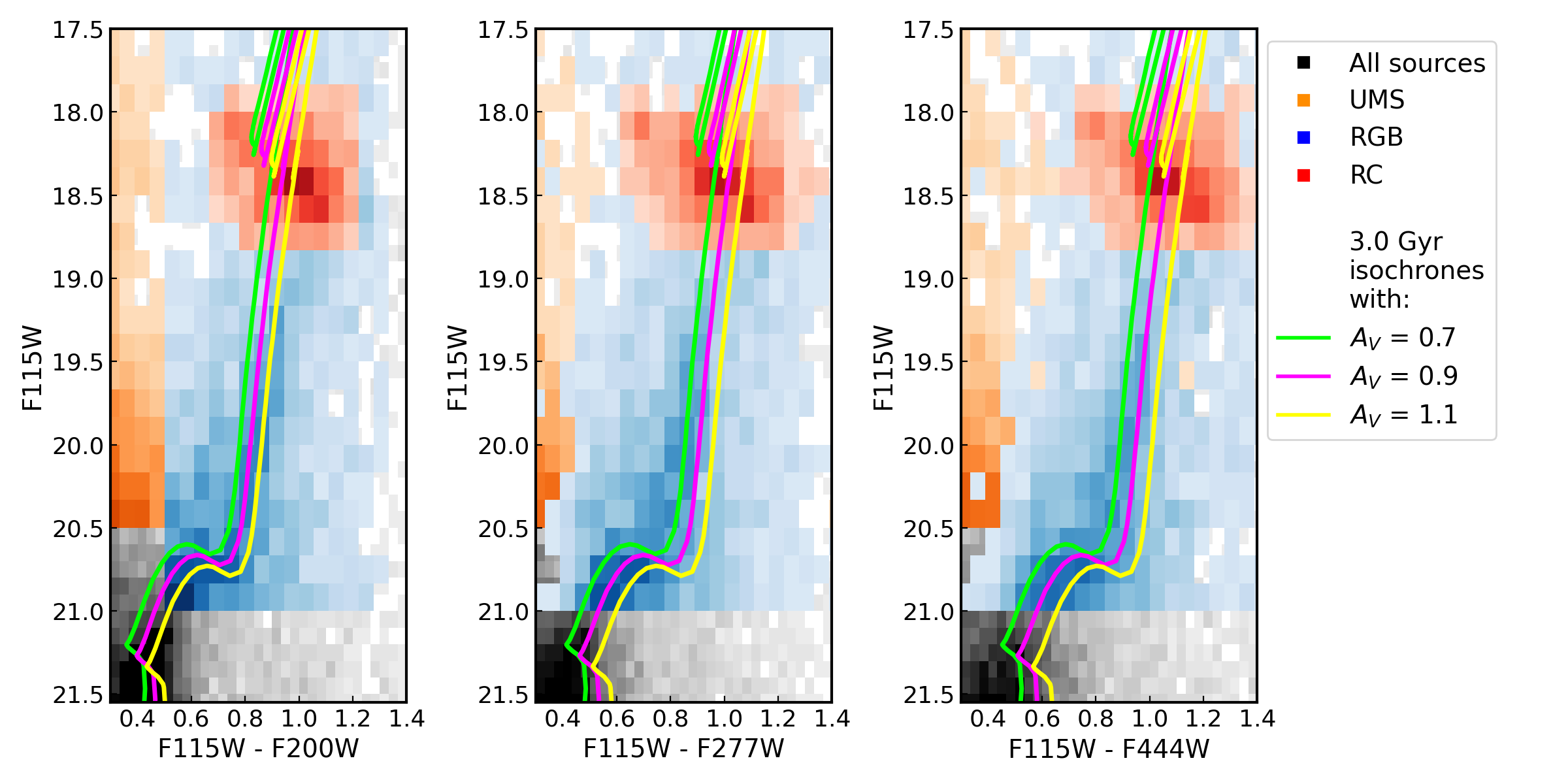}} \par
\subfloat[Three isochrones with an age of 3.0 Myr, metallicity Z = 0.004 and distance modulus (m - M) = 18.96~($\pm$~0.035) are shown in each panel.
The isochrones have different values for $\mathrm{A_V}$: 1.0 (blue), 1.5 (black) and 2.0 (green).]{\label{fig:isochrones_UMS} \includegraphics[width=0.99\linewidth]{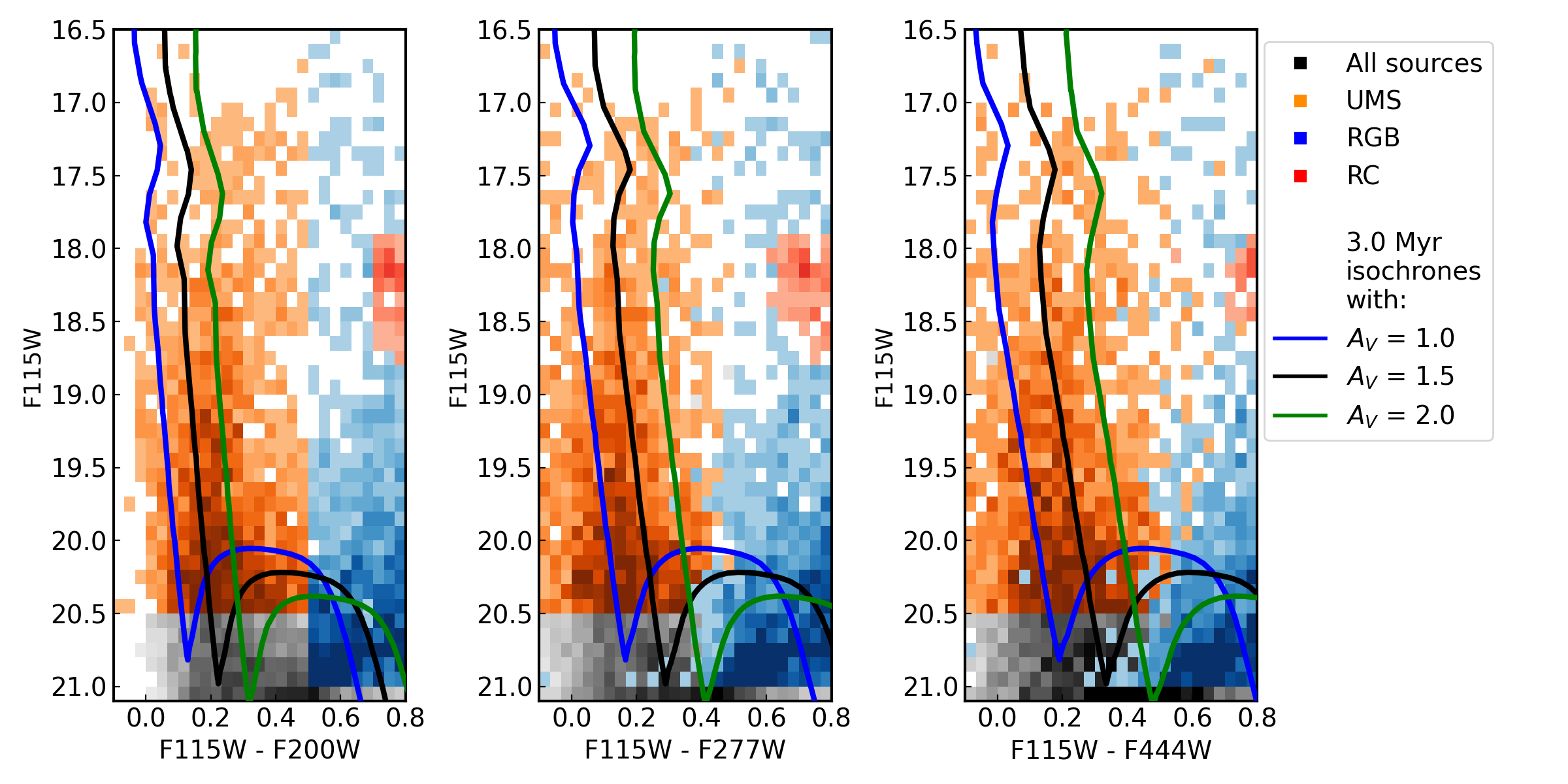}}
\caption{Three CMDs in Hess format, focused on the RGB (top panels) and UMS (bottom panels): left is F115W vs F115W-F200W, middle is F115W vs F115W-F277W, and right is F115W vs F115W-F444W.
The UMS, RGB and RC are shown in orange, blue and red, respectively.
For the older populations, we assumed $\mathrm{A_V}$~=~0.9, while for the young populations we applied $\mathrm{A_V}$~=~1.5.}
\label{fig:isochrones}
\end{figure*}

\subsubsection{Extinction of different populations} \label{subsubsec:extinction_populations}
We determined the extinction of both the younger and older population by utilising three NIRCam CMDs: F115W-F200W vs F115W, F115W-F277W vs F115W and F115W-F444W vs F115W.
Fig. \ref{fig:isochrones_RGB} shows the three RGB CMDs in Hess diagram format.
We over-plotted multiple 3.0 Gyr isochrones, with a metallicity Z~=~0.004 (equivalent to 0.2~$\mathrm{Z_\odot}$), distance modulus (m - M)~=~18.96~($\pm$~0.035) and varying extinction values between $\mathrm{A_V}$ = 0.7~-~1.1.
We then determined by eye which isochrone appeared to be the best match.
We chose 3.0 Gyr isochrones, because these simultaneously fitted the main sequence turn-off (MSTO), RC and RGB better than using an age of 4.3~($\pm$~0.1)~Gyr, which is the age of the intermediate-age star cluster BS~90 \citep{Sabbi2007}.
We used the density of sources in the MSTO and RGB to constrain the extinction.
The 3.0 Gyr isochrone with an $\mathrm{A_V}$ of 0.7 shows a reasonable match in the F115W-F277W vs F115W CMD, but was too low for the other two CMDs, whereas using an $\mathrm{A_V}$ of 1.1 resulted in a reasonable match in the F115W-F200W vs F115W CMD, but was too high for the other two CMDs.
For the older population, we chose the isochrone with an $\mathrm{A_V}$ of 0.9.
This provided the best match in both the F115W-F200W vs F115W and F115W-F444W vs F115W CMDs, as it traces the densest parts of the MSTO and RGB and is located close to the centre of the RC (see Fig. \ref{fig:isochrones_RGB}).
In addition, it was a reasonable match for the F115W-F277W vs F115W CMD, since it follows the densest part of the MSTO and ended up inside the densest part of the RC, while lying slightly to the right of the densest part of the RGB.\newline

\indent The slanted black lines cutting through the UMS in Fig. \ref{fig:extinction_arrows} correspond to reddening vectors drawn from a zero-age main sequence (ZAMS) isochrone for Z = 0.004, in steps of 0.5 mag for stars brighter than F115W = 20.5.
The length of these slanted lines corresponds to a range in $\mathrm{A_V}$ of 2.0.
The lines show the possible displacement of UMS stars, for a specific F115W magnitude, due to varying extinction values ($\mathrm{A_V}$ = 0.0 - 2.0).
For every 0.5 mag in the F115W filter below 20.5 mag, these possibly similar UMS stars are coloured blue or red in Fig. \ref{fig:extinction_arrows}, showing over 3000 stars.
Not all of these red and blue coloured stars ended up in our selection of UMS stars, as we used a cut-off F115W magnitude of 20.5 for UMS stars, as is explained in Section \ref{subsec:characterisation}. \newline
\indent To determine the extinction of the younger population, we used the three CMDs shown in Fig. \ref{fig:isochrones_UMS}.
We used the same approach as for the RGB, plotting multiple 3.0 Myr isochrones, which \citet{Sabbi2007} found to be the age of multiple subclusters of NGC~346, with varying $\mathrm{A_V}$ (see Fig. \ref{fig:isochrones_UMS}).
An $\mathrm{A_V}$ of 1.0 gave a reasonable match for the F115W-F277W vs F115W and F115W-F444W vs F115W CMDs, but was shown to be too low for the F115W-F200W vs F115W CMD.
Similarly, an $\mathrm{A_V}$ of 2.0 gave a reasonable match for the F115W-F200W vs F115W CMD, but was too high for the other two CMDs.
Using an $\mathrm{A_V}$ of 1.5 gave the most consistent match to all three CMDs.
In this work, we therefore adopted the value for $\mathrm{A_V}$ = 1.5 as the reddening towards younger populations.

\subsection{Source characterisation}\label{subsec:characterisation}
To identify different stellar populations, we used multiple combinations of CMDs, analysed separately for NIRCam and MIRI.
We add two columns to the catalogue, assigning a flag to each source based on the source characterisation using NIRCam and MIRI CMDs independently.
These columns are named ``\textit{Flag\_NIRCam}'' and ``\textit{Flag\_MIRI}''.
In addition, we utilise the F115W-F200W vs F115W-F187N CCD to select sources with Pa$\mathrm{\alpha}$-excess, adding an additional column (named ``\textit{Pa$\mathrm{\alpha}$ excess}'') to the catalogue. 
The source characterisation and our flagging system are described in detail below. \newline

\begin{deluxetable}{l|l|c|c|c}
\tablewidth{0pt}
\tablecaption{UMS, RGB and RC populations identified using NIRCam CMD F115W-F200W vs F115W for the NGC~346 regions and the whole NIRCam field. The expected number of these sources in the whole field based on the deep exposure `backgrounds' is shown as the `contamination estimation'. Note that the expected number of RC sources is based on the shallow exposure `background' areas as no RC stars were detected in the deep exposure `background' areas, due to their smaller size. \label{tab:Selection_UMS_RGB_RC}} 
\tablehead{
\colhead{Population} & \colhead{Colour selection} & \colhead{\# of sources in} & \colhead{\# of sources} & \colhead{Contamination}\\
\colhead{ } & \colhead{} & \colhead{NGC~346 regions} & \colhead{in entire field} & \colhead{estimation}}
\startdata
RC & 0.70 $<$ F115W-F200W $<$ 1.20 & 72  & 742 & 615 \\ 
 & and 17.95 $<$ F115W $<$ 18.75 & & & \\ \hline
RGB & 0.50 $<$ F115W-F200W $<$ 1.30 & 539 & 2,755 & 1,650 \\
 & and F115W $<$ 21.0 & & & \\ \hline
UMS & -0.10 $<$ F115W-F200W $<$ 0.50 & 543 & 2,024 & 1281 \\
 &  and F115W $<$ 20.5 & & &  \\
\enddata
\end{deluxetable}

\subsubsection{Colour-magnitude diagrams}\label{subsubsec:CMD}
\begin{figure*}
    \centering
    \includegraphics[width=1\linewidth]{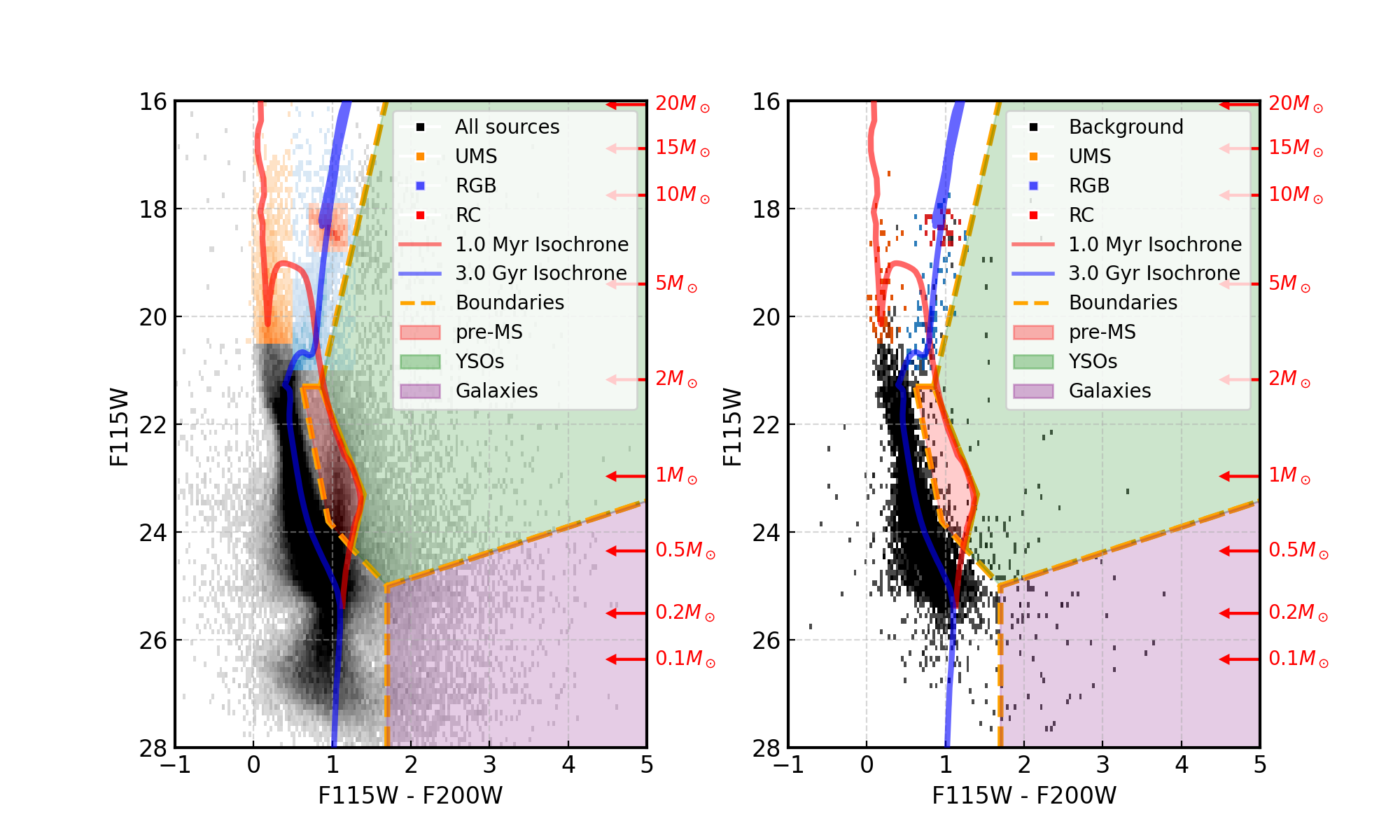}
    \caption{The F115W vs F115W-F200W CMD of the whole mosaic (left) and `background' (right).
    The UMS, RGB and RC are shown in orange, blue and red, respectively, with the rest of the sources shown in black.
    Stellar masses are shown with red arrows on the right side, and boundaries for the pre-MS (red shaded), YSO (green shaded) and galaxy region (purple shaded) are shown with orange dashed lines.
    The boundary between the pre-MS and YSO regions follows the 1 Myr isochrone (adopting $\mathrm{A_V}$=1.5), which is shown as a red line.
    The 3.0 Gyr isochrone, with $\mathrm{A_V}$=0.9 is also shown.}
    \label{fig:F115W-F200W_whole_with_bkg}
\end{figure*}

\begin{figure*}
    \centering
    \includegraphics[width=1\linewidth]{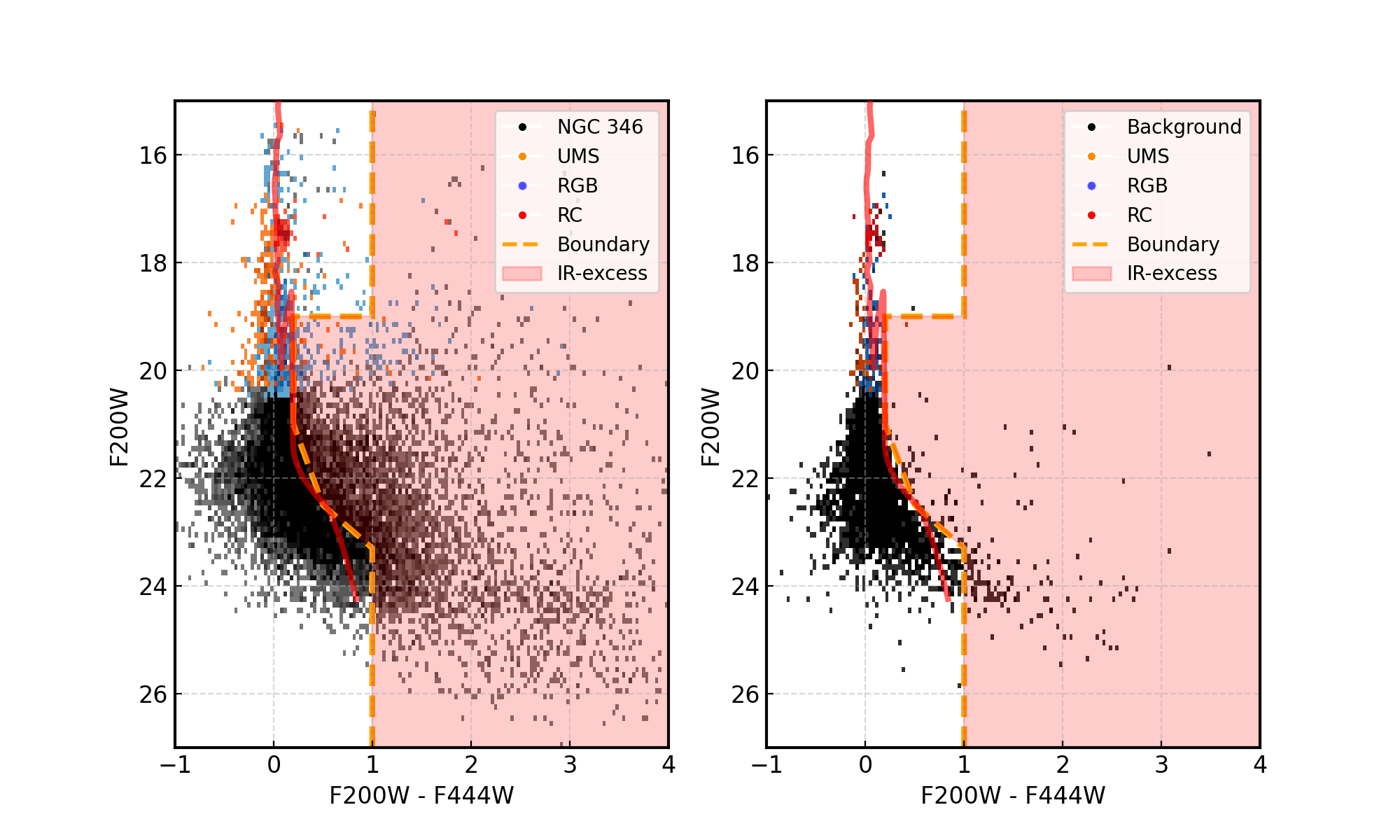}
    \caption{The F200W vs F200W-F444W CMD of the NGC~346 region (left) and `background' (right).
    The UMS, RGB and RC stars are shown in orange, blue and red.
    The 1 Myr isochrone (same as in Fig. \ref{fig:F115W-F200W_whole_with_bkg}) is shown as a red line.
    Boundaries for the IR-excess region are shown with orange dashed lines, and the IR-excess region is shaded in red.}
    \label{fig:F200W-F444W_with_bkg}
\end{figure*}

From the NIRCam F115W-F200W vs F115W CMD (see Fig. \ref{fig:F115W-F200W_whole_with_bkg}) we selected the UMS, RGB and RC stars, as this CMD contains the most sources and these two filters are most sensitive to the stellar contribution.
The characterisation criteria for the UMS, RGB and RC are shown in Table \ref{tab:Selection_UMS_RGB_RC}.
We chose the cut-off magnitude of 21.0 in F115W for the RGB to incorporate the MSTO starting around F115W-F200W = 0.5 and F115W = 21.2 (see Fig. \ref{fig:isochrones_RGB}).
We set the cut-off magnitude for the UMS in the F115W filter as 20.5 mag, because this is $\sim$0.5 magnitude above the drop in density of the MS, and we aimed to minimise the contamination of the UMS. 
In addition, setting the cut-off to this magnitude ensures only stars with $\mathrm{M_*} >\ \sim 3\ \mathrm{M_\odot}$ are counted as UMS stars. 
We showed in Sect.~\ref{subsubsec:extinction_populations} that an $\mathrm{A_V}$ of 2.0 for the young population was the upper value for a reasonable match (see Fig. \ref{fig:isochrones_UMS}).
A range in $\mathrm{A_V}$ of 0~-~2.0 results in a range from 0 to 0.5 in F115W-F200W colour, as shown in Fig. \ref{fig:extinction_arrows} with the black slanted lines.
Therefore, the boundary between the UMS and RGB was set to a F115W-F200W colour of 0.5.
The left boundary of the UMS (F115W-F200W = 0.0) and the right boundary of the RGB (F115W-F200W = 1.3) were drawn by eye.
The boundaries of the RC were drawn to encompass the over-density of sources in the upper region of the RGB in the F115W-F200W vs F115W CMD.
We assigned these sources the corresponding flag UMS, RGB or RC in the catalogue, regardless of where these sources are located in other CMDs. \newline

\indent For all the remaining sources, the assigned flag consists of ten characters based on the source characterisations in the ten utilised NIRCam CMDs, separating the flag assigned for each CMD with a hyphen.
The first separation of sources in all ten CMDs is between the MS + RGB and IR-excess sources.
For each CMD, we created a background CMD containing only the sources from the background areas.
In these, we drew a boundary by eye to the right of the MS and RGB (see right panels in Fig. \ref{fig:F115W-F200W_whole_with_bkg} and Fig. \ref{fig:F200W-F444W_with_bkg}).
Everything to the right of this boundary is defined as IR-excess. \newline
\indent The CMDs that include the F115W filter (F115W-F200W vs F115W, F115W-F277W vs F115W, F115W-F335M vs F115W and F115W-F444W vs F115W) were used to further differentiate the IR-excess sources in three populations: pre-MS stars, YSOs and galaxies.
We do not consider AGB stars in the IR-excess region in this work because the AGB stars with the most IR-excess (Carbon-rich AGB stars) are orders of magnitude brighter than RC stars \citep[see e.g. Fig. 4 in ][]{Nally2024} and these would be saturated in our data.
In these four CMDs the MS is well separated from the younger populations (pre-MS stars + YSOs) and we are able to separate the pre-MS stars from the YSOs.
The boundaries between the MS, pre-MS, YSO and galaxy regions were determined as follows, and shown in Fig. \ref{fig:F115W-F200W_whole_with_bkg}:
\begin{itemize}
    \item The boundary between the MS and pre-MS is determined by the boundary drawn right of the MS in the background CMD.
    \item The upper boundary of the pre-MS region is chosen slightly below the RGB, to minimise contamination.
    \item The boundary between the pre-MS and YSO region is drawn by following the 1 Myr isochrone, for which we assumed an $\mathrm{A_V = 1.5}$, as determined in Section \ref{subsubsec:extinction_populations}. The left upper boundary of the YSO region is drawn parallel to the RGB.
    \item We drew the boundary between the MS and galaxy region vertically down from the point where the lower-MS stopped in the background CMD.
    \item The boundary between the YSO and galaxy region also starts from the point where the lower-MS stopped in the background CMD.
    It is drawn towards the right, going slightly upwards, following the approach in \citet{Zeidler2024}, resulting in a more conservative selection of YSOs than applying a horizontal boundary.
\end{itemize} 


\indent For the CMDs not including the F115W filter, the pre-MS, YSOs and galaxies are harder to differentiate as all filters are, to some extent, sensitive to the dust continuum and less sensitive to the stellar continuum.
In addition, the 1 Myr isochrone in these CMDs lies directly next to the drop-off in density of the MS (see Fig. \ref{fig:F200W-F444W_with_bkg}), which prevents us from separating the pre-MS stars from the YSOs.
Therefore, we used the remaining six NIRCam CMDs (F200W-F277W vs F200W, F200W-F335M vs F200W, F200W-F277W vs F200W, F277W-F335M vs F277W, F277W-F444W vs F277W and F335M-F444W vs F335M) only to differentiate IR-excess sources from the MS + RGB. \newline

\indent Each source is flagged according to the category it belongs to in each of the ten CMDs, creating a flag for the NIRCam source characterisation that is ten characters long.
The first character of each source, that is not categorised as UMS, RGB or RC, is from the F115W-F200W vs F115W CMD, the second character from the F115W-F277W vs F115W CMD, continuing until the last character from the F335M-F444W vs F335M CMD.
The possible characters in the catalogue column ``\textit{Flag\_NIRCam}'' are:
\begin{itemize}
    \item ``0'' = not found in either one or both of the filters of the specific CMD.
    \item ``P'' = source located in the pre-MS region in the specific CMD (containing the F115W filter).
    \item ``Y'' = source located in the YSO region in the specific CMD (containing the F115W filter).
    \item ``G'' = source located in the galaxies region in the specific CMD (containing the F115W filter).
    \item ``I'' = source located in the IR-excess region in the specific CMD (not containing the F115W filter).
    \item ``M'' = main-sequence star in the specific CMD.
\end{itemize}

For the MIRI source selection, we used all ten possible CMD combinations of the F770W, F1000W, F1130W, F1500W and F2100W filters, creating a flag that is ten characters long.
Any source that is selected as UMS, RGB or RC in the NIRCam source selection will receive this flag for the MIRI source selection as well.
For the remaining sources, we used the MIRI CMDs and drew a line on the right of the MS, characterising any source on the right of this line as IR-excess.
The possible characters that make up the flag in the column ``\textit{Flag\_MIRI}'' in the catalogue are:
\begin{itemize}
    \item ``0'' = not found in either one or both of the filters of the specific CMD.
    \item ``I'' = source located in the IR-excess region in the specific CMD.
    \item ``M'' = main sequence star in the specific CMD.
\end{itemize}
In some cases, a source can receive a double character for a specific CMD, due to the uncertainty in magnitude and colour.
Using the uncertainty in colour as the width and the uncertainty in the magnitude as the height, we created a rectangle around each source.
Any source near the boundary of two regions can receive a double character for that specific CMD.
For example, a source that is located in the pre-MS region, but its uncertainty overlaps with the YSO region, will receive the character `Py' in the flag.
The capital letter shows in which region the source is located, whereas the lower-case letter shows the regions overlapped when including the uncertainty. 

\begin{figure}
    \centering
    \includegraphics[width=1.0\linewidth]{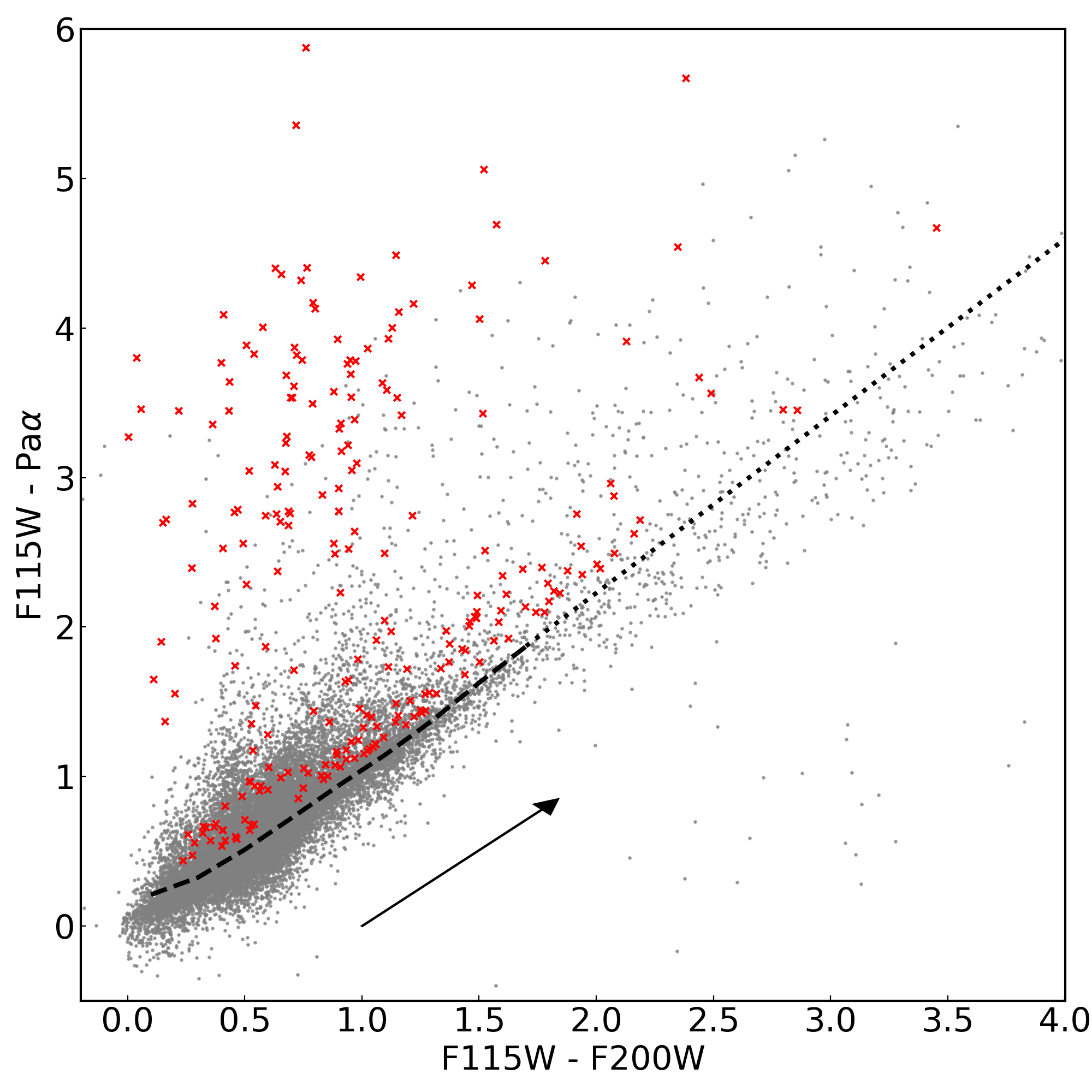}
    \caption{The F115W-F187N vs F115W-F200W CCD of all sources (gray dots).
    The dashed line represents the running median F115W-Pa$\alpha$ colour, which is extrapolated after F115W-F200W = 1.7 (represented by the dotted line).
    A reddening vector for $\mathrm{A_V}$ = 3 is shown for illustrative purposes.
    A total of 239 non-spurious Pa$\alpha$ excess sources are shown as red crosses.}
    \label{fig:CCD_Paschen}
\end{figure}

\subsubsection{Colour-Colour diagram}\label{subsubsec:CCD}
For the final assigned flag in the column ``\textit{Pa$\alpha$ excess}'', we utilised the F115W-F200W vs F115W-F187N CCD to select sources that have Pa$\mathrm{\alpha}$ excess, following a similar approach as \citet{DeMarchi2011}. 
The Pa$\mathrm{\alpha}$ recombination line at 1.875 $\textrm{\micron}$ is a tracer for accretion, so using this line allowed us to select pre-MS and YSO candidates that are actively accreting.
In a typical stellar field, the majority of stars do not have excess Pa$\alpha$ emission and therefore the median value of the F115W-Pa$\alpha$ colour index at a given effective temperature, $T_{eff}$ defines a spectral reference template for all stars with that $T_{eff}$.
To determine the reference template, we used the F115W-F200W vs F115W-F187N CCD, containing 23,902 sources, and selected sources whose mean error $\delta_3$ in the three filters does not exceed 0.1 mag, where
\begin{equation}
\delta_3 = \sqrt{\frac{\delta_{115}^2 + \delta_{187}^2 + \delta_{200}^2}{3}}
\end{equation}
\noindent and $\mathrm{\delta_{F115W}}$, $\mathrm{\delta_{F187N}}$ and $\mathrm{\delta_{F200W}}$ are the photometric uncertainties in each individual band.
The value of $\delta_3$ is dominated by the uncertainty on the Pa$\alpha$ magnitude, with a median value of $\mathrm{\delta_{F187N}}$ = 0.088, while the median value of the uncertainty in the other two bands is $\mathrm{\delta_{F115W}}$ = 0.019 and $\mathrm{\delta_{F200W}}$ = 0.015.
In total 4,668 sources satisfy this condition.
We used sigma clipping on these sources to calculate the mean F115W-Pa$\alpha$ colour per bin of 0.2 width in F115W-F200W colour.
The running median is shown in Fig. \ref{fig:CCD_Paschen} as the dashed black line, which is extrapolated after a F115W-F200W colour of 1.7, shown as the dotted black line.
We tested each source for Pa$\alpha$ excess against this reference template.
Any source with a F115W-Pa$\alpha$ colour that exceeds the reference template by more than three times the uncertainty on their F115W-Pa$\alpha$ colour and is more than 0.1 mag above the line, was selected as a source with Pa$\alpha$ excess.
We chose this value of 0.1 mag above the reference template to exclude sources that might have Pa$\alpha$ emission from chromospheric activity \citep{White2003}.
In total, 376 sources were selected this way, which were all checked by eye against the F115W image, and 137 sources were detected within the PSF wings of bright sources.
In the catalogue, these were given the flag ``S'' for spurious detection in the ``Pa$\alpha$ excess'' column, whereas the remaining 239 with Pa$\alpha$ excess received the flag ``Y''.
These non-spurious Pa$\alpha$ excess sources are shown as red crosses in Fig. \ref{fig:CCD_Paschen} and the distribution of these is shown in Fig. \ref{fig:Pa-alpha_distribution}.
Every other source was given the flag ``N''.

\section{Results \& Discussion}\label{sec:results}

\subsection{Source catalogue} \label{subsec:source_catalogue}

\begin{deluxetable*}{l c c r c c c r r c}
\tablewidth{0pt}
\tablecaption{Sample of our photometric catalogue. The columns of all the filters and their uncertainty between F115W and eF2100W are omitted, but follow the same pattern as the example columns shown. RA and DEC are in the ICRS coordinate system, and the magnitudes are given in the Vega system. \label{tab:catalogue}}
\tablehead{
\colhead{Catalogue} & \colhead{RA} & \colhead{DEC} & \colhead{flag} & \colhead{F115W} & \colhead{...} & \colhead{eF2100W} & \colhead{Flag}  & \colhead{Flag} & \colhead{Pa$\alpha$} \\
\colhead{Number} & \colhead{[ICRS]} & \colhead{[ICRS]} & \colhead{} & \colhead{} & \colhead{} & \colhead{} & \colhead{NIRCam} & \colhead{MIRI} & \colhead{excess}
}
\startdata
CN0  & 14.7769194 & -72.1601907 & 12 & 24.140 & ... & -- & My-0-0-0-0-0-0-0-0-0 & 0-0-0-0-0-0-0-0-0-0 & N \\
CN1 & 14.7700272 & -72.1600488 & 12 & 17.072 & ... & -- & Ym-Y-Y-Y-M-M-M-M-M-M & M-M-M-0-M-M-0-M-0-0 & Y \\
CN2 & 14.7771317 & -72.1595729 & 4 & 22.242 & ... & -- & Mp-M-Mp-M-M-M-M-M-M-M & 0-0-0-0-0-0-0-0-0-0 & N \\
CN3 & 14.7771922 & -72.1591325 & 8 & 22.975 & ... & -- & M-M-M-M-M-M-M-M-M-M & 0-0-0-0-0-0-0-0-0-0 & N \\
CN4 & 14.7721221 & -72.1586775 & 0 & 17.472 & ... & -- & M-M-M-M-M-M-M-M-M-M & M-M-M-0-M-M-0-M-0-0 & N \\
CN5 & 14.7307299 & -72.1586006 & 0 & 18.027 & ... & -- & Y-Y-Y-Y-M-M-M-M-M-M & M-M-0-0-M-0-0-0-0-0 & N \\
CN6 & 14.7593734 & -72.1574489 & 0 & 17.412 & ... & -- & M-M-M-M-M-M-M-M-M-M & M-M-M-0-M-M-0-M-0-0 & N \\
CN7 & 14.7776764 & -72.1573902 & 0 & 20.498 & ... & -- & Y-Y-Y-Y-M-M-M-M-M-M & M-0-0-0-0-0-0-0-0-0 & N \\
CN8 & 14.7776223 & -72.1572986 & 8 & 22.725 & ... & -- & M-M-0-M-M-0-0-0-0-0 & 0-0-0-0-0-0-0-0-0-0 & N \\
CN9 & 14.7509363 & -72.1572292 & 0 & 18.392 & ... & -- & RC & RC & N \\
CN10 & 14.7773772 & -72.1571516 & 4 & 23.966 & ... & -- & M-M-0-0-M-0-0-0-0-0 & 0-0-0-0-0-0-0-0-0-0 & N \\
CN11 & 14.7771422 & -72.1571212 & 8 & 23.365 & ... & -- & M-M-0-M-M-0-M-0-M-0 & 0-0-0-0-0-0-0-0-0-0 & N \\
CN12 & 14.7755658 & -72.1571119 & 0 & 22.409 & ... & -- & M-M-M-M-M-M-M-M-M-M & 0-0-0-0-0-0-0-0-0-0 & N \\
CN13 & 14.7769806 & -72.1570322 & 0 & 22.177 & ... & -- & M-M-Mp-M-M-Mi-M-I-I-M & 0-0-0-0-0-0-0-0-0-0 & N \\
CN14 & 14.7763641 & -72.1570352 & 8 & 22.439 & ... & -- & M-M-M-M-M-M-M-M-M-M & 0-0-0-0-0-0-0-0-0-0 & N \\
CN15 & 14.7773993 & -72.1570269 & 4 & 23.358 & ... & -- & M-M-Mp-0-M-M-0-Mi-0-0 & 0-0-0-0-0-0-0-0-0-0 & N \\
CN16 & 14.7753100 & -72.1570291 & 12 & 23.571 & ... & -- & M-M-M-0-M-M-0-M-0-0 & 0-0-0-0-0-0-0-0-0-0 & N \\
CN17 & 14.7768338 & -72.1570097 & 0 & 22.816 & ... & -- & M-M-0-0-M-0-0-0-0-0 & 0-0-0-0-0-0-0-0-0-0 & N \\
CN18 & 14.7748916 & -72.1570120 & 0 & 21.816 & ... & -- & M-M-M-M-M-M-M-M-M-M & 0-0-0-0-0-0-0-0-0-0 & N \\
CN19 & 14.7728764 & -72.1570006 & 12 & 22.918 & ... & -- & M-M-M-0-M-M-0-M-0-0 & 0-0-0-0-0-0-0-0-0-0 & N \\
\enddata
\end{deluxetable*}

The full photometric catalogue has twenty-nine columns. 
For a sample of sources, nine of the columns are shown in Table \ref{tab:catalogue}, including the catalogue number, right ascension and declination in the ICRS coordinate system, the F115W magnitude in Vegamag \footnote{\url{https://jwst-docs.stsci.edu/jwst-near-infrared-camera/nircam-performance/nircam-absolute-flux-calibration-and-zeropoints\#NIRCamAbsoluteFluxCalibrationandZeropoints-Vegamagnitudes\&gsc.tab=0}}, the uncertainty in F2100W magnitude (eF2100W) in Vegamag and several flags.
The columns from F187N until eF1500W are omitted from the table for style purposes, as these are all similar to the F115W and eF2100W columns.
The column `flag' (column 4) shows the flag that is added during the PSF photometry of \texttt{STARBUGII} \citep{Starbug}, which can have four different values:
\begin{itemize}
    \item 0: Source OK.
    \item 4: Source has an asymmetric flux distribution between the photometry of individual dithers (mean and median more than 5\% different).
    \item 8: Source has PSF photometry with a forced position.
    \item 12: Source has both an asymmetric flux distribution between matches and PSF photometry with a forced position.
\end{itemize}
The last three columns show the flags added during the source characterisation, as discussed in Sect.~\ref{subsec:characterisation}. \newline

\subsubsection{Upper Main Sequence} \label{sec:ums}
Fig \ref{fig:F115W-F200W_whole_with_bkg} shows the F115W-F200W vs F115W CMDs in Hess diagram format of all the sources detected in these two filters (left) and of the SMC field sources of both the deep and shallow exposure `background' areas combined (right).
Differences between populations of all sources and only the SMC field sources are evident.
First, there is a clear over-density of sources fainter than a F115W magnitude of $\sim$25.5-26.0 for the whole mosaic with respect to the SMC field sources.
This is because the deeper exposure areas only cover a part of the whole mosaic, and over 95\% of these dim sources are located within these deeper exposure areas.
Another difference is that there are a significant number of pre-MS and YSO candidates located in the region with respect to the SMC field.
In the F115W-F200W vs F115W CMD of the whole mosaic, we characterised 3,927 sources as pre-MS and 2,883 as YSO, whereas we would expect $\sim$1,164 pre-MS stars and $\sim$490 YSOs in this CMD based on the deep exposure background areas (see Table \ref{tab:NIRCam_pre-MS-YSO-Gal_selection_total}).
The last clear difference can be seen in the UMS population, shown in the top left of the CMDs of Fig \ref{fig:F115W-F200W_whole_with_bkg} as the orange branch located at -0.1 $<$ F115W-F200W $<$ 0.5 and F115W $<$ 20.5.
Based on the high exposure backgrounds we would expect a total of $\sim$1281 UMS stars in the whole field, in comparison to the 2,024 characterised (see Table \ref{tab:Selection_UMS_RGB_RC}).
Most of these `background' UMS stars have masses $\leq \mathrm{5\ M_\odot}$.
This indicates that young and high-mass MS stars are not as common in the field population, which is what we expect if our chosen `backgrounds' are representative of the SMC field population.

We detected high-mass UMS stars (M$_* \geq$ 8 M$_\odot$) in multiple polygon regions from Fig \ref{fig:4_filter_images} (top left), with $\sim$72\% of them located in region 2, including the most massive and brightest UMS stars.
The 30 O-type stars with masses 35-100 M$_\odot$ \citep{Massey1989, Evans2006, Dufton2019} are saturated in our data and thus not included in our study.
Polygon region 3 contains only three UMS stars above 5\ M$_\odot$, one of which is above 8\ M$_\odot$\ and polygon region 6 contains five UMS stars above 5\ M$_\odot$\, without any above 8\ M$_\odot$. 
When comparing the number density of high-mass UMS stars in the different polygon regions, polygon regions 2, 4 and 5 stand out with 0.16, 0.15 and 0.10 high-mass UMS stars $\mathrm{pc^{-2}}$.
The other regions 1, 3 and 6 have high-mass UMS number densities of respectively 0.01, 0.04, and 0.0 $\mathrm{pc^{-2}}$.
A reason for the low number density of high-mass UMS stars in polygon region 3 could be that it is co-located with the highest velocity CO clumps in the region \citep{Rubio2000}.
This could indicate that stellar winds have quenched star formation \citep{Sabbi2007}. \newline
\indent Polygon region 1 contains a part of the filamentary structure of the Main Arc (see Fig. \ref{fig:4_filter_images}), but it has one of the lowest high-mass UMS star number densities of all the regions, as well as the lowest number density of UMS stars between 5~-~8\ M$_\odot$.
This, in combination with the lack of stellar clusters found in this region by \citet{Sabbi2007}, could indicate that there is less (high-mass) star formation occurring in this region.
Another explanation for the low number density of high-mass UMS in this region was proposed by \citet{Cignoni2011}, who found a lack of high-mass stars compared to the number of pre-MS stars detected with \hst\ in this region, suggesting that the high-mass stars are still embedded.
This suggestion is strengthened by the relatively low number of NIRCam high-confidence pre-MS and YSO candidates (see Fig. \ref{fig:F444W_NIRCam_tests} and Sections \ref{subsubsec:pre-MS_and_YSO} and \ref{subsubsec:IR-excess}), while simultaneously having a relatively high number of MIRI high-confidence YSO candidates in this region (see Fig. \ref{fig:F770W_MIRI_tests} and Section \ref{subsubsec:IR-excess}), indicating that this region is more embedded and consists of younger stars.


\subsubsection{Red Giant Branch and Red Clump}\label{sec:rgb_rc}

\begin{figure*}
    \centering
    \includegraphics[width=0.95\textwidth]{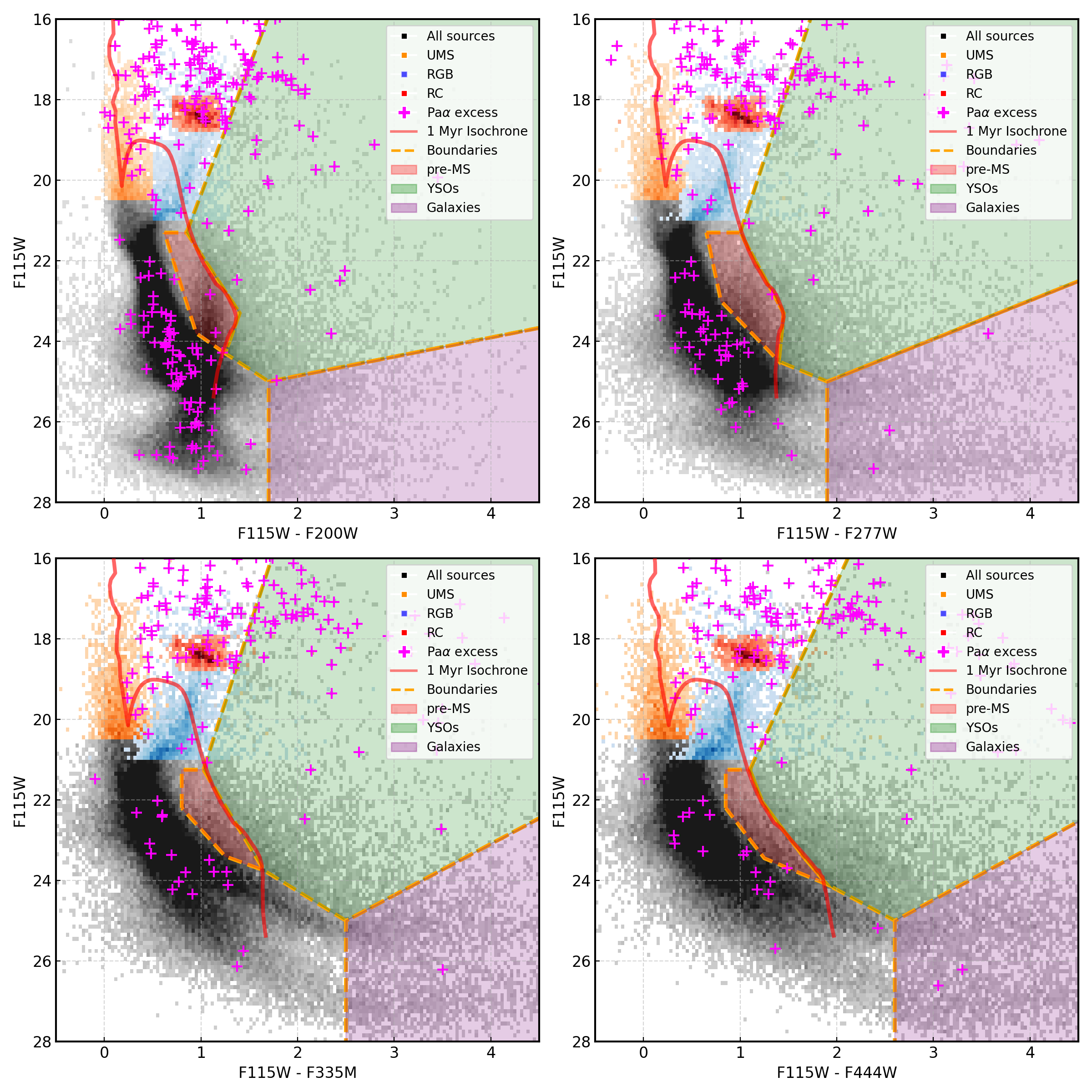}
    \caption{Four NIRCam CMDs, showing all sources (black) with the F115W filter on the vertical axis: F115W-F200W top left, F115W-F277W top right,
    F115W-F335M bottom left and F115W-F444W bottom right.
    The boundaries between the pre-MS, YSO and Galaxy regions are shown in orange dashed lines.
    The border between the pre-MS and YSO region is determined by the 1 Myr isochrone (red solid line).
    The UMS, RGB and RC stars are shown in respectively orange, blue and red.
    Sources with Pa$\alpha$ excess are shown in magenta.}
    \label{fig:NGC346_F115W-F...W_CMDs}
\end{figure*}

The RGB consists of stars that have left the MS and started H-shell burning and is located at 0.5 $<$ F115W-F200W $<$ 1.3 and F115W $<$ 21.0, shown as the blue branch in Fig \ref{fig:F115W-F200W_whole_with_bkg}.
In single-aged populations the MSTO appears as a tight track connecting the MS and RGB, but in the F115W-F200W vs F115W CMD of Fig \ref{fig:F115W-F200W_whole_with_bkg} it shows as a broadened track.
This indicates that the SMC has had a prolonged episode of star formation, as found by \citet{Sabbi2007} and \citet{Cignoni2011}.
They found a major episode of star formation between 3 - 5 Gyr ago, with an almost instantaneous burst of star formation around $\sim$ 4.3 Gyr ago, forming the stellar cluster BS~90 (see Fig. \ref{fig:4_filter_images} top left).
They also showed that the SMC field contains stars up to ages of 10 Gyr and that there has been an increase in the star formation rate in the last $\sim$100-150 Myr.
This all adds to the MSTO being broadened in our CMD.
Another reason for the broadening of the MSTO is that the RGB stars are widely spread over the mosaic, with a clear over-density in the stellar cluster BS~90 and a slight over-density within the NGC~346 main body, and these regions have different values of extinction as shown in Section \ref{subsubsec:extinction_populations}. \newline
\indent The RGB itself has a spread in colour space as well, even though the location of the RGB in a CMD is fairly independent of the SFH.
For a fixed value of metallicity, younger RGB stars appear slightly bluer and older RGB stars appear slightly redder.
In addition, metal-poorer stars appear bluer than metal-rich stars, for a specified age, an effect known as the age-metallicity degeneracy \citep{Worthey1999}.
It is therefore difficult to determine with certainty if the position on the RGB of an individual star is caused by either the age or metallicity.
We note that due to our choice not to leave a gap between the UMS and RGB, the RGB can be contaminated with UMS stars that seem redder due to dust extinction.

Within the upper part of the RGB, there is an over-density of sources located at 0.7 $<$ F115W-F200W $<$ 1.2 and 17.95 $<$ F115W $<$ 18.75, shown in red in Fig \ref{fig:F115W-F200W_whole_with_bkg}: the RC.
In this work, we characterised a total of 742 RC stars and 2,755 RGB stars (see Table \ref{tab:Selection_UMS_RGB_RC}).

\subsubsection{Pre-MS, YSO and Galaxies}\label{subsubsec:pre-MS_and_YSO}
\begin{deluxetable}{l c c c c c c}
\tablewidth{0pt}
\tablecaption{NIRCam CMD characterisation of pre-MS stars, YSOs and Galaxies for the whole NIRCam field. The estimated number of SMC field sources, based on the deeper exposure `background' areas, for each population is also shown. \label{tab:NIRCam_pre-MS-YSO-Gal_selection_total}}
\tablehead{
\colhead{CMD} & \colhead{pre-MS} & \colhead{YSO} & \colhead{Galaxies} & \colhead{Contamination} & \colhead{Contamination} & \colhead{Contamination} \\
\colhead{} & \colhead{count} & \colhead{count} & \colhead{count} & \colhead{pre-MS estimation} & \colhead{YSO estimation} & \colhead{galaxies estimation}
}
\startdata
F115W-F200W  & 3,927 & 2,883 & 1,630 & 1,397 & 466 & 1,048 \\
F115W-F277W  & 3,595 & 2,926 & 3,209 & 1,164 & 582 & 1,048 \\
F115W-F335M  & 1,193 & 2,717 & 2,989 & 582 & 466 & 1,397 \\
F115W-F444W  & 1,519 & 3,686 & 3,457 & 815 & 699 & 2,096 \\ \hline
Total Sources & 6,274 & 7,350  & 8,441 & - & - & - \\
\enddata
\end{deluxetable}

\begin{deluxetable}{l r r r}
\tablewidth{0pt}
\tablecaption{Total number of IR-excess sources from the whole NIRCam field for six NIRCam CMDs. The number of estimated contaminating IR-excess sources, based on the deeper exposure `background areas', is given, as well as the contamination percentage. \label{tab:NIRCam_IR_selection_total}}

\tablehead{
\colhead{CMD} & \colhead{IR-excess} & \colhead{Contamination} & \colhead{\%} \\ \colhead{} & \colhead{count} & \colhead{IR-excess} & \colhead{conta-} \\
\colhead{} & \colhead{} & \colhead{estimation} & \colhead{mination} 
}
\startdata
F200W-F277W  & 6,941 & 1,863 & 26.8 \\
F200W-F335M  & 6,607 & 2,445 & 37.0 \\
F200W-F444W  & 10,093 & 2,678 & 26.5 \\ \hline
F277W-F335M  & 6,381 & 932 & 14.6 \\ 
F277W-F444W  & 8,198 & 1,630 & 19.9 \\ \hline
F335M-F444W  & 2,957 & 466 & 15.8 \\  \hline 
Total Sources &  23,819 & - & - \\
\enddata
\end{deluxetable}

Table \ref{tab:NIRCam_pre-MS-YSO-Gal_selection_total} shows the number of characterised pre-MS stars, YSOs and galaxies, with the estimated contamination from the SMC field, for each of the four utilised CMDs shown in Fig. \ref{fig:NGC346_F115W-F...W_CMDs}.
6,274 sources were characterised as pre-MS, 7,350 sources as YSO, and 8,441 sources as galaxy in at least one of the four CMDs (see Table \ref{tab:NIRCam_pre-MS-YSO-Gal_selection_total}).
Based on our NIRCam deep exposure `backgrounds', we would expect 4,090~$\pm$~751 unique sources to be characterised as galaxies in the whole mosaic, whereas we find 8,441 of these sources.
This is a factor of $\sim$2 times higher than expected, which cannot be explained without accounting for numerous possible low-mass YSOs that are characterised as galaxies.
We characterised $\sim$~5~-~6 times more YSOs than expected based on the deep exposure backgrounds in the four CMDs shown in Fig. \ref{fig:NGC346_F115W-F...W_CMDs}.
In comparison, we characterised only $\sim$~2~-~3 times more pre-MS stars than expected.
This shows that there is a clear presence of sources younger than 1~Myr in NGC~346, as these YSOs are all located on the right of the 1~Myr isochrone used to separate the pre-MS stars from the YSOs (see Section \ref{subsubsec:extinction_populations}).
In addition we selected 4,225 sources as either pre-MS with uncertainty overlapping the YSO region or as YSO with uncertainty overlapping the pre-MS region.
For the pre-MS candidates, YSO candidates or a combination of the two, we counted how many of these sources were characterised at least three times as such in the four CMDs.
This way, we characterised 1,004 pre-MS candidates, 1,476 YSO candidates and 209 pre-MS or YSO candidates.

\subsubsection{IR-excess sources}\label{subsubsec:IR-excess}
\begin{figure*}
    \centering
    \includegraphics[width=0.95\textwidth]{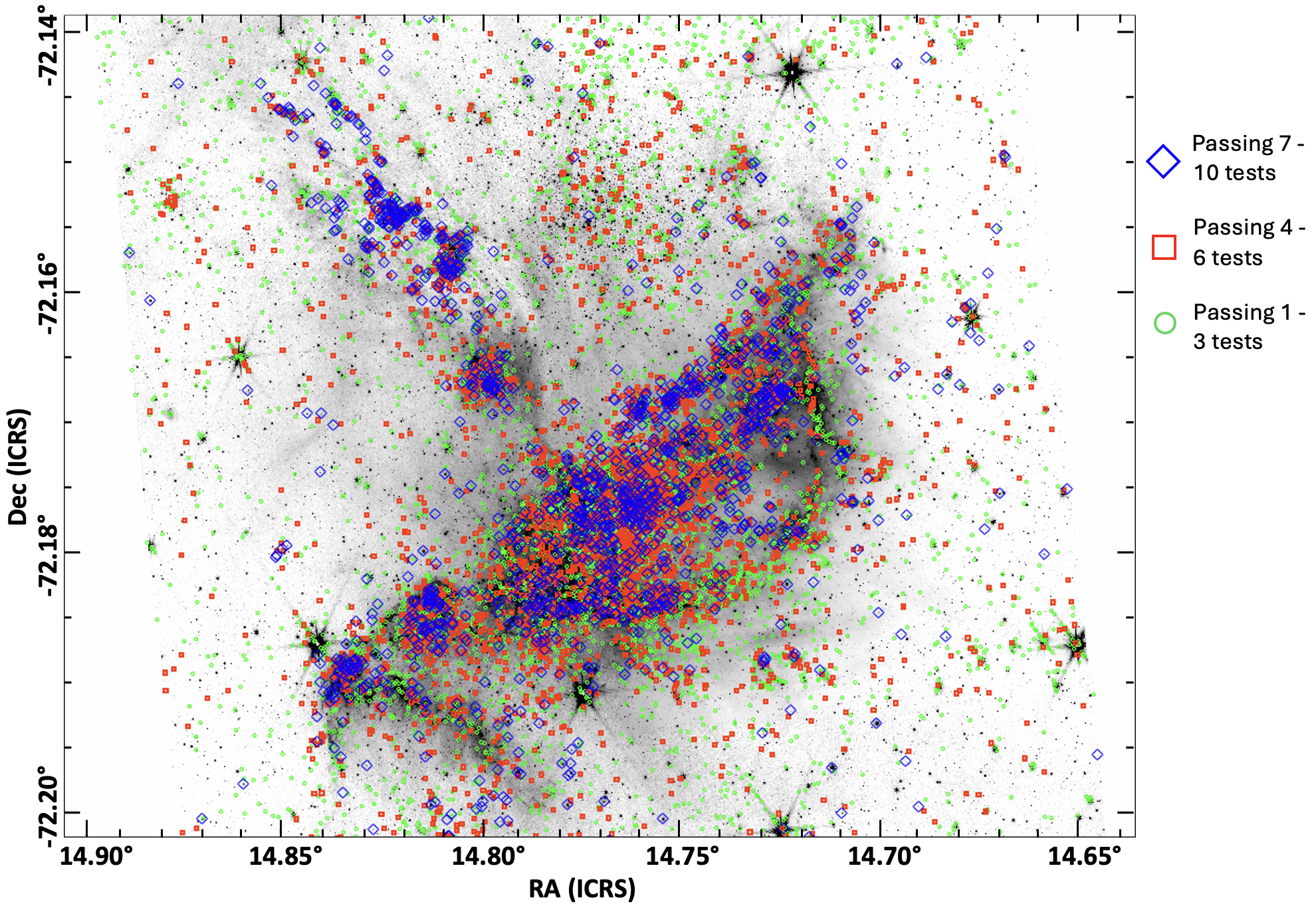}
    \caption{The F444W filter image, zoomed in on the main body of NGC~346, overlaid with all the sources that are selected as pre-MS, YSO or NIRCam IR-excess 7-10 times (blue diamonds), 4-6 times (red squares) and 1-3 times (green small circles).}
    \label{fig:F444W_NIRCam_tests}
\end{figure*}

Using the remaining six NIRCam CMDs, we characterised 23,819 sources with an IR-excess using the method described in Section \ref{subsec:characterisation}.
Table \ref{tab:NIRCam_IR_selection_total} shows the number of NIRCam IR-excess sources per CMD, with the expected number of SMC field NIRCam IR-excess sources and the contamination percentage. \newline
\indent We then combined the characterisation of pre-MS and YSO candidates with the NIRCam IR-excess sources, to select the strongest pre-MS and YSO candidates.
This gave a total of ten `tests', where a pass means that the source is selected as either pre-MS, YSO, a combination of pre-MS and YSO, or IR-excess. 
Sources that passed 7-10 tests are high-confidence candidates, 4-6 tests are mid-confidence candidates, and 1-3 tests are low-confidence candidates.
This resulted in 7992 low-confidence candidates, 3,761 mid-confidence candidates and 1,583 high-confidence candidates, which are shown on top of the F444W filter image, zoomed in on the main body of NGC~346, in Fig \ref{fig:F444W_NIRCam_tests}.
Note that we excluded sources characterised as galaxies in the count of the four NIRCam CMDs containing the F115W filter, even though there can be low-mass YSOs categorised as galaxies.
The high- and mid-confidence candidates are predominantly found in the dusty filaments.
The low-confidence sources mostly follow these as well, but there is also a slight over-density of these in the BS~90 cluster.
This can partly be explained by reddened RGB stars since some of these sources are located slightly on the right of the RGB branch in the F115W-F200W vs F115W CMD.
It could also be due to a misclassification of BS~90 sources as pre-MS stars, YSOs or IR-excess sources in one to three of the NIRCam CMDs, i.e. our low confidence sources. If any extra reddening is occurring in the BS~90 area, as suggested by the reddened RGB stars, then this could explain this over-density of low-confidence NIRCam candidates.

\begin{deluxetable}{l r r r r}
\tablewidth{0pt}
\tablecaption{MIRI CMD colour cuts for IR-excess source selection. The number of IR-excess sources per CMD is given, as well as the estimated number of contaminating IR-excess sources and the contamination percentage. \label{tab:MIRI_IR_selection_total}}
\tablehead{
\colhead{CMD} & \colhead{Colour cut} & \colhead{IR-} & \colhead{Contamination} & \colhead{\%} \\
\colhead{} & \colhead{[VegaMag]} & \colhead{excess} & \colhead{IR-excess } &
\colhead{contamination}
}
\startdata
F770W-F1000W  & $>$ 0.35 & 1,598 & 425 & 26.6 \\
F770W-F1130W  & $>$ 1.00 & 3,164 & 267 & 8.4 \\
F770W-F1500W  & $>$ 0.75 & 1,495 & 267 & 17.9 \\
F770W-F2100W  & $>$ 1.50 & 314 & 85 & 27.1 \\ \hline
F1000W-F1130W  & $>$ 0.85 & 2,273 & 134 & 5.9 \\
F1000W-F1500W  & $>$ 0.65 & 1,330 & 182 & 13.7 \\
F1000W-F2100W  & $>$ 1.20 & 302 & 85 & 28.1 \\ \hline
F1130W-F1500W  & $>$ 0.00 & 340 & 109 & 32.1 \\
F1130W-F2100W  & $>$ 0.40 & 286 & 85 & 29.7 \\ \hline
F1500W-F2100W  & $>$ 0.50 & 290 & 73 & 25.2 \\ \hline
Total Sources &   & 5,303 & - & - \\ 
\enddata
\end{deluxetable}

\begin{figure*}
    \centering
    \includegraphics[width=0.95\textwidth]{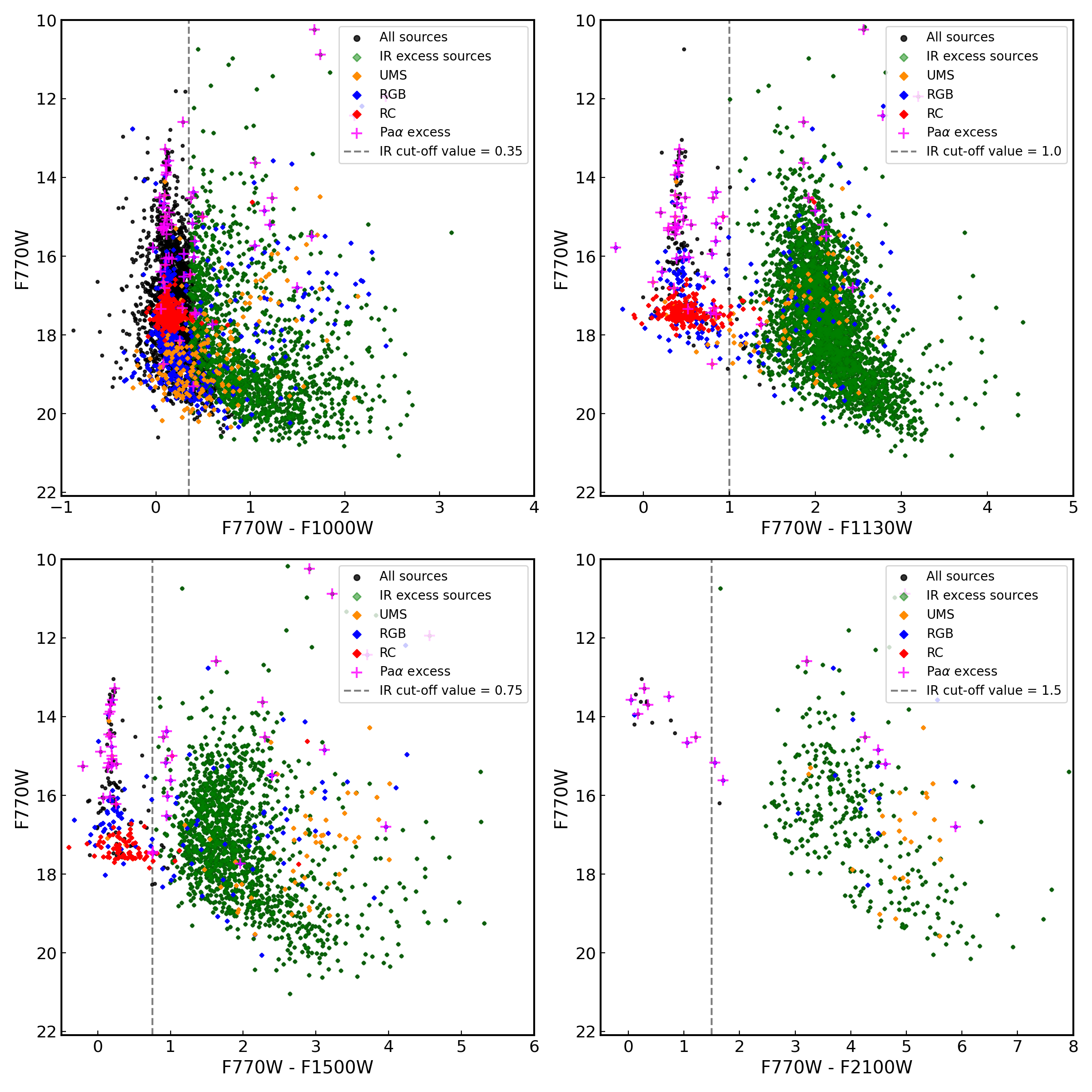}
    \caption{Four MIRI CMDS of the full catalogue (black dots) with the F770W filter on the vertical axis and the colour on the horizontal axis: F770W-F1000W top left, F770W-F1130W top right, F770W-F1500W bottom left and F770W-F2100W bottom right.
    The IR-excess cut-off value for each CMD is shown (gray dashed line), as well as sources that are categorised as UMS (yellow), RGB (blue) or RC (red) stars.
    Sources that have Pa$\alpha$ excess are shown with a magenta +.}
    \label{fig:multiple_MIRI_CMDs}
\end{figure*}

\begin{figure*}
    \centering
    \includegraphics[width=0.95\textwidth]{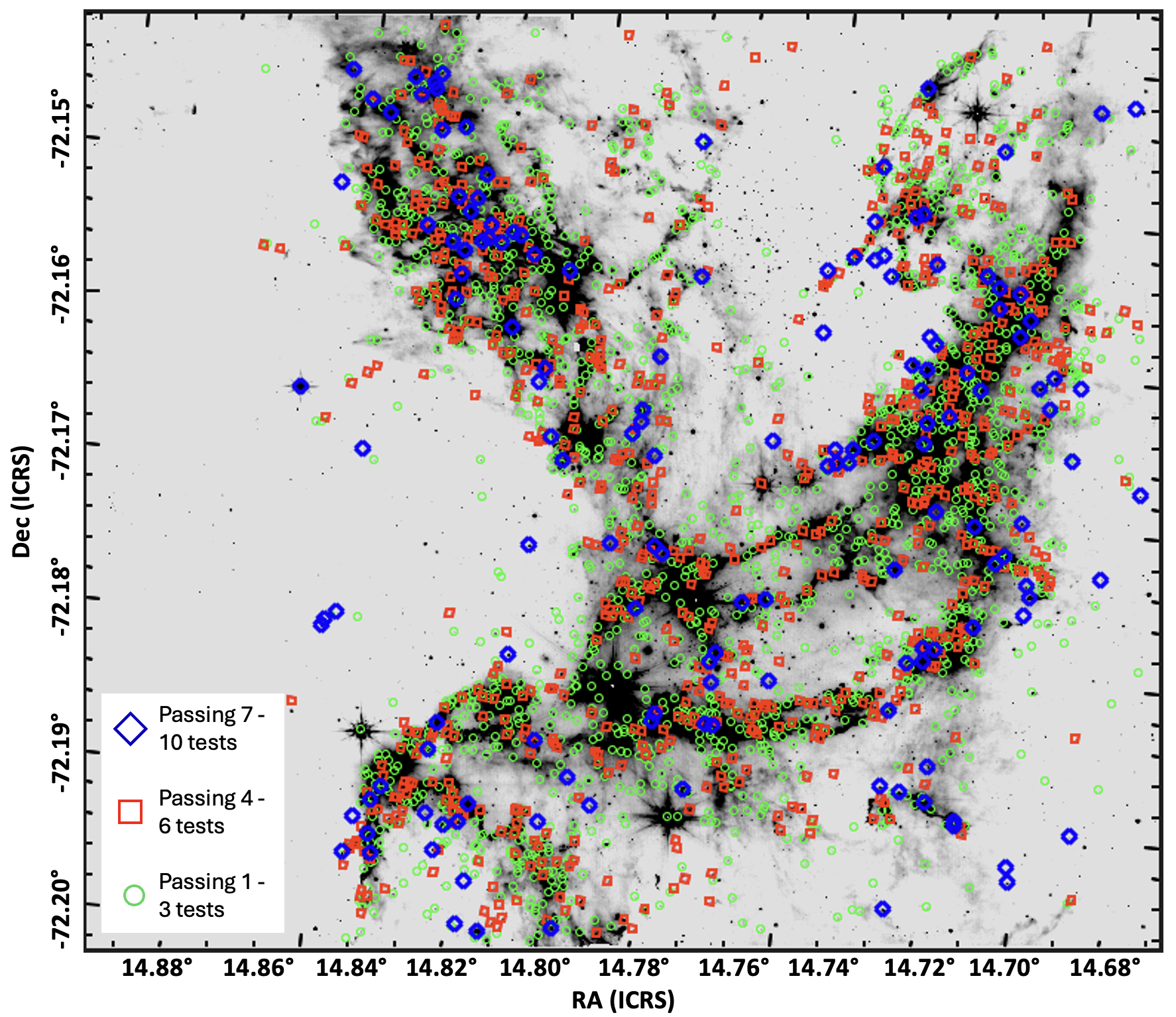}
    \caption{The F770W filter image overlaid with all the sources that are selected as MIRI IR-excess 7-10 times (blue diamonds), 4-6 times (red squares) and 1-3 times (green small circles).}
    \label{fig:F770W_MIRI_tests}
\end{figure*}

Utilising this selection method for high-, mid- and low-confidence candidates means that all high-confidence candidates are selected as either pre-MS, YSO or a combination of the two, but the mid- and low-confidence candidates can have passed the tests based on the IR-excess and be categorised as galaxies in some of the CMDs with F115W. \newline
\indent Table \ref{tab:MIRI_IR_selection_total} shows the number of MIRI IR-excess sources in each CMD, together with the estimated number of SMC field MIRI IR-excess sources and the contamination percentage. 
A total of 5,303 individual sources were categorised as IR-excess across all the MIRI CMDs.
Fig. \ref{fig:multiple_MIRI_CMDs} shows the four MIRI CMDs that include the F770W filter.
The UMS, RGB and RC stars are shown in orange, blue and red respectively, and Pa$\alpha$ excess sources as a pink `+'.
A small percentage of UMS sources were detected in MIRI and show IR-excess in the MIRI CMDs.
For the UMS stars that show a small amount of IR-excess in the F770W-F1000W vs F770W CMD, this IR-excess can be attributed to optically thin dust emission or free-free emission in a hot, dense ionised wind of OB type stars \citep[e.g.; ][]{Hartmann1977, Hovhannessian2001, Siebenmorgen2018, Deng2022}.
The UMS stars that show a significant IR-excess in the CMDs shown in Fig. \ref{fig:multiple_MIRI_CMDs} have steep rising SEDs in the MIRI wavelengths that cannot be explained by this optically thin dust emission or free-free emission.
This amount of IR-excess could be explained by either cirrus hot spots \citep[e.g. ][]{VanBuren1988, Adams2013} or disks.
Cirrus hot spots are the more likely of the two, because the vast majority of the SEDs peak at 21~\micron\ in our data, which shows that the dust is colder than what is expected for a disk.
\newline
\indent We combined the ten MIRI CMD characterisations to select the sources that were most often categorised as IR-excess.
Sources categorised as IR-excess in 1-3 CMDs are low-confidence, in 4-6 CMDs are mid-confidence and in 7-10 are high-confidence MIRI candidates.
This resulted in 1,749 low-confidence MIRI candidates, 730 mid-confidence MIRI candidates and 153 high-confidence MIRI candidates, which are shown on top of the F770W filter image in Fig. \ref{fig:F770W_MIRI_tests}. 54.9\% of the 153 high-confidence MIRI candidates are also high- or mid-confidence NIRCam candidates and 16.4\% of the 730 mid-confidence MIRI candidates are also high- or mid-confidence NIRCam candidates.
Besides the difference in the number of sources between the NIRCam and MIRI candidates, Fig \ref{fig:F444W_NIRCam_tests} and \ref{fig:F770W_MIRI_tests} show another difference: The MIRI IR-excess candidates follow almost exclusively the dust ridges, creating a gap between the Centre Arc and Main Arc (see Fig \ref{fig:4_filter_images} - bottom left panel), whereas the NIRCam IR-excess candidates do populate this area.
This is because the region between the Main Arc and Centre Arc is less dusty, which makes the stellar contribution more important for the detection.
The shortest NIRCam wavelength bands are more sensitive to this stellar contribution and therefore NIRCam is better equipped to detect sources in this area.
We will conduct a more detailed study of these IR-excess candidates in a forthcoming paper, performing SED model fitting to constrain the physical properties (e.g. stellar mass and temperature, disk and envelope masses and sizes) of each candidate, and we will perform comparative studies with other young stellar populations in the Milky Way and Magellanic Clouds.

\subsubsection{Paschen alpha excess} \label{subsec:paschen_alpha_results}
\begin{figure}
    \centering
    \includegraphics[width=1.0\linewidth]{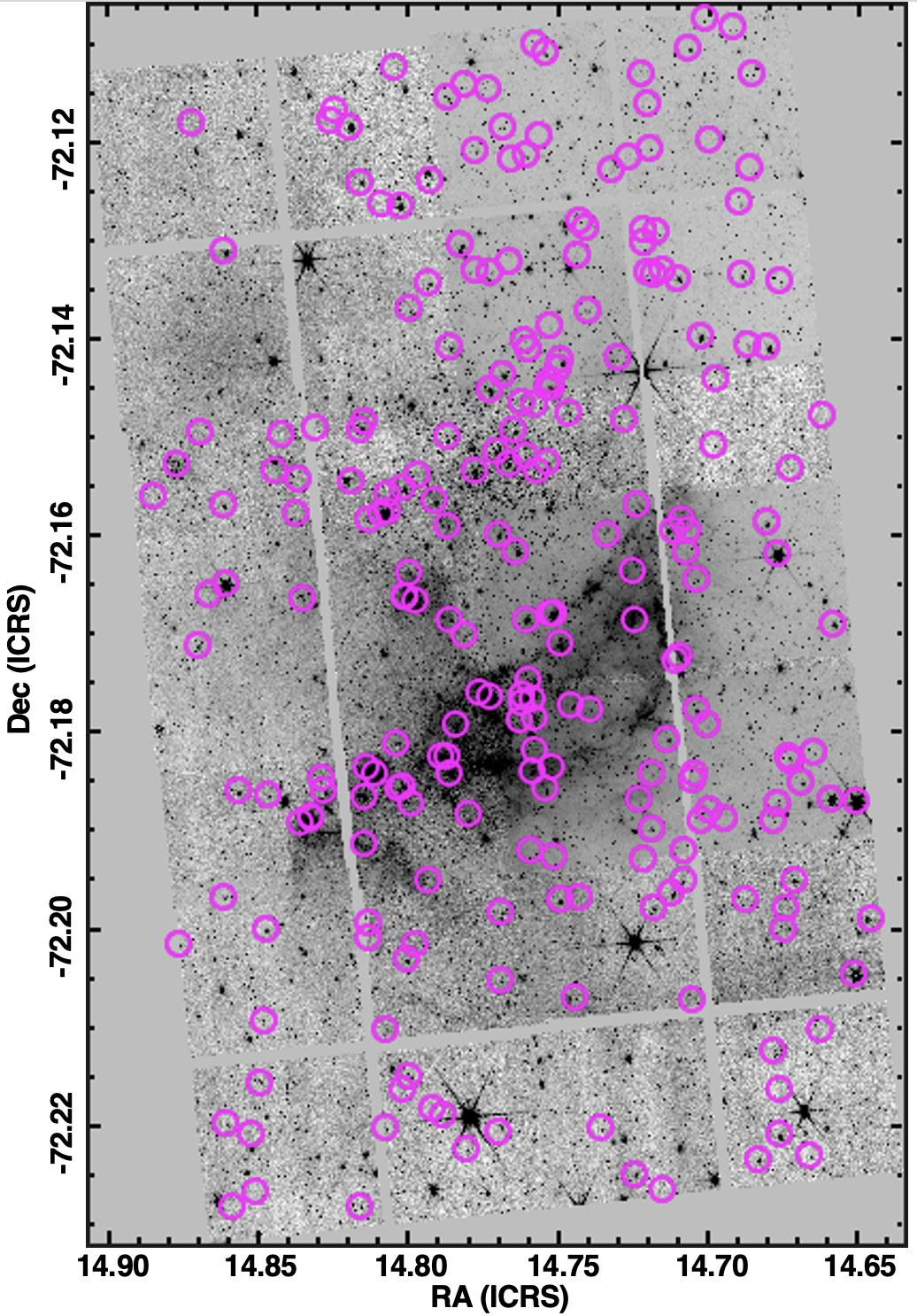}
    \caption{Showing the F200W NIRCam image with the 239 non-spurious Pa$\alpha$ excess sources overlaid in purple circles.
    The Pa$\alpha$ excess sources are more concentrated around the diffuse emission and in the deep exposure regions.}
    \label{fig:Pa-alpha_distribution}
\end{figure}

Fig \ref{fig:CCD_Paschen} shows the 239 non-spurious sources that are characterised as Pa$\alpha$ excess, as well as the reddening vector for an $\mathrm{A_V}$ = 3 for illustration purposes.
The reddening vector is almost parallel to the reference template, which implies that the selection of stars with Pa$\alpha$ excess would not be significantly affected by even relatively large uncertainties in the reddening correction.

The distribution of the non-spurious Pa$\alpha$ excess sources is shown in Fig. \ref{fig:Pa-alpha_distribution}.
The Pa$\alpha$ excess sources seem more concentrated around the diffuse emission and practically all sub-clusters found by \citet{Sabbi2007} contain at least one Pa$\alpha$ excess source, even though most of these sub-clusters are located in the shallow exposure areas.
This indicates presence of ongoing star formation in NGC~346.
Fig. \ref{fig:Pa-alpha_distribution} also shows that the distribution of the Pa$\alpha$ excess sources is favoured towards the deeper exposure areas, due to the reasonably poor performance of the F187N narrow band filter in the low exposure regions.
It is especially visible in the number of Pa$\alpha$ excess sources in the top left quadrant versus the top right quadrant.
This becomes even more clear when comparing the number of sources detected in the F187N filter and the number of sources expected in the whole field based on the deep exposure backgrounds (see Table \ref{tab:background_NIRCam_2}), as the expected number of sources is higher.

As is shown in the four NIRCam CMDs in Fig. \ref{fig:NGC346_F115W-F...W_CMDs}, most of the Pa$\alpha$ excess sources are found in the upper part of the CMDs.
The Pa$\alpha$ recombination line is characteristic of young stars undergoing mass accretion \citep{Jones2023}.
These bright sources with Pa$\alpha$ excess could be massive YSOs or pre-MS stars that are still transiting through the CMD.
However, another possibility for the Pa$\alpha$ excess sources within the UMS region is that these are Herbig Ae/Be objects.
These are young intermediate mass objects that are still forming and about to leave the MS, where the most massive of them become supergiants \citep{Brittain2023}.
These stars have been shown to contain Hydrogen recombination lines in their spectra, including Pa$\alpha$, as \citet{Rogers2024} found for several of their intermediate-mass pre-MS stars.
The more massive Herbig Ae/Be objects (Herbig Be) have ionising winds that create hydrogen recombination lines, which could also be a reason for the Pa$\alpha$ excess of these sources.
We note that \citet{Herbig1985}, \citet{Young1989}, and \citet{Strassmeier1990} showed that recombination lines can be present in MS stars through chromospheric emission, although the majority of those stars were of lower mass than Herbig Ae/Be stars.
Another possibility for the Pa$\alpha$ excess sources within the upper part of the RGB region are red giants (RGs), since these stars experience mass-loss and this material is ionised by the stars around them \citep[e.g. ][]{Wright2014}, as it is unlikely that the central star itself is hot enough to ionise the material.
\newline
\indent Some of the Pa$\alpha$ excess sources lie on the MS.
\citet{DeMarchi2024} showed that older pre-MS candidates having H$\alpha$ excess emission (with also excess in Paschen and Brackett line emission) appeared close to the MS in their V-I vs V CMD.
We therefore expect that the MS sources with Pa$\alpha$ excess are older pre-MS stars.

Our strict selection criteria for Pa$\alpha$ excess sources made it difficult to identify fainter pre-MS stars with Pa$\alpha$ excess, as these pre-MS stars are located in or near the dusty, filamentary regions, and the photometry in those regions is less accurate.
These sources therefore do not have $\mathrm{\delta_3}$ $<$ 0.10, which is why we only detected a small number of these, unlike \citet{Habel2024}, who used a different method for finding Pa$\alpha$ excess sources.

\subsection{Six different regions in NGC~346} \label{subsec:stellar_population_NGC346_polygons}
\begin{figure*}
    \centering
    \includegraphics[width=0.95\linewidth]{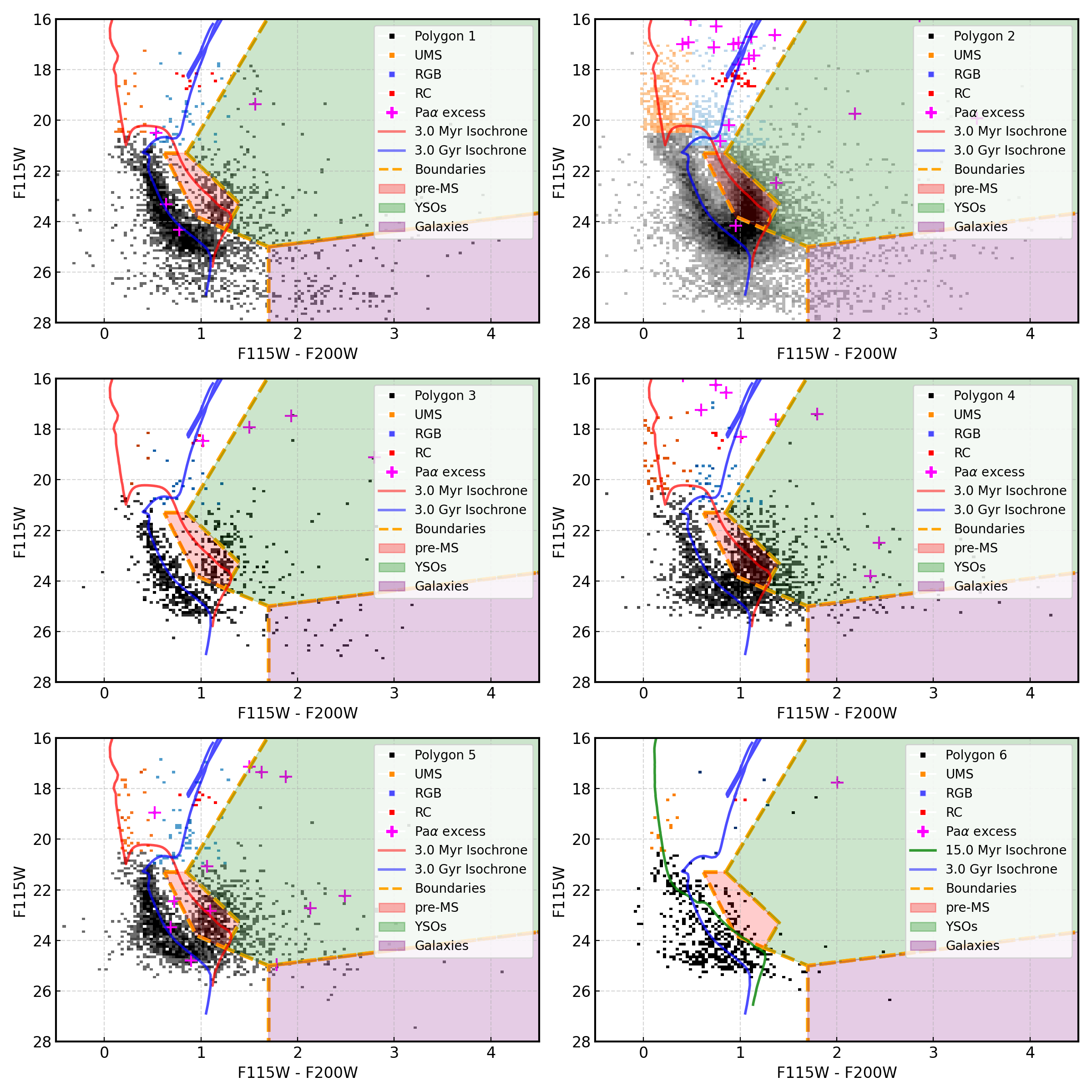}
    \caption{Showing the F115W vs F115W-F200W  CMDs for the six different regions (see Fig \ref{fig:4_filter_images} - top left panel): Region 1 (top left), Region 2 (top right), Region 3 (middle left), Region 4 (middle right), Region 5 (bottom left) and Region 6 (bottom right). 
    UMS, RGB and RC stars are shown in orange, blue and red, respectively. Pa$\alpha$ sources are shown with a pink +.
    The areas of the CMD for the categorisation of pre-MS (red shaded), YSOs (green shaded) and galaxies (purple shaded) are shown as well.
    Panel 1-5 shows the 3.0 Myr isochrone, as per the ages of the subclusters \citep{Sabbi2007} contained within the different regions and the 3.0 Gyr isochrone, the best fit for simultaneously the MSTO, RGB and RC (see Sect. \ref{subsubsec:extinction_populations}).
    For the same reason, panel 6 shows the 15.0 Myr and 3.0 Gyr isochrone.
    Regions 1 to 5 all have sources within the pre-MS and YSO region, while region 6 only has a small population in the pre-MS region.}
    \label{fig:F115W_F200W_6_polygons} 
\end{figure*}

In this section, we compare the stellar populations in the six regions from the top right panel of Fig \ref{fig:4_filter_images}.
The regions 1-6 have areas of $\sim$161.2 pc$^2$, $\sim$370.0 pc$^2$, $\sim$25.9 pc$^2$, $\sim$67.3 pc$^2$, $\sim$98.3 pc$^2$ and $\sim$14.5 pc$^2$, respectively.
The stellar density for each of the six regions is: $\sim$37.2 pc$^{-2}$, $\sim$58.0 pc$^{-2}$, $\sim$47.5 pc$^{-2}$, $\sim$47.8 pc$^{-2}$, $\sim$39.2 pc$^{-2}$ and $\sim$46.3 pc$^{-2}$.
In comparison, the stellar densities for the NIRCam shallower and deeper exposure background areas are respectively 13.5 $\pm$ 0.7 pc$^{-2}$ and 17.4 $\pm$ 0.5 pc$^{-2}$, so each of the six regions has between $\sim$~2.1~-~3.3 times higher source density than the high exposure background and $\sim$~2.8~-~4.3 times higher source density than the low exposure background.

For each of the six regions, the F115W-F200W vs F115W CMD is shown in Fig. \ref{fig:F115W_F200W_6_polygons}.
Two isochrones are also shown in each panel, one for the ages found by \citet{Sabbi2007} of the subclusters each region contains and one with an age of 3.0 Gyr for the older populations (see Section \ref{subsubsec:extinction_populations}).
A couple of differences are noticeable.
First, regions 1, 2 and 3 have a number of sources with magnitudes fainter than 26, because these regions are partly located in the deeper exposure areas.
Second, regions 1-5 all have a clear over-density of pre-MS and YSO sources, with respect to the SMC field population (see Fig \ref{fig:F115W-F200W_whole_with_bkg} - right panel), indicating that these regions are young and have ongoing star formation, in line with previous studies \citep[e.g. ][]{Rubio2000, Sabbi2007, Simon2007, DeMarchi2011, Jones2022, Habel2024}.
Region 6 only has a slight over-density of pre-MS sources, close to the MS, indicating that this region is more evolved than the others and that no new stars are being formed. \newline
\indent The other five regions do have Pa$\alpha$ excess sources, especially regions 2, 3, 4 and 5.
Since region 2 is the largest region and is located in the centre of NGC~346, where the most massive stars are located, it is expected to see plenty Pa$\alpha$ excess sources in this region, especially bright sources (see Fig \ref{fig:F115W_F200W_6_polygons} - top right panel).
Regions 3, 4 and 5 are smaller than region 1, with areas only having $\sim$16.1\%\, $\sim$41.7\%\ and $\sim$61.0\% of the size of region 1, but containing respectively the same number, $\sim$2.5 times and $\sim$3 times the number of Pa$\alpha$ excess sources.
This could indicate that the star formation rate in region 1 is lower than in regions 2, 3, 4 and 5, even though the region shows an appreciable amount of nebulosity and dust (see Fig. \ref{fig:4_filter_images}).
Another explanation is that the high-mass stars are still embedded, as suggested by \citet{Cignoni2011}.
Since our Pa$\alpha$ excess selection is more sensitive towards these high-mass sources in the dusty regions, the lack of high-mass sources in this region leads directly to a lower number of Pa$\alpha$ excess sources. \newline
\indent The stellar density of region 1 is only at $\sim$64.1-80.4\% of the stellar densities of regions 2, 3, 4 and 6, even though it is completely located within the deeper exposure area, while it is on par with the stellar density of region 5.
This can have a couple of reasons.
First, region 5 includes areas with lower stellar densities between the subclusters Sc13-Sc15 from \citet{Sabbi2007}, lowering the overall stellar density in this region.
Second, region 5 is entirely located in the shallow exposure area, resulting in a generally lower stellar density.
Finally, the slightly lower stellar density of region 1 can also be caused by the sources in this region being more embedded.

\section{Conclusion}\label{sec:conclusion}
In this work, we present the full \jwst\ NIRCam and MIRI photometric catalogue of the low-metallicity star formation region NGC~346 in the SMC.
Utilizing the spatial resolution and sensitivity of \jwst\ we characterise multiple stellar populations in the near- and mid-IR in the region.

\begin{enumerate}
    \item We produce a NIRCam and MIRI combined \jwst\ photometric catalogue of NGC~346 with \texttt{STARBUGII}, with a spatial resolution $\sim$10$\times$ higher than the previous Spitzer observations.
    Our \jwst\ observations also reach over 10 magnitudes below the Spitzer observations for the youngest, most embedded objects at comparable wavelengths, as well as 2 magnitudes below \hst\ for more evolved pre-MS sources, reaching $\sim$0.1 \msun.
    \item We characterise the UMS, RGB and RC populations using the F115W-F200W vs F115W CMD and show the location of these sources in other NIRCam and MIRI CMDs.
    \item We characterise pre-MS stars and YSOs using multiple NIRCam CMDs. We further characterise sources with an IR-excess in NIRCam and MIRI CMDs.
    This way, we select candidate pre-MS stars and YSOs, resulting in 3,761 mid-confidence and 1,583 high-confidence candidates in NIRCam, as well as 730 mid-confidence and 153 high-confidence MIRI candidates.
    In a future paper, we will study these candidate pre-MS stars and YSOs in more detail, using SED modelling to constrain the physical parameters of each candidate.
    \item We show that there is a significant number of sources younger than 1 Myr in NGC~346, besides also containing older populations, showing the ongoing and complex star formation history.
    \item We characterise 239 non-spurious sources with Pa$\alpha$ excess using the F115W-F200W vs F115W-F187N CCD, showing that there is ongoing accretion.
\end{enumerate}

\begin{acknowledgments}
The authors thank the anonymous referee whose detailed and constructive suggestions strengthened the paper.
This work is based on observations made with the NASA/ESA/CSA James Webb Space Telescope. The data were obtained from the Mikulski Archive for Space Telescopes at the Space Telescope Science Institute, which is operated by the Association of Universities for Research in Astronomy, Inc., under NASA contract NAS 5-03127 for JWST. These observations are associated with program \#1227.
All of the data presented in this paper were obtained from the Mikulski Archive for Space Telescopes (MAST) at the Space Telescope Science Institute. The specific observations analyzed can be accessed via doi:10.17909/sz47-c239.

JJ and PJK acknowledge support from the Research Ireland Pathway programme under Grant Number 21/PATH-S/9360.
OCJ acknowledge support from an STFC Webb fellowship. 
MM and NH acknowledge support through a NASA/JWST task plan 71-209636, and MM and LL acknowledge support from the NSF through grant 2054178.
MM acknowledges that a portion of her research was carried out at the Jet Propulsion Laboratory, California Institute of Technology, under a contract with the National Aeronautics and Space Administration (80NM0018D0004).
\end{acknowledgments}



\bibliography{refs.bib}

@ARTICLE{Habel2024,
       author = {{Habel}, Nolan and {Nally}, Conor and {Lenki{\'c}}, Laura and {Meixner}, Margaret and {De Marchi}, Guido and {Kavanagh}, Patrick J. and {Fahrion}, Katja and {Nayak}, Omnarayani and {Hirschauer}, Alec S. and {Jones}, Olivia C. and {Biazzo}, Katia and {Brandl}, Bernhard R. and {Jaspers}, J. and {Pontoppidan}, Klaus M. and {Robberto}, Massimo and {Rogers}, C. and {Sabbi}, E. and {Sargent}, B.~A. and {Soderblom}, David R. and {Zeidler}, Peter},
        title = "{Young Stellar Objects in NGC 346: A JWST NIRCam/MIRI Imaging Survey}",
      journal = {\apj},
         year = 2024,
        month = aug,
       volume = {971},
       number = {1},
          eid = {108},
        pages = {108},
          doi = {10.3847/1538-4357/ad5343},
archivePrefix = {arXiv},
       eprint = {2404.16242},
 primaryClass = {astro-ph.SR},
       adsurl = {https://ui.adsabs.harvard.edu/abs/2024ApJ...971..108H}
}

@ARTICLE{deGrijs2015,
       author = {{de Grijs}, Richard and {Bono}, Giuseppe},
        title = "{Clustering of Local Group Distances: Publication Bias or Correlated Measurements? III. The Small Magellanic Cloud}",
      journal = {\aj},
         year = 2015,
        month = jun,
       volume = {149},
       number = {6},
          eid = {179},
        pages = {179},
          doi = {10.1088/0004-6256/149/6/179},
archivePrefix = {arXiv},
       eprint = {1504.00417},
 primaryClass = {astro-ph.SR},
       adsurl = {https://ui.adsabs.harvard.edu/abs/2015AJ....149..179D}
}

@ARTICLE{Russell1992,
       author = {{Russell}, Stephen C. and {Dopita}, Michael A.},
        title = "{Abundances of the Heavy Elements in the Magellanic Clouds. III. Interpretation of Results}",
      journal = {\apj},
         year = 1992,
        month = jan,
       volume = {384},
        pages = {508},
          doi = {10.1086/170893},
       adsurl = {https://ui.adsabs.harvard.edu/abs/1992ApJ...384..508R}
}

@ARTICLE{Hony2015,
       author = {{Hony}, S. and {Gouliermis}, D.~A. and {Galliano}, F. and {Galametz}, M. and {Cormier}, D. and {Chen}, C. -H.~R. and {Dib}, S. and {Hughes}, A. and {Klessen}, R.~S. and {Roman-Duval}, J. and {Smith}, L. and {Bernard}, J. -P. and {Bot}, C. and {Carlson}, L. and {Gordon}, K. and {Indebetouw}, R. and {Lebouteiller}, V. and {Lee}, M. -Y. and {Madden}, S.~C. and {Meixner}, M. and {Oliveira}, J. and {Rubio}, M. and {Sauvage}, M. and {Wu}, R.},
        title = "{Star formation rates from young-star counts and the structure of the ISM across the NGC 346/N66 complex in the SMC}",
      journal = {\mnras},
         year = 2015,
        month = apr,
       volume = {448},
       number = {2},
        pages = {1847-1862},
          doi = {10.1093/mnras/stv107},
archivePrefix = {arXiv},
       eprint = {1501.03634},
 primaryClass = {astro-ph.GA},
       adsurl = {https://ui.adsabs.harvard.edu/abs/2015MNRAS.448.1847H}
}

@ARTICLE{Choudhury2020,
       author = {{Choudhury}, Samyaday and {de Grijs}, Richard and {Rubele}, Stefano and {Bekki}, Kenji and {Cioni}, Maria-Rosa L. and {Ivanov}, Valentin D. and {van Loon}, Jacco Th and {Niederhofer}, Florian and {Oliveira}, Joana M. and {Ripepi}, Vincenzo},
        title = "{The VMC survey - XXXIX. Mapping metallicity trends in the Small Magellanic Cloud using near-infrared passbands}",
      journal = {\mnras},
         year = 2020,
        month = sep,
       volume = {497},
       number = {3},
        pages = {3746-3760},
          doi = {10.1093/mnras/staa2140},
archivePrefix = {arXiv},
       eprint = {2007.08753},
 primaryClass = {astro-ph.GA},
       adsurl = {https://ui.adsabs.harvard.edu/abs/2020MNRAS.497.3746C}
}

@ARTICLE{Madau2014,
       author = {{Madau}, Piero and {Dickinson}, Mark},
        title = "{Cosmic Star-Formation History}",
      journal = {\araa},
         year = 2014,
        month = aug,
       volume = {52},
        pages = {415-486},
          doi = {10.1146/annurev-astro-081811-125615},
archivePrefix = {arXiv},
       eprint = {1403.0007},
 primaryClass = {astro-ph.CO},
       adsurl = {https://ui.adsabs.harvard.edu/abs/2014ARA&A..52..415M}
}

@ARTICLE{DeMarchi2011,
       author = {{De Marchi}, Guido and {Panagia}, Nino and {Romaniello}, Martino and {Sabbi}, Elena and {Sirianni}, Marco and {Prada Moroni}, Pier Giorgio and {Degl'Innocenti}, Scilla},
        title = "{Photometric Determination of the Mass Accretion Rates of Pre-main-sequence Stars. II. NGC 346 in the Small Magellanic Cloud}",
      journal = {\apj},
         year = 2011,
        month = oct,
       volume = {740},
       number = {1},
          eid = {11},
        pages = {11},
          doi = {10.1088/0004-637X/740/1/11},
archivePrefix = {arXiv},
       eprint = {1104.4494},
 primaryClass = {astro-ph.SR},
       adsurl = {https://ui.adsabs.harvard.edu/abs/2011ApJ...740...11D}
}

@ARTICLE{DeMarchi2011b,
       author = {{De Marchi}, Guido and {Panagia}, Nino and {Sabbi}, Elena},
        title = "{Clues to the Star Formation in NGC 346 across Time and Space}",
      journal = {\apj},
         year = 2011,
        month = oct,
       volume = {740},
       number = {1},
          eid = {10},
        pages = {10},
          doi = {10.1088/0004-637X/740/1/10},
archivePrefix = {arXiv},
       eprint = {1106.5780},
 primaryClass = {astro-ph.SR},
       adsurl = {https://ui.adsabs.harvard.edu/abs/2011ApJ...740...10D}
}

@ARTICLE{Sabbi2007,
       author = {{Sabbi}, E. and {Sirianni}, M. and {Nota}, A. and {Tosi}, M. and {Gallagher}, J. and {Meixner}, M. and {Oey}, M.~S. and {Walterbos}, R. and {Pasquali}, A. and {Smith}, L.~J. and {Angeretti}, L.},
        title = "{Past and Present Star Formation in the SMC: NGC 346 and its Neighborhood}",
      journal = {\aj},
         year = 2007,
        month = jan,
       volume = {133},
       number = {1},
        pages = {44-57},
          doi = {10.1086/509257},
archivePrefix = {arXiv},
       eprint = {astro-ph/0609330},
 primaryClass = {astro-ph},
       adsurl = {https://ui.adsabs.harvard.edu/abs/2007AJ....133...44S}
}

@ARTICLE{Bouret2003,
       author = {{Bouret}, J. -C. and {Lanz}, T. and {Hillier}, D.~J. and {Heap}, S.~R. and {Hubeny}, I. and {Lennon}, D.~J. and {Smith}, L.~J. and {Evans}, C.~J.},
        title = "{Quantitative Spectroscopy of O Stars at Low Metallicity: O Dwarfs in NGC 346}",
      journal = {\apj},
         year = 2003,
        month = oct,
       volume = {595},
       number = {2},
        pages = {1182-1205},
          doi = {10.1086/377368},
archivePrefix = {arXiv},
       eprint = {astro-ph/0301454},
 primaryClass = {astro-ph},
       adsurl = {https://ui.adsabs.harvard.edu/abs/2003ApJ...595.1182B}
}

@misc{Willot2022,
    author       = {Willott, C.},
    title        = {{jwst - Tools for processing and analyzing JWST data}},
    month        = aug,
    year         = 2022,
    url          = {https://github.com/chriswillott/jwst}
    }

@INPROCEEDINGS{Rieke2005,
       author = {{Rieke}, Marcia J. and {Kelly}, Douglas and {Horner}, Scott},
        title = "{Overview of James Webb Space Telescope and NIRCam's Role}",
    booktitle = {Cryogenic Optical Systems and Instruments XI},
         year = 2005,
       editor = {{Heaney}, James B. and {Burriesci}, Lawrence G.},
       series = {Society of Photo-Optical Instrumentation Engineers (SPIE) Conference Series},
       volume = {5904},
        month = aug,
        pages = {1-8},
          doi = {10.1117/12.615554},
       adsurl = {https://ui.adsabs.harvard.edu/abs/2005SPIE.5904....1R}
}

@misc{Rest2023,
    author       = {Rest, A.},
    title        = {{JWST/HST Alignment Tool}},
    month        = mar,
    year         = 2023,
    url          = {https://github.com/arminrest/jhat}
    }

@ARTICLE{Schlawin2020,
       author = {{Schlawin}, Everett and {Leisenring}, Jarron and {Misselt}, Karl and {Greene}, Thomas P. and {McElwain}, Michael W. and {Beatty}, Thomas and {Rieke}, Marcia},
        title = "{JWST Noise Floor. I. Random Error Sources in JWST NIRCam Time Series}",
      journal = {\aj},
         year = 2020,
        month = nov,
       volume = {160},
       number = {5},
          eid = {231},
        pages = {231},
          doi = {10.3847/1538-3881/abb811},
archivePrefix = {arXiv},
       eprint = {2010.03564},
 primaryClass = {astro-ph.IM},
       adsurl = {https://ui.adsabs.harvard.edu/abs/2020AJ....160..231S}
}

@software{Starbug,
       author = {{Nally}, Conor},
        title = "{StarbugII: JWST PSF photometry for crowded fields}",
 howpublished = {Astrophysics Source Code Library, record ascl:2309.012},
         year = 2023,
        month = sep,
          eid = {ascl:2309.012},
       adsurl = {https://ui.adsabs.harvard.edu/abs/2023ascl.soft09012N}
}

@ARTICLE{Nally2024,
       author = {{Nally}, Conor and {Jones}, Olivia C. and {Lenki{\'c}}, Laura and {Habel}, Nolan and {Hirschauer}, Alec S. and {Meixner}, Margaret and {Kavanagh}, P.~J. and {Boyer}, Martha L. and {Ferguson}, Annette M.~N. and {Sargent}, B.~A. and {Nayak}, Omnarayani and {Temim}, Tea},
        title = "{JWST MIRI and NIRCam unveil previously unseen infrared stellar populations in NGC 6822}",
      journal = {\mnras},
         year = 2024,
        month = jun,
       volume = {531},
       number = {1},
        pages = {183-198},
          doi = {10.1093/mnras/stae1163},
archivePrefix = {arXiv},
       eprint = {2309.13521},
 primaryClass = {astro-ph.GA},
       adsurl = {https://ui.adsabs.harvard.edu/abs/2024MNRAS.531..183N}
}

@INPROCEEDINGS{Perrin2014,
       author = {{Perrin}, Marshall D. and {Sivaramakrishnan}, Anand and {Lajoie}, Charles-Philippe and {Elliott}, Erin and {Pueyo}, Laurent and {Ravindranath}, Swara and {Albert}, Lo{\"\i}c.},
        title = "{Updated point spread function simulations for JWST with WebbPSF}",
    booktitle = {Space Telescopes and Instrumentation 2014: Optical, Infrared, and Millimeter Wave},
         year = 2014,
       editor = {{Oschmann}, Jacobus M., Jr. and {Clampin}, Mark and {Fazio}, Giovanni G. and {MacEwen}, Howard A.},
       series = {Society of Photo-Optical Instrumentation Engineers (SPIE) Conference Series},
       volume = {9143},
        month = aug,
          eid = {91433X},
        pages = {91433X},
          doi = {10.1117/12.2056689},
       adsurl = {https://ui.adsabs.harvard.edu/abs/2014SPIE.9143E..3XP}
}

@INPROCEEDINGS{TOPCAT,
       author = {{Taylor}, M.~B.},
        title = "{TOPCAT \& STIL: Starlink Table/VOTable Processing Software}",
    booktitle = {Astronomical Data Analysis Software and Systems XIV},
         year = 2005,
       editor = {{Shopbell}, P. and {Britton}, M. and {Ebert}, R.},
       series = {Astronomical Society of the Pacific Conference Series},
       volume = {347},
        month = dec,
        pages = {29},
       adsurl = {https://ui.adsabs.harvard.edu/abs/2005ASPC..347...29T}
}

@ARTICLE{Rieke2023,
       author = {{Rieke}, Marcia J. and {Kelly}, Douglas M. and {Misselt}, Karl and {Stansberry}, John and {Boyer}, Martha and {Beatty}, Thomas and {Egami}, Eiichi and {Florian}, Michael and {Greene}, Thomas P. and {Hainline}, Kevin and {Leisenring}, Jarron and {Roellig}, Thomas and {Schlawin}, Everett and {Sun}, Fengwu and {Tinnin}, Lee and {Williams}, Christina C. and {Willmer}, Christopher N.~A. and {Wilson}, Debra and {Clark}, Charles R. and {Rohrbach}, Scott and {Brooks}, Brian and {Canipe}, Alicia and {Correnti}, Matteo and {DiFelice}, Audrey and {Gennaro}, Mario and {Girard}, Julian and {Hartig}, George and {Hilbert}, Bryan and {Koekemoer}, Anton M. and {Nikolov}, Nikolay K. and {Pirzkal}, Norbert and {Rest}, Armin and {Robberto}, Massimo and {Sunnquist}, Ben and {Telfer}, Randal and {Wu}, Chi Rai and {Ferry}, Malcolm and {Lewis}, Dan and {Baum}, Stefi and {Beichman}, Charles and {Doyon}, Ren{\'e} and {Dressler}, Alan and {Eisenstein}, Daniel J. and {Ferrarese}, Laura and {Hodapp}, Klaus and {Horner}, Scott and {Jaffe}, Daniel T. and {Johnstone}, Doug and {Krist}, John and {Martin}, Peter and {McCarthy}, Donald W. and {Meyer}, Michael and {Rieke}, George H. and {Trauger}, John and {Young}, Erick T.},
        title = "{Performance of NIRCam on JWST in Flight}",
      journal = {\pasp},
         year = 2023,
        month = feb,
       volume = {135},
       number = {1044},
          eid = {028001},
        pages = {028001},
          doi = {10.1088/1538-3873/acac53},
archivePrefix = {arXiv},
       eprint = {2212.12069},
 primaryClass = {astro-ph.IM},
       adsurl = {https://ui.adsabs.harvard.edu/abs/2023PASP..135b8001R}
}

@ARTICLE{Rieke2015,
       author = {{Rieke}, G.~H. and {Wright}, G.~S. and {B{\"o}ker}, T. and {Bouwman}, J. and {Colina}, L. and {Glasse}, Alistair and {Gordon}, K.~D. and {Greene}, T.~P. and {G{\"u}del}, Manuel and {Henning}, Th. and {Justtanont}, K. and {Lagage}, P. -O. and {Meixner}, M.~E. and {N{\o}rgaard-Nielsen}, H. -U. and {Ray}, T.~P. and {Ressler}, M.~E. and {van Dishoeck}, E.~F. and {Waelkens}, C.},
        title = "{The Mid-Infrared Instrument for the James Webb Space Telescope, I: Introduction}",
      journal = {\pasp},
         year = 2015,
        month = jul,
       volume = {127},
       number = {953},
        pages = {584},
          doi = {10.1086/682252},
archivePrefix = {arXiv},
       eprint = {1508.02294},
 primaryClass = {astro-ph.IM},
       adsurl = {https://ui.adsabs.harvard.edu/abs/2015PASP..127..584R}
}

@ARTICLE{Wright2023,
       author = {{Wright}, Gillian S. and {Rieke}, George H. and {Glasse}, Alistair and {Ressler}, Michael and {Garc{\'\i}a Mar{\'\i}n}, Macarena and {Aguilar}, Jonathan and {Alberts}, Stacey and {{\'A}lvarez-M{\'a}rquez}, Javier and {Argyriou}, Ioannis and {Banks}, Kimberly and {Baudoz}, Pierre and {Boccaletti}, Anthony and {Bouchet}, Patrice and {Bouwman}, Jeroen and {Brandl}, Bernard R. and {Breda}, David and {Bright}, Stacey and {Cale}, Steven and {Colina}, Luis and {Cossou}, Christophe and {Coulais}, Alain and {Cracraft}, Misty and {De Meester}, Wim and {Dicken}, Daniel and {Engesser}, Michael and {Etxaluze}, Mireya and {Fox}, Ori D. and {Friedman}, Scott and {Fu}, Henry and {Gasman}, Danny and {G{\'a}sp{\'a}r}, Andr{\'a}s and {Gastaud}, Ren{\'e} and {Geers}, Vincent and {Glauser}, Adrian Michael and {Gordon}, Karl D. and {Greene}, Thomas and {Greve}, Thomas R. and {Grundy}, Timothy and {G{\"u}del}, Manuel and {Guillard}, Pierre and {Haderlein}, Peter and {Hashimoto}, Ryan and {Henning}, Thomas and {Hines}, Dean and {Holler}, Bryan and {Detre}, {\"O}rs Hunor and {Jahromi}, Amir and {James}, Bryan and {Jones}, Olivia C. and {Justtanont}, Kay and {Kavanagh}, Patrick and {Kendrew}, Sarah and {Klaassen}, Pamela and {Krause}, Oliver and {Labiano}, Alvaro and {Lagage}, Pierre-Olivier and {Lambros}, Scott and {Larson}, Kirsten and {Law}, David and {Lee}, David and {Libralato}, Mattia and {Lorenzo Alverez}, Jose and {Meixner}, Margaret and {Morrison}, Jane and {Mueller}, Migo and {Murray}, Katherine and {Mycroft}, Matthew and {Myers}, Richard and {Nayak}, Omnarayani and {Naylor}, Bret and {Nickson}, Bryony and {Noriega-Crespo}, Alberto and {{\"O}stlin}, G{\"o}ran and {O'Sullivan}, Brian and {Ottens}, Richard and {Patapis}, Polychronis and {Penanen}, Konstantin and {Pietraszkiewicz}, Martin and {Ray}, Tom and {Regan}, Michael and {Roteliuk}, Anthony and {Royer}, Pierre and {Samara-Ratna}, Piyal and {Samuelson}, Bridget and {Sargent}, Beth A. and {Scheithauer}, Silvia and {Schneider}, Analyn and {Schreiber}, J{\"u}rgen and {Shaughnessy}, Bryan and {Sheehan}, Even and {Shivaei}, Irene and {Sloan}, G.~C. and {Tamas}, Laszlo and {Teague}, Kelly and {Temim}, Tea and {Tikkanen}, Tuomo and {Tustain}, Samuel and {van Dishoeck}, Ewine F. and {Vandenbussche}, Bart and {Weilert}, Mark and {Whitehouse}, Paul and {Wolff}, Schuyler},
        title = "{The Mid-infrared Instrument for JWST and Its In-flight Performance}",
      journal = {\pasp},
         year = 2023,
        month = apr,
       volume = {135},
       number = {1046},
          eid = {048003},
        pages = {048003},
          doi = {10.1088/1538-3873/acbe66},
       adsurl = {https://ui.adsabs.harvard.edu/abs/2023PASP..135d8003W}
}

@ARTICLE{Knuth2006,
       author = {{Knuth}, Kevin H.},
        title = "{Optimal Data-Based Binning for Histograms}",
      journal = {arXiv e-prints},
         year = 2006,
        month = may,
          eid = {physics/0605197},
        pages = {physics/0605197},
          doi = {10.48550/arXiv.physics/0605197},
archivePrefix = {arXiv},
       eprint = {physics/0605197},
 primaryClass = {physics.data-an},
       adsurl = {https://ui.adsabs.harvard.edu/abs/2006physics...5197K}
}

@ARTICLE{Cardelli1989,
       author = {{Cardelli}, Jason A. and {Clayton}, Geoffrey C. and {Mathis}, John S.},
        title = "{The Relationship between Infrared, Optical, and Ultraviolet Extinction}",
      journal = {\apj},
         year = 1989,
        month = oct,
       volume = {345},
        pages = {245},
          doi = {10.1086/167900},
       adsurl = {https://ui.adsabs.harvard.edu/abs/1989ApJ...345..245C}
}

@ARTICLE{White2003,
       author = {{White}, Russel J. and {Basri}, Gibor},
        title = "{Very Low Mass Stars and Brown Dwarfs in Taurus-Auriga}",
      journal = {\apj},
         year = 2003,
        month = jan,
       volume = {582},
       number = {2},
        pages = {1109-1122},
          doi = {10.1086/344673},
archivePrefix = {arXiv},
       eprint = {astro-ph/0209164},
 primaryClass = {astro-ph},
       adsurl = {https://ui.adsabs.harvard.edu/abs/2003ApJ...582.1109W}
}

@ARTICLE{Simon2007,
       author = {{Simon}, Joshua D. and {Bolatto}, Alberto D. and {Whitney}, Barbara A. and {Robitaille}, Thomas P. and {Shah}, Ronak Y. and {Makovoz}, David and {Stanimirovi{\'c}}, Sne{\v{z}}ana and {Barb{\'a}}, Rodolfo H. and {Rubio}, M{\'o}nica},
        title = "{The Spitzer Survey of the Small Magellanic Cloud: Discovery of Embedded Protostars in the H II Region NGC 346}",
      journal = {\apj},
         year = 2007,
        month = nov,
       volume = {669},
       number = {1},
        pages = {327-336},
          doi = {10.1086/521544},
archivePrefix = {arXiv},
       eprint = {0707.3998},
 primaryClass = {astro-ph},
       adsurl = {https://ui.adsabs.harvard.edu/abs/2007ApJ...669..327S}
}

@ARTICLE{Bica1995,
       author = {{Bica}, Eduardo L.~D. and {Schmitt}, Henrique R.},
        title = "{A Revised and Extended Catalog of Magellanic System Clusters, Associations, and Emission Nebulae. I. Small Magellanic Cloud and Bridge}",
      journal = {\apjs},
         year = 1995,
        month = nov,
       volume = {101},
        pages = {41},
          doi = {10.1086/192233},
       adsurl = {https://ui.adsabs.harvard.edu/abs/1995ApJS..101...41B}
}

@ARTICLE{Gaia2021,
       author = {{Gaia Collaboration} and {Brown}, A.~G.~A. and {Vallenari}, A. and {Prusti}, T. and {de Bruijne}, J.~H.~J. and {Babusiaux}, C. and {Biermann}, M. and {Creevey}, O.~L. and {Evans}, D.~W. and {Eyer}, L. and {Hutton}, A. and {Jansen}, F. and {Jordi}, C. and {Klioner}, S.~A. and {Lammers}, U. and {Lindegren}, L. and {Luri}, X. and {Mignard}, F. and {Panem}, C. and {Pourbaix}, D. and {Randich}, S. and {Sartoretti}, P. and {Soubiran}, C. and {Walton}, N.~A. and {Arenou}, F. and {Bailer-Jones}, C.~A.~L. and {Bastian}, U. and {Cropper}, M. and {Drimmel}, R. and {Katz}, D. and {Lattanzi}, M.~G. and {van Leeuwen}, F. and {Bakker}, J. and {Cacciari}, C. and {Casta{\~n}eda}, J. and {De Angeli}, F. and {Ducourant}, C. and {Fabricius}, C. and {Fouesneau}, M. and {Fr{\'e}mat}, Y. and {Guerra}, R. and {Guerrier}, A. and {Guiraud}, J. and {Jean-Antoine Piccolo}, A. and {Masana}, E. and {Messineo}, R. and {Mowlavi}, N. and {Nicolas}, C. and {Nienartowicz}, K. and {Pailler}, F. and {Panuzzo}, P. and {Riclet}, F. and {Roux}, W. and {Seabroke}, G.~M. and {Sordo}, R. and {Tanga}, P. and {Th{\'e}venin}, F. and {Gracia-Abril}, G. and {Portell}, J. and {Teyssier}, D. and {Altmann}, M. and {Andrae}, R. and {Bellas-Velidis}, I. and {Benson}, K. and {Berthier}, J. and {Blomme}, R. and {Brugaletta}, E. and {Burgess}, P.~W. and {Busso}, G. and {Carry}, B. and {Cellino}, A. and {Cheek}, N. and {Clementini}, G. and {Damerdji}, Y. and {Davidson}, M. and {Delchambre}, L. and {Dell'Oro}, A. and {Fern{\'a}ndez-Hern{\'a}ndez}, J. and {Galluccio}, L. and {Garc{\'\i}a-Lario}, P. and {Garcia-Reinaldos}, M. and {Gonz{\'a}lez-N{\'u}{\~n}ez}, J. and {Gosset}, E. and {Haigron}, R. and {Halbwachs}, J. -L. and {Hambly}, N.~C. and {Harrison}, D.~L. and {Hatzidimitriou}, D. and {Heiter}, U. and {Hern{\'a}ndez}, J. and {Hestroffer}, D. and {Hodgkin}, S.~T. and {Holl}, B. and {Jan{\ss}en}, K. and {Jevardat de Fombelle}, G. and {Jordan}, S. and {Krone-Martins}, A. and {Lanzafame}, A.~C. and {L{\"o}ffler}, W. and {Lorca}, A. and {Manteiga}, M. and {Marchal}, O. and {Marrese}, P.~M. and {Moitinho}, A. and {Mora}, A. and {Muinonen}, K. and {Osborne}, P. and {Pancino}, E. and {Pauwels}, T. and {Petit}, J. -M. and {Recio-Blanco}, A. and {Richards}, P.~J. and {Riello}, M. and {Rimoldini}, L. and {Robin}, A.~C. and {Roegiers}, T. and {Rybizki}, J. and {Sarro}, L.~M. and {Siopis}, C. and {Smith}, M. and {Sozzetti}, A. and {Ulla}, A. and {Utrilla}, E. and {van Leeuwen}, M. and {van Reeven}, W. and {Abbas}, U. and {Abreu Aramburu}, A. and {Accart}, S. and {Aerts}, C. and {Aguado}, J.~J. and {Ajaj}, M. and {Altavilla}, G. and {{\'A}lvarez}, M.~A. and {{\'A}lvarez Cid-Fuentes}, J. and {Alves}, J. and {Anderson}, R.~I. and {Anglada Varela}, E. and {Antoja}, T. and {Audard}, M. and {Baines}, D. and {Baker}, S.~G. and {Balaguer-N{\'u}{\~n}ez}, L. and {Balbinot}, E. and {Balog}, Z. and {Barache}, C. and {Barbato}, D. and {Barros}, M. and {Barstow}, M.~A. and {Bartolom{\'e}}, S. and {Bassilana}, J. -L. and {Bauchet}, N. and {Baudesson-Stella}, A. and {Becciani}, U. and {Bellazzini}, M. and {Bernet}, M. and {Bertone}, S. and {Bianchi}, L. and {Blanco-Cuaresma}, S. and {Boch}, T. and {Bombrun}, A. and {Bossini}, D. and {Bouquillon}, S. and {Bragaglia}, A. and {Bramante}, L. and {Breedt}, E. and {Bressan}, A. and {Brouillet}, N. and {Bucciarelli}, B. and {Burlacu}, A. and {Busonero}, D. and {Butkevich}, A.~G. and {Buzzi}, R. and {Caffau}, E. and {Cancelliere}, R. and {C{\'a}novas}, H. and {Cantat-Gaudin}, T. and {Carballo}, R. and {Carlucci}, T. and {Carnerero}, M.~I. and {Carrasco}, J.~M. and {Casamiquela}, L. and {Castellani}, M. and {Castro-Ginard}, A. and {Castro Sampol}, P. and {Chaoul}, L. and {Charlot}, P. and {Chemin}, L. and {Chiavassa}, A. and {Cioni}, M. -R.~L. and {Comoretto}, G. and {Cooper}, W.~J. and {Cornez}, T. and {Cowell}, S. and {Crifo}, F. and {Crosta}, M. and {Crowley}, C. and {Dafonte}, C. and {Dapergolas}, A. and {David}, M. and {David}, P.},
        title = "{Gaia Early Data Release 3. Summary of the contents and survey properties}",
      journal = {\aap},
         year = 2021,
        month = may,
       volume = {649},
          eid = {A1},
        pages = {A1},
          doi = {10.1051/0004-6361/202039657},
archivePrefix = {arXiv},
       eprint = {2012.01533},
 primaryClass = {astro-ph.GA},
       adsurl = {https://ui.adsabs.harvard.edu/abs/2021A&A...649A...1G}
}

@ARTICLE{Fabricius2021,
       author = {{Fabricius}, C. and {Luri}, X. and {Arenou}, F. and {Babusiaux}, C. and {Helmi}, A. and {Muraveva}, T. and {Reyl{\'e}}, C. and {Spoto}, F. and {Vallenari}, A. and {Antoja}, T. and {Balbinot}, E. and {Barache}, C. and {Bauchet}, N. and {Bragaglia}, A. and {Busonero}, D. and {Cantat-Gaudin}, T. and {Carrasco}, J.~M. and {Diakit{\'e}}, S. and {Fabrizio}, M. and {Figueras}, F. and {Garcia-Gutierrez}, A. and {Garofalo}, A. and {Jordi}, C. and {Kervella}, P. and {Khanna}, S. and {Leclerc}, N. and {Licata}, E. and {Lambert}, S. and {Marrese}, P.~M. and {Masip}, A. and {Ramos}, P. and {Robichon}, N. and {Robin}, A.~C. and {Romero-G{\'o}mez}, M. and {Rubele}, S. and {Weiler}, M.},
        title = "{Gaia Early Data Release 3. Catalogue validation}",
      journal = {\aap},
         year = 2021,
        month = may,
       volume = {649},
          eid = {A5},
        pages = {A5},
          doi = {10.1051/0004-6361/202039834},
archivePrefix = {arXiv},
       eprint = {2012.06242},
 primaryClass = {astro-ph.GA},
       adsurl = {https://ui.adsabs.harvard.edu/abs/2021A&A...649A...5F}
}

@ARTICLE{Rubio2000,
       author = {{Rubio}, M. and {Contursi}, A. and {Lequeux}, J. and {Probst}, R. and {Barb{\'a}}, R. and {Boulanger}, F. and {Cesarsky}, D. and {Maoli}, R.},
        title = "{Multiwavelength observations of N 66 in the SMC: unveiling photodissociation interfaces and star formation}",
      journal = {\aap},
         year = 2000,
        month = jul,
       volume = {359},
        pages = {1139-1146},
       adsurl = {https://ui.adsabs.harvard.edu/abs/2000A&A...359.1139R}
}

@ARTICLE{Cignoni2011,
       author = {{Cignoni}, M. and {Tosi}, M. and {Sabbi}, E. and {Nota}, A. and {Gallagher}, J.~S.},
        title = "{History and Modes of Star Formation in the Most Active Region of the Small Magellanic Cloud, NGC 346}",
      journal = {\aj},
         year = 2011,
        month = feb,
       volume = {141},
       number = {2},
          eid = {31},
        pages = {31},
          doi = {10.1088/0004-6256/141/2/31},
archivePrefix = {arXiv},
       eprint = {1010.0340},
 primaryClass = {astro-ph.GA},
       adsurl = {https://ui.adsabs.harvard.edu/abs/2011AJ....141...31C}
}

@ARTICLE{Bildsten2012,
       author = {{Bildsten}, Lars and {Paxton}, Bill and {Moore}, Kevin and {Macias}, Phillip J.},
        title = "{Acoustic Signatures of the Helium Core Flash}",
      journal = {\apjl},
         year = 2012,
        month = jan,
       volume = {744},
       number = {1},
          eid = {L6},
        pages = {L6},
          doi = {10.1088/2041-8205/744/1/L6},
archivePrefix = {arXiv},
       eprint = {1111.6867},
 primaryClass = {astro-ph.SR},
       adsurl = {https://ui.adsabs.harvard.edu/abs/2012ApJ...744L...6B}
}

@PROCEEDINGS{Habing2003,
        title = "{Asymptotic giant branch stars}",
    booktitle = {Asymptotic giant branch stars},
         year = 2003,
       editor = {{Habing}, Harm J. and {Olofsson}, Hans},
        month = jan,
       adsurl = {https://ui.adsabs.harvard.edu/abs/2003agbs.conf.....H}
}

@INPROCEEDINGS{Worthey1999,
       author = {{Worthey}, G.},
        title = "{The Age-Metallicity Degeneracy}",
    booktitle = {Spectrophotometric Dating of Stars and Galaxies},
         year = 1999,
       editor = {{Hubeny}, Ivan and {Heap}, Sally and {Cornett}, Robert},
       series = {Astronomical Society of the Pacific Conference Series},
       volume = {192},
        month = jan,
        pages = {283},
       adsurl = {https://ui.adsabs.harvard.edu/abs/1999ASPC..192..283W}
}

@ARTICLE{DeMarchi2014,
       author = {{De Marchi}, Guido and {Panagia}, Nino and {Girardi}, L{\'e}o},
        title = "{Probing interstellar extinction near the 30 Doradus nebula with red giant stars}",
      journal = {\mnras},
         year = 2014,
        month = feb,
       volume = {438},
       number = {1},
        pages = {513-528},
          doi = {10.1093/mnras/stt2233},
archivePrefix = {arXiv},
       eprint = {1311.3659},
 primaryClass = {astro-ph.SR},
       adsurl = {https://ui.adsabs.harvard.edu/abs/2014MNRAS.438..513D}
}

@ARTICLE{Bressan2012,
       author = {{Bressan}, Alessandro and {Marigo}, Paola and {Girardi}, L{\'e}o. and {Salasnich}, Bernardo and {Dal Cero}, Claudia and {Rubele}, Stefano and {Nanni}, Ambra},
        title = "{PARSEC: stellar tracks and isochrones with the PAdova and TRieste Stellar Evolution Code}",
      journal = {\mnras},
         year = 2012,
        month = nov,
       volume = {427},
       number = {1},
        pages = {127-145},
          doi = {10.1111/j.1365-2966.2012.21948.x},
archivePrefix = {arXiv},
       eprint = {1208.4498},
 primaryClass = {astro-ph.SR},
       adsurl = {https://ui.adsabs.harvard.edu/abs/2012MNRAS.427..127B}
}

@ARTICLE{Tang2014,
       author = {{Tang}, Jing and {Bressan}, Alessandro and {Rosenfield}, Philip and {Slemer}, Alessandra and {Marigo}, Paola and {Girardi}, L{\'e}o and {Bianchi}, Luciana},
        title = "{New PARSEC evolutionary tracks of massive stars at low metallicity: testing canonical stellar evolution in nearby star-forming dwarf galaxies}",
      journal = {\mnras},
         year = 2014,
        month = dec,
       volume = {445},
       number = {4},
        pages = {4287-4305},
          doi = {10.1093/mnras/stu2029},
archivePrefix = {arXiv},
       eprint = {1410.1745},
 primaryClass = {astro-ph.SR},
       adsurl = {https://ui.adsabs.harvard.edu/abs/2014MNRAS.445.4287T}
}

@ARTICLE{Chen2014,
       author = {{Chen}, Yang and {Girardi}, L{\'e}o and {Bressan}, Alessandro and {Marigo}, Paola and {Barbieri}, Mauro and {Kong}, Xu},
        title = "{Improving PARSEC models for very low mass stars}",
      journal = {\mnras},
         year = 2014,
        month = nov,
       volume = {444},
       number = {3},
        pages = {2525-2543},
          doi = {10.1093/mnras/stu1605},
archivePrefix = {arXiv},
       eprint = {1409.0322},
 primaryClass = {astro-ph.SR},
       adsurl = {https://ui.adsabs.harvard.edu/abs/2014MNRAS.444.2525C}
}

@ARTICLE{Chen2015,
       author = {{Chen}, Yang and {Bressan}, Alessandro and {Girardi}, L{\'e}o and {Marigo}, Paola and {Kong}, Xu and {Lanza}, Antonio},
        title = "{PARSEC evolutionary tracks of massive stars up to 350 M$_{{\ensuremath{\odot}}}$ at metallicities 0.0001 {\ensuremath{\leq}} Z {\ensuremath{\leq}} 0.04}",
      journal = {\mnras},
         year = 2015,
        month = sep,
       volume = {452},
       number = {1},
        pages = {1068-1080},
          doi = {10.1093/mnras/stv1281},
archivePrefix = {arXiv},
       eprint = {1506.01681},
 primaryClass = {astro-ph.SR},
       adsurl = {https://ui.adsabs.harvard.edu/abs/2015MNRAS.452.1068C}
}

@ARTICLE{Marigo2013,
       author = {{Marigo}, Paola and {Bressan}, Alessandro and {Nanni}, Ambra and {Girardi}, L{\'e}o and {Pumo}, Maria Letizia},
        title = "{Evolution of thermally pulsing asymptotic giant branch stars - I. The COLIBRI code}",
      journal = {\mnras},
         year = 2013,
        month = sep,
       volume = {434},
       number = {1},
        pages = {488-526},
          doi = {10.1093/mnras/stt1034},
archivePrefix = {arXiv},
       eprint = {1305.4485},
 primaryClass = {astro-ph.SR},
       adsurl = {https://ui.adsabs.harvard.edu/abs/2013MNRAS.434..488M}
}

@ARTICLE{Rosenfield2016,
       author = {{Rosenfield}, Philip and {Marigo}, Paola and {Girardi}, L{\'e}o and {Dalcanton}, Julianne J. and {Bressan}, Alessandro and {Williams}, Benjamin F. and {Dolphin}, Andrew},
        title = "{Evolution of Thermally Pulsing Asymptotic Giant Branch Stars. V. Constraining the Mass Loss and Lifetimes of Intermediate-mass, Low-metallicity AGB Stars}",
      journal = {\apj},
         year = 2016,
        month = may,
       volume = {822},
       number = {2},
          eid = {73},
        pages = {73},
          doi = {10.3847/0004-637X/822/2/73},
archivePrefix = {arXiv},
       eprint = {1603.05283},
 primaryClass = {astro-ph.SR},
       adsurl = {https://ui.adsabs.harvard.edu/abs/2016ApJ...822...73R}
}

@ARTICLE{Pastorelli2019,
       author = {{Pastorelli}, Giada and {Marigo}, Paola and {Girardi}, L{\'e}o and {Chen}, Yang and {Rubele}, Stefano and {Trabucchi}, Michele and {Aringer}, Bernhard and {Bladh}, Sara and {Bressan}, Alessandro and {Montalb{\'a}n}, Josefina and {Boyer}, Martha L. and {Dalcanton}, Julianne J. and {Eriksson}, Kjell and {Groenewegen}, Martin A.~T. and {H{\"o}fner}, Susanne and {Lebzelter}, Thomas and {Nanni}, Ambra and {Rosenfield}, Philip and {Wood}, Peter R. and {Cioni}, Maria-Rosa L.},
        title = "{Constraining the thermally pulsing asymptotic giant branch phase with resolved stellar populations in the Small Magellanic Cloud}",
      journal = {\mnras},
         year = 2019,
        month = jun,
       volume = {485},
       number = {4},
        pages = {5666-5692},
          doi = {10.1093/mnras/stz725},
archivePrefix = {arXiv},
       eprint = {1903.04499},
 primaryClass = {astro-ph.SR},
       adsurl = {https://ui.adsabs.harvard.edu/abs/2019MNRAS.485.5666P}
}

@ARTICLE{Pastorelli2020,
       author = {{Pastorelli}, Giada and {Marigo}, Paola and {Girardi}, L{\'e}o and {Aringer}, Bernhard and {Chen}, Yang and {Rubele}, Stefano and {Trabucchi}, Michele and {Bladh}, Sara and {Boyer}, Martha L. and {Bressan}, Alessandro and {Dalcanton}, Julianne J. and {Groenewegen}, Martin A.~T. and {Lebzelter}, Thomas and {Mowlavi}, Nami and {Chubb}, Katy L. and {Cioni}, Maria-Rosa L. and {de Grijs}, Richard and {Ivanov}, Valentin D. and {Nanni}, Ambra and {van Loon}, Jacco Th and {Zaggia}, Simone},
        title = "{Constraining the thermally pulsing asymptotic giant branch phase with resolved stellar populations in the Large Magellanic Cloud}",
      journal = {\mnras},
         year = 2020,
        month = nov,
       volume = {498},
       number = {3},
        pages = {3283-3301},
          doi = {10.1093/mnras/staa2565},
archivePrefix = {arXiv},
       eprint = {2008.08595},
 primaryClass = {astro-ph.SR},
       adsurl = {https://ui.adsabs.harvard.edu/abs/2020MNRAS.498.3283P}
}

@ARTICLE{DeMarchi2024,
       author = {{De Marchi}, Guido and {Giardino}, Giovanna and {Biazzo}, Katia and {Panagia}, Nino and {Sabbi}, Elena and {Beck}, Tracy L. and {Robberto}, Massimo and {Zeidler}, Peter and {Jones}, Olivia C. and {Meixner}, Margaret and {Fahrion}, Katja and {Habel}, Nolan and {Nally}, Conor and {Hirschauer}, Alec S. and {Soderblom}, David R. and {Nayak}, Omnarayani and {Lenki{\'c}}, Laura and {Rogers}, Ciaran and {Brandl}, Bernhard and {Keyes}, Charles D.},
        title = "{Protoplanetary Disks around Sun-like Stars Appear to Live Longer When the Metallicity is Low}",
      journal = {\apj},
         year = 2024,
        month = dec,
       volume = {977},
       number = {2},
          eid = {214},
        pages = {214},
          doi = {10.3847/1538-4357/ad7a63},
archivePrefix = {arXiv},
       eprint = {2412.10361},
 primaryClass = {astro-ph.SR},
       adsurl = {https://ui.adsabs.harvard.edu/abs/2024ApJ...977..214D}
}

@ARTICLE{Jones2022,
       author = {{Jones}, O.~C. and {Reiter}, M. and {Sanchez-Janssen}, R. and {Evans}, C.~J. and {Robertson}, C.~S. and {Meixner}, M. and {Ochsendorf}, B.},
        title = "{Near-infrared spectroscopy of embedded protostars in the massive metal-poor star-forming region NGC 346}",
      journal = {\mnras},
         year = 2022,
        month = nov,
       volume = {517},
       number = {1},
        pages = {1518-1537},
          doi = {10.1093/mnras/stac2491},
archivePrefix = {arXiv},
       eprint = {2209.00040},
 primaryClass = {astro-ph.SR},
       adsurl = {https://ui.adsabs.harvard.edu/abs/2022MNRAS.517.1518J}
}

@ARTICLE{Jones2023,
       author = {{Jones}, Olivia C. and {Nally}, Conor and {Habel}, Nolan and {Lenki{\'c}}, Laura and {Fahrion}, Katja and {Hirschauer}, Alec S. and {Chu}, Laurie E.~U. and {Meixner}, Margaret and {De Marchi}, Guido and {Nayak}, Omnarayani and {Robberto}, Massimo and {Sabbi}, Elena and {Zeidler}, Peter and {Alves de Oliveira}, Catarina and {Beck}, Tracy and {Biazzo}, Katia and {Brandl}, Bernhard and {Giardino}, Giovanna and {Jerabkova}, Teresa and {Keyes}, Charles and {Muzerolle}, James and {Panagia}, Nino and {Pontoppidan}, Klaus and {Rogers}, Ciaran and {Sargent}, B.~A. and {Soderblom}, David},
        title = "{JWST/NIRCam detections of dusty subsolar-mass young stellar objects in the Small Magellanic Cloud}",
      journal = {Nature Astronomy},
         year = 2023,
        month = jun,
       volume = {7},
        pages = {694-701},
          doi = {10.1038/s41550-023-01945-7},
archivePrefix = {arXiv},
       eprint = {2301.03932},
 primaryClass = {astro-ph.SR},
       adsurl = {https://ui.adsabs.harvard.edu/abs/2023NatAs...7..694J}
}

@ARTICLE{Brittain2023,
       author = {{Brittain}, Sean D. and {Kamp}, Inga and {Meeus}, Gwendolyn and {Oudmaijer}, Ren{\'e} D. and {Waters}, L.~B.~F.~M.},
        title = "{Herbig Stars}",
      journal = {\ssr},
         year = 2023,
        month = feb,
       volume = {219},
       number = {1},
          eid = {7},
        pages = {7},
          doi = {10.1007/s11214-023-00949-z},
archivePrefix = {arXiv},
       eprint = {2301.01165},
 primaryClass = {astro-ph.SR},
       adsurl = {https://ui.adsabs.harvard.edu/abs/2023SSRv..219....7B}
}

@ARTICLE{Rogers2024,
       author = {{Rogers}, Ciar{\'a}n and {de Marchi}, Guido and {Brandl}, Bernhard},
        title = "{Determining stellar accretion rates from Pa$_{{\ensuremath{\alpha}}}$ and Br$_{{\ensuremath{\beta}}}$ emission lines with JWST NIRSpec. Accretion of pre-main-sequence stars in NGC 3603}",
      journal = {\aap},
         year = 2024,
        month = apr,
       volume = {684},
          eid = {L8},
        pages = {L8},
          doi = {10.1051/0004-6361/202449282},
archivePrefix = {arXiv},
       eprint = {2403.09568},
 primaryClass = {astro-ph.SR},
       adsurl = {https://ui.adsabs.harvard.edu/abs/2024A&A...684L...8R}
}

@ARTICLE{Herbig1985,
       author = {{Herbig}, G.~H.},
        title = "{Chromospheric H alpha emission in F8-G3 dwarfs and its connection with the T Tauri stars.}",
      journal = {\apj},
         year = 1985,
        month = feb,
       volume = {289},
        pages = {269-278},
          doi = {10.1086/162887},
       adsurl = {https://ui.adsabs.harvard.edu/abs/1985ApJ...289..269H}
}

@ARTICLE{Strassmeier1990,
       author = {{Strassmeier}, Klaus G. and {Fekel}, Francis C. and {Bopp}, Bernard W. and {Dempsey}, Robert C. and {Henry}, Gregory W.},
        title = "{Chromospheric CA II H and K and H alpha Emission in Single and Binary Stars of Spectra Types F6--M2}",
      journal = {\apjs},
         year = 1990,
        month = jan,
       volume = {72},
        pages = {191},
          doi = {10.1086/191414},
       adsurl = {https://ui.adsabs.harvard.edu/abs/1990ApJS...72..191S}
}

@ARTICLE{Young1989,
       author = {{Young}, Arthur and {Skumanich}, Andrew and {Stauffer}, John R. and {Bopp}, Bernard W. and {Harlan}, Eugene},
        title = "{A Study of Excess H alpha Emission in Chromospherically Active M Dwarf Stars}",
      journal = {\apj},
         year = 1989,
        month = sep,
       volume = {344},
        pages = {427},
          doi = {10.1086/167810},
       adsurl = {https://ui.adsabs.harvard.edu/abs/1989ApJ...344..427Y}
}

@ARTICLE{Gordon2011,
       author = {{Gordon}, K.~D. and {Meixner}, M. and {Meade}, M.~R. and {Whitney}, B. and {Engelbracht}, C. and {Bot}, C. and {Boyer}, M.~L. and {Lawton}, B. and {Sewi{\l}o}, M. and {Babler}, B. and {Bernard}, J. -P. and {Bracker}, S. and {Block}, M. and {Blum}, R. and {Bolatto}, A. and {Bonanos}, A. and {Harris}, J. and {Hora}, J.~L. and {Indebetouw}, R. and {Misselt}, K. and {Reach}, W. and {Shiao}, B. and {Tielens}, X. and {Carlson}, L. and {Churchwell}, E. and {Clayton}, G.~C. and {Chen}, C. -H.~R. and {Cohen}, M. and {Fukui}, Y. and {Gorjian}, V. and {Hony}, S. and {Israel}, F.~P. and {Kawamura}, A. and {Kemper}, F. and {Leroy}, A. and {Li}, A. and {Madden}, S. and {Marble}, A.~R. and {McDonald}, I. and {Mizuno}, A. and {Mizuno}, N. and {Muller}, E. and {Oliveira}, J.~M. and {Olsen}, K. and {Onishi}, T. and {Paladini}, R. and {Paradis}, D. and {Points}, S. and {Robitaille}, T. and {Rubin}, D. and {Sandstrom}, K. and {Sato}, S. and {Shibai}, H. and {Simon}, J.~D. and {Smith}, L.~J. and {Srinivasan}, S. and {Vijh}, U. and {Van Dyk}, S. and {van Loon}, J. Th. and {Zaritsky}, D.},
        title = "{Surveying the Agents of Galaxy Evolution in the Tidally Stripped, Low Metallicity Small Magellanic Cloud (SAGE-SMC). I. Overview}",
      journal = {\aj},
         year = 2011,
        month = oct,
       volume = {142},
       number = {4},
          eid = {102},
        pages = {102},
          doi = {10.1088/0004-6256/142/4/102},
archivePrefix = {arXiv},
       eprint = {1107.4313},
 primaryClass = {astro-ph.CO},
       adsurl = {https://ui.adsabs.harvard.edu/abs/2011AJ....142..102G}
}

@ARTICLE{Peimbert2000,
       author = {{Peimbert}, Manuel and {Peimbert}, Antonio and {Ruiz}, Mar{\'\i}a Teresa},
        title = "{The Chemical Composition of the Small Magellanic Cloud H II Region NGC 346 and the Primordial Helium Abundance}",
      journal = {\apj},
         year = 2000,
        month = oct,
       volume = {541},
       number = {2},
        pages = {688-700},
          doi = {10.1086/309485},
archivePrefix = {arXiv},
       eprint = {astro-ph/0003154},
 primaryClass = {astro-ph},
       adsurl = {https://ui.adsabs.harvard.edu/abs/2000ApJ...541..688P}
}

@ARTICLE{Massey1989,
       author = {{Massey}, Philip and {Parker}, Joel W. and {Garmany}, Catharine D.},
        title = "{The Stellar Content of NGC 346: A Plethora of O Stars in the SMC}",
      journal = {\aj},
         year = 1989,
        month = oct,
       volume = {98},
        pages = {1305},
          doi = {10.1086/115217},
       adsurl = {https://ui.adsabs.harvard.edu/abs/1989AJ.....98.1305M}
}

@ARTICLE{Evans2006,
       author = {{Evans}, C.~J. and {Lennon}, D.~J. and {Smartt}, S.~J. and {Trundle}, C.},
        title = "{The VLT-FLAMES survey of massive stars: observations centered on the Magellanic Cloud clusters NGC 330, NGC 346, NGC 2004, and the N11 region}",
      journal = {\aap},
         year = 2006,
        month = sep,
       volume = {456},
       number = {2},
        pages = {623-638},
          doi = {10.1051/0004-6361:20064988},
archivePrefix = {arXiv},
       eprint = {astro-ph/0606405},
 primaryClass = {astro-ph},
       adsurl = {https://ui.adsabs.harvard.edu/abs/2006A&A...456..623E}
}

@ARTICLE{Dufton2019,
       author = {{Dufton}, P.~L. and {Evans}, C.~J. and {Hunter}, I. and {Lennon}, D.~J. and {Schneider}, F.~R.~N.},
        title = "{A census of massive stars in NGC 346. Stellar parameters and rotational velocities}",
      journal = {\aap},
         year = 2019,
        month = jun,
       volume = {626},
          eid = {A50},
        pages = {A50},
          doi = {10.1051/0004-6361/201935415},
archivePrefix = {arXiv},
       eprint = {1905.03359},
 primaryClass = {astro-ph.SR},
       adsurl = {https://ui.adsabs.harvard.edu/abs/2019A&A...626A..50D}
}

@ARTICLE{Nota2006,
       author = {{Nota}, A. and {Sirianni}, M. and {Sabbi}, E. and {Tosi}, M. and {Clampin}, M. and {Gallagher}, J. and {Meixner}, M. and {Oey}, M.~S. and {Pasquali}, A. and {Smith}, L.~J. and {Walterbos}, R. and {Mack}, J.},
        title = "{Discovery of a Population of Pre-Main-Sequence Stars in NGC 346 from Deep Hubble Space Telescope ACS Images}",
      journal = {\apjl},
         year = 2006,
        month = mar,
       volume = {640},
       number = {1},
        pages = {L29-L33},
          doi = {10.1086/503301},
archivePrefix = {arXiv},
       eprint = {astro-ph/0602218},
 primaryClass = {astro-ph},
       adsurl = {https://ui.adsabs.harvard.edu/abs/2006ApJ...640L..29N}
}

@ARTICLE{Hennekemper2008,
       author = {{Hennekemper}, Eva and {Gouliermis}, Dimitrios A. and {Henning}, Thomas and {Brandner}, Wolfgang and {Dolphin}, Andrew E.},
        title = "{NGC 346 in the Small Magellanic Cloud. III. Recent Star Formation and Stellar Clustering Properties in the Bright H II Region N66}",
      journal = {\apj},
         year = 2008,
        month = jan,
       volume = {672},
       number = {2},
        pages = {914-929},
          doi = {10.1086/524105},
archivePrefix = {arXiv},
       eprint = {0710.0774},
 primaryClass = {astro-ph},
       adsurl = {https://ui.adsabs.harvard.edu/abs/2008ApJ...672..914H}
}

@ARTICLE{Bolatto2007,
       author = {{Bolatto}, Alberto D. and {Simon}, Joshua D. and {Stanimirovi{\'c}}, Sne{\v{z}}ana and {van Loon}, Jacco Th. and {Shah}, Ronak Y. and {Venn}, Kim and {Leroy}, Adam K. and {Sandstrom}, Karin and {Jackson}, James M. and {Israel}, Frank P. and {Li}, Aigen and {Staveley-Smith}, Lister and {Bot}, Caroline and {Boulanger}, Francois and {Rubio}, M{\'o}nica},
        title = "{The Spitzer Survey of the Small Magellanic Cloud: S$^{3}$MC Imaging and Photometry in the Mid- and Far-Infrared Wave Bands}",
      journal = {\apj},
         year = 2007,
        month = jan,
       volume = {655},
       number = {1},
        pages = {212-232},
          doi = {10.1086/509104},
archivePrefix = {arXiv},
       eprint = {astro-ph/0608561},
 primaryClass = {astro-ph},
       adsurl = {https://ui.adsabs.harvard.edu/abs/2007ApJ...655..212B}
}

@ARTICLE{Meixner2013,
       author = {{Meixner}, M. and {Panuzzo}, P. and {Roman-Duval}, J. and {Engelbracht}, C. and {Babler}, B. and {Seale}, J. and {Hony}, S. and {Montiel}, E. and {Sauvage}, M. and {Gordon}, K. and {Misselt}, K. and {Okumura}, K. and {Chanial}, P. and {Beck}, T. and {Bernard}, J. -P. and {Bolatto}, A. and {Bot}, C. and {Boyer}, M.~L. and {Carlson}, L.~R. and {Clayton}, G.~C. and {Chen}, C. -H.~R. and {Cormier}, D. and {Fukui}, Y. and {Galametz}, M. and {Galliano}, F. and {Hora}, J.~L. and {Hughes}, A. and {Indebetouw}, R. and {Israel}, F.~P. and {Kawamura}, A. and {Kemper}, F. and {Kim}, S. and {Kwon}, E. and {Lebouteiller}, V. and {Li}, A. and {Long}, K.~S. and {Madden}, S.~C. and {Matsuura}, M. and {Muller}, E. and {Oliveira}, J.~M. and {Onishi}, T. and {Otsuka}, M. and {Paradis}, D. and {Poglitsch}, A. and {Reach}, W.~T. and {Robitaille}, T.~P. and {Rubio}, M. and {Sargent}, B. and {Sewi{\l}o}, M. and {Skibba}, R. and {Smith}, L.~J. and {Srinivasan}, S. and {Tielens}, A.~G.~G.~M. and {van Loon}, J. Th. and {Whitney}, B.},
        title = "{The HERSCHEL Inventory of The Agents of Galaxy Evolution in the Magellanic Clouds, a Herschel Open Time Key Program}",
      journal = {\aj},
         year = 2013,
        month = sep,
       volume = {146},
       number = {3},
          eid = {62},
        pages = {62},
          doi = {10.1088/0004-6256/146/3/62},
       adsurl = {https://ui.adsabs.harvard.edu/abs/2013AJ....146...62M}
}

@ARTICLE{Sewilo2013,
       author = {{Sewi{\l}o}, M. and {Carlson}, L.~R. and {Seale}, J.~P. and {Indebetouw}, R. and {Meixner}, M. and {Whitney}, B.~A. and {Robitaille}, T.~P. and {Oliveira}, J.~M. and {Gordon}, K. and {Meade}, M.~R. and {Babler}, B.~L. and {Hora}, J.~L. and {Block}, M. and {Misselt}, K. and {van Loon}, J. Th. and {Chen}, C. -H.~R. and {Churchwell}, E. and {Shiao}, B.},
        title = "{Surveying the Agents of Galaxy Evolution in the Tidally Stripped, Low Metallicity Small Magellanic Cloud (SAGE-SMC). III. Young Stellar Objects}",
      journal = {\apj},
         year = 2013,
        month = nov,
       volume = {778},
       number = {1},
          eid = {15},
        pages = {15},
          doi = {10.1088/0004-637X/778/1/15},
       adsurl = {https://ui.adsabs.harvard.edu/abs/2013ApJ...778...15S}
}

@ARTICLE{Seale2014,
       author = {{Seale}, Jonathan P. and {Meixner}, Margaret and {Sewi{\l}o}, Marta and {Babler}, Brian and {Engelbracht}, Charles W. and {Gordon}, Karl and {Hony}, Sacha and {Misselt}, Karl and {Montiel}, Edward and {Okumura}, Koryo and {Panuzzo}, Pasquale and {Roman-Duval}, Julia and {Sauvage}, Marc and {Boyer}, Martha L. and {Chen}, C. -H. Rosie and {Indebetouw}, Remy and {Matsuura}, Mikako and {Oliveira}, Joana M. and {Srinivasan}, Sundar and {van Loon}, Jacco Th. and {Whitney}, Barbara and {Woods}, Paul M.},
        title = "{Herschel Key Program Heritage: a Far-Infrared Source Catalog for the Magellanic Clouds}",
      journal = {\aj},
         year = 2014,
        month = dec,
       volume = {148},
       number = {6},
          eid = {124},
        pages = {124},
          doi = {10.1088/0004-6256/148/6/124},
       adsurl = {https://ui.adsabs.harvard.edu/abs/2014AJ....148..124S}
}

@ARTICLE{Gouliermis2014,
       author = {{Gouliermis}, Dimitrios A. and {Hony}, Sacha and {Klessen}, Ralf S.},
        title = "{The complex distribution of recently formed stars. Bimodal stellar clustering in the star-forming region NGC 346}",
      journal = {\mnras},
         year = 2014,
        month = apr,
       volume = {439},
       number = {4},
        pages = {3775-3789},
          doi = {10.1093/mnras/stu228},
archivePrefix = {arXiv},
       eprint = {1402.0078},
 primaryClass = {astro-ph.GA},
       adsurl = {https://ui.adsabs.harvard.edu/abs/2014MNRAS.439.3775G}
}

@ARTICLE{Sabbi2008,
       author = {{Sabbi}, E. and {Sirianni}, M. and {Nota}, A. and {Tosi}, M. and {Gallagher}, J. and {Smith}, L.~J. and {Angeretti}, L. and {Meixner}, M. and {Oey}, M.~S. and {Walterbos}, R. and {Pasquali}, A.},
        title = "{The Stellar Mass Distribution in the Giant Star Forming Region NGC 346}",
      journal = {\aj},
         year = 2008,
        month = jan,
       volume = {135},
       number = {1},
        pages = {173-181},
          doi = {10.1088/0004-6256/135/1/173},
archivePrefix = {arXiv},
       eprint = {0710.0558},
 primaryClass = {astro-ph},
       adsurl = {https://ui.adsabs.harvard.edu/abs/2008AJ....135..173S}
}

@ARTICLE{Rubio2018,
       author = {{Rubio}, M. and {Barb{\'a}}, R.~H. and {Kalari}, V.~M.},
        title = "{Massive young stellar objects in the N 66/NGC 346 region of the SMC}",
      journal = {\aap},
         year = 2018,
        month = jul,
       volume = {615},
          eid = {A121},
        pages = {A121},
          doi = {10.1051/0004-6361/201730487},
archivePrefix = {arXiv},
       eprint = {1803.10833},
 primaryClass = {astro-ph.GA},
       adsurl = {https://ui.adsabs.harvard.edu/abs/2018A&A...615A.121R}
}

@ARTICLE{Contursi2000,
       author = {{Contursi}, A. and {Lequeux}, J. and {Cesarsky}, D. and {Boulanger}, F. and {Rubio}, M. and {Hanus}, M. and {Sauvage}, M. and {Tran}, D. and {Bosma}, A. and {Madden}, S. and {Vigroux}, L.},
        title = "{Mid-infrared imaging and spectrophotometry of N 66 in the SMC with ISOCAM}",
      journal = {\aap},
         year = 2000,
        month = oct,
       volume = {362},
        pages = {310-324},
          doi = {10.48550/arXiv.astro-ph/0006185},
archivePrefix = {arXiv},
       eprint = {astro-ph/0006185},
 primaryClass = {astro-ph},
       adsurl = {https://ui.adsabs.harvard.edu/abs/2000A&A...362..310C}
}

@ARTICLE{Sabbi2022,
       author = {{Sabbi}, E. and {Zeidler}, P. and {Marel}, R.~P. van der and {Nota}, A. and {Anderson}, J. and {Gallagher}, J.~S. and {Lennon}, D.~J. and {Smith}, L.~J. and {Gennaro}, M.},
        title = "{The Internal Proper Motion Kinematics of NGC 346: Past Formation and Future Evolution}",
      journal = {\apj},
         year = 2022,
        month = sep,
       volume = {936},
       number = {2},
          eid = {135},
        pages = {135},
          doi = {10.3847/1538-4357/ac8005},
archivePrefix = {arXiv},
       eprint = {2209.03215},
 primaryClass = {astro-ph.GA},
       adsurl = {https://ui.adsabs.harvard.edu/abs/2022ApJ...936..135S}
}

@ARTICLE{Zeidler2022,
       author = {{Zeidler}, Peter and {Sabbi}, Elena and {Nota}, Antonella},
        title = "{The Internal Line-of-Sight Kinematics of NGC 346: The Rotation of the Core Region}",
      journal = {\apj},
         year = 2022,
        month = sep,
       volume = {936},
       number = {2},
          eid = {136},
        pages = {136},
          doi = {10.3847/1538-4357/ac8004},
archivePrefix = {arXiv},
       eprint = {2209.03237},
 primaryClass = {astro-ph.GA},
       adsurl = {https://ui.adsabs.harvard.edu/abs/2022ApJ...936..136Z}
}

@ARTICLE{Zeidler2024,
       author = {{Zeidler}, Peter and {Sabbi}, Elena and {Nota}, Antonella and {Manjavacas}, Elena and {Jones}, Olivia C. and {Pacifici}, Camilla},
        title = "{Discovering Subsolar Metallicity Brown Dwarf Candidates in the Small Magellanic Cloud}",
      journal = {\apj},
         year = 2024,
        month = nov,
       volume = {975},
       number = {1},
          eid = {18},
        pages = {18},
          doi = {10.3847/1538-4357/ad779e},
       adsurl = {https://ui.adsabs.harvard.edu/abs/2024ApJ...975...18Z}
}

@ARTICLE{Hovhannessian2001,
       author = {{Hovhannessian}, R. Kh. and {Hovhannessian}, E.~R.},
        title = "{Gas{\textemdash}Dust Shells around Some Early-Type Stars with an IR Excess (of Emission)}",
      journal = {Astrophysics},
         year = 2001,
        month = oct,
       volume = {44},
       number = {4},
        pages = {454-462},
          doi = {10.1023/A:1014244720865},
       adsurl = {https://ui.adsabs.harvard.edu/abs/2001Ap.....44..454H}
}

@ARTICLE{Siebenmorgen2018,
       author = {{Siebenmorgen}, R. and {Scicluna}, P. and {Kre{\l}owski}, J.},
        title = "{Far-infrared emission of massive stars}",
      journal = {\aap},
         year = 2018,
        month = nov,
       volume = {620},
          eid = {A32},
        pages = {A32},
          doi = {10.1051/0004-6361/201833546},
archivePrefix = {arXiv},
       eprint = {1809.06658},
 primaryClass = {astro-ph.SR},
       adsurl = {https://ui.adsabs.harvard.edu/abs/2018A&A...620A..32S}
}

@ARTICLE{Deng2022,
       author = {{Deng}, Dingshan and {Sun}, Yang and {Wang}, Tianding and {Wang}, Yuxi and {Jiang}, Biwei},
        title = "{Infrared Excess of a Large OB Star Sample}",
      journal = {\apj},
         year = 2022,
        month = aug,
       volume = {935},
       number = {2},
          eid = {175},
        pages = {175},
          doi = {10.3847/1538-4357/ac8168},
archivePrefix = {arXiv},
       eprint = {2207.06961},
 primaryClass = {astro-ph.SR},
       adsurl = {https://ui.adsabs.harvard.edu/abs/2022ApJ...935..175D}
}

@ARTICLE{Hartmann1977,
       author = {{Hartmann}, L. and {Cassinelli}, J.~P.},
        title = "{The structure of the winds from Wolf-Rayet stars as determined from observations of the infrared continua.}",
      journal = {\apj},
         year = 1977,
        month = jul,
       volume = {215},
        pages = {155-158},
          doi = {10.1086/155342},
       adsurl = {https://ui.adsabs.harvard.edu/abs/1977ApJ...215..155H}
}

@ARTICLE{Yanchulova2021,
       author = {{Yanchulova Merica-Jones}, Petia and {Sandstrom}, Karin M. and {Johnson}, L. Clifton and {Dolphin}, Andrew E. and {Dalcanton}, Julianne J. and {Gordon}, Karl and {Roman-Duval}, Julia and {Weisz}, Daniel R. and {Williams}, Benjamin F.},
        title = "{Three-dimensional Structure and Dust Extinction in the Small Magellanic Cloud}",
      journal = {\apj},
         year = 2021,
        month = jan,
       volume = {907},
       number = {1},
          eid = {50},
        pages = {50},
          doi = {10.3847/1538-4357/abc48b},
archivePrefix = {arXiv},
       eprint = {2010.11181},
 primaryClass = {astro-ph.GA},
       adsurl = {https://ui.adsabs.harvard.edu/abs/2021ApJ...907...50Y}
}

@ARTICLE{Murray2024,
       author = {{Murray}, Claire E. and {Hasselquist}, Sten and {Peek}, Joshua E.~G. and {Lindberg}, Christina Willecke and {Almeida}, Andres and {Choi}, Yumi and {Craig}, Jessica E.~M. and {D{\'e}nes}, Helga and {Dickey}, John M. and {Di Teodoro}, Enrico M. and {Federrath}, Christoph and {Gerrard}, Isabella. A. and {Gibson}, Steven J. and {Leahy}, Denis and {Lee}, Min-Young and {Lynn}, Callum and {Ma}, Yik Ki and {Marchal}, Antoine and {McClure-Griffiths}, N.~M. and {Nidever}, David and {Nguyen}, Hiep and {Pingel}, Nickolas M. and {Tarantino}, Elizabeth and {Uscanga}, Lucero and {van Loon}, Jacco Th.},
        title = "{A Galactic Eclipse: The Small Magellanic Cloud Is Forming Stars in Two Superimposed Systems}",
      journal = {\apj},
         year = 2024,
        month = feb,
       volume = {962},
       number = {2},
          eid = {120},
        pages = {120},
          doi = {10.3847/1538-4357/ad1591},
archivePrefix = {arXiv},
       eprint = {2312.07750},
 primaryClass = {astro-ph.GA},
       adsurl = {https://ui.adsabs.harvard.edu/abs/2024ApJ...962..120M}
}

@ARTICLE{Wright2014,
       author = {{Wright}, Nicholas J. and {Wesson}, Roger and {Drew}, Janet E. and {Barentsen}, Geert and {Barlow}, Michael J. and {Walsh}, Jeremy R. and {Zijlstra}, Albert and {Drake}, Jeremy J. and {Eisl{\"o}ffel}, Jochen and {Farnhill}, Hywel J.},
        title = "{The ionized nebula surrounding the red supergiant W26 in Westerlund 1}",
      journal = {\mnras},
         year = 2014,
        month = jan,
       volume = {437},
       number = {1},
        pages = {L1-L5},
          doi = {10.1093/mnrasl/slt127},
archivePrefix = {arXiv},
       eprint = {1309.4086},
 primaryClass = {astro-ph.SR},
       adsurl = {https://ui.adsabs.harvard.edu/abs/2014MNRAS.437L...1W}
}

@ARTICLE{Rogers2013,
       author = {{Rogers}, H. and {Pittard}, J.~M.},
        title = "{Feedback from winds and supernovae in massive stellar clusters - I. Hydrodynamics}",
      journal = {\mnras},
         year = 2013,
        month = may,
       volume = {431},
       number = {2},
        pages = {1337-1351},
          doi = {10.1093/mnras/stt255},
archivePrefix = {arXiv},
       eprint = {1302.2443},
 primaryClass = {astro-ph.SR},
       adsurl = {https://ui.adsabs.harvard.edu/abs/2013MNRAS.431.1337R}
}

@ARTICLE{Raptis2025,
       author = {{Raptis}, Menelaos and {Rudie}, Gwen C. and {Trainor}, Ryan F. and {Rogers}, Noah S.~J. and {Strom}, Allison L. and {Korhonen Cuestas}, Nathalie A. and {von Raesfeld}, Caroline and {Lin}, Ye and {Ojodomo Abraham}, Ojima and {Chapman}, Christopher and {Steidel}, Charles C. and {Maseda}, Michael V.},
        title = "{CECILIA: The Mass-Metallicity Relation of Low-Mass Galaxies at Cosmic Noon}",
      journal = {arXiv e-prints},
         year = 2025,
        month = nov,
          eid = {arXiv:2512.00162},
        pages = {arXiv:2512.00162},
          doi = {10.48550/arXiv.2512.00162},
archivePrefix = {arXiv},
       eprint = {2512.00162},
 primaryClass = {astro-ph.GA},
       adsurl = {https://ui.adsabs.harvard.edu/abs/2025arXiv251200162R}
}

@ARTICLE{Rubele2018,
       author = {{Rubele}, Stefano and {Pastorelli}, Giada and {Girardi}, L{\'e}o and {Cioni}, Maria-Rosa L. and {Zaggia}, Simone and {Marigo}, Paola and {Bekki}, Kenji and {Bressan}, Alessandro and {Clementini}, Gisella and {de Grijs}, Richard and {Emerson}, Jim and {Groenewegen}, Martin A.~T. and {Ivanov}, Valentin D. and {Muraveva}, Tatiana and {Nanni}, Ambra and {Oliveira}, Joana M. and {Ripepi}, Vincenzo and {Sun}, Ning-Chen and {van Loon}, Jacco Th},
        title = "{The VMC survey - XXXI: The spatially resolved star formation history of the main body of the Small Magellanic Cloud}",
      journal = {\mnras},
         year = 2018,
        month = aug,
       volume = {478},
       number = {4},
        pages = {5017-5036},
          doi = {10.1093/mnras/sty1279},
archivePrefix = {arXiv},
       eprint = {1805.04516},
 primaryClass = {astro-ph.GA},
       adsurl = {https://ui.adsabs.harvard.edu/abs/2018MNRAS.478.5017R}
}

@ARTICLE{VanBuren1988,
       author = {{van Buren}, Dave and {McCray}, Richard},
        title = "{Bow Shocks and Bubbles Are Seen around Hot Stars by IRAS}",
      journal = {\apjl},
         year = 1988,
        month = jun,
       volume = {329},
        pages = {L93},
          doi = {10.1086/185184},
       adsurl = {https://ui.adsabs.harvard.edu/abs/1988ApJ...329L..93V}
}

@ARTICLE{Adams2013,
       author = {{Adams}, Joshua J. and {Simon}, Joshua D. and {Bolatto}, Alberto D. and {Sloan}, G.~C. and {Sandstrom}, Karin M. and {Schmiedeke}, Anika and {van Loon}, Jacco Th. and {Oliveira}, Joana M. and {Keller}, Luke D.},
        title = "{Dusty OB Stars in the Small Magellanic Cloud. II. Extragalactic Disks or Examples of the Pleiades Phenomenon?}",
      journal = {\apj},
         year = 2013,
        month = jul,
       volume = {771},
       number = {2},
          eid = {112},
        pages = {112},
          doi = {10.1088/0004-637X/771/2/112},
archivePrefix = {arXiv},
       eprint = {1305.4954},
 primaryClass = {astro-ph.SR},
       adsurl = {https://ui.adsabs.harvard.edu/abs/2013ApJ...771..112A}
}

@ARTICLE{Bailer-Jones2021,
       author = {{Bailer-Jones}, C.~A.~L. and {Rybizki}, J. and {Fouesneau}, M. and {Demleitner}, M. and {Andrae}, R.},
        title = "{Estimating Distances from Parallaxes. V. Geometric and Photogeometric Distances to 1.47 Billion Stars in Gaia Early Data Release 3}",
      journal = {\aj},
         year = 2021,
        month = mar,
       volume = {161},
       number = {3},
          eid = {147},
        pages = {147},
          doi = {10.3847/1538-3881/abd806},
archivePrefix = {arXiv},
       eprint = {2012.05220},
 primaryClass = {astro-ph.SR},
       adsurl = {https://ui.adsabs.harvard.edu/abs/2021AJ....161..147B}
}
\bibliographystyle{aasjournalv7}

\newpage
\appendix
\restartappendixnumbering
\section{Supplementary tables \& figure} \label{sec:appendix}

\begin{deluxetable}{ c  c  c  c  c  c  c }[!h]
\tablewidth{0pt}
\tablecaption{NIRCam Observation Parameters. Note that in this work, we refer to ``Mosaic Part 1'' as the shallow exposure and ``Mosaic Part 2'' as the deep exposure. \label{tab:NIRCam_obs_param}}
\tablehead{
\colhead{Filter} & \colhead{} & \colhead{Read mode} & \colhead{Groups/Int} & \colhead{Integrations/Exp} & \colhead{Total Exposure Time (s)} & \colhead{Number of Tiles}
}
\startdata
F115W & Mosaic Part 1 & BRIGHT2 & 2 & 1 & 171.8 & 3 \\
 & Mosaic Part 2 & BRIGHT2 & 7 & 1 & 601.3 & 1 \\ \hline
F187N & Mosaic Part 1 & BRIGHT2 & 2 & 1 & 171.8 & 3 \\
 & Mosaic Part 2 & BRIGHT2 & 7 & 1 & 601.3 & 1 \\ \hline
F200W & Mosaic Part 1 & BRIGHT2 & 2 & 1 & 171.8 & 3 \\
 & Mosaic Part 2 & BRIGHT2 & 7 & 1 & 601.3 & 1 \\ \hline
F277W & Mosaic Part 1 & BRIGHT2 & 2 & 1 & 171.8 & 3 \\
 & Mosaic Part 2 & BRIGHT2 & 7 & 1 & 601.3 & 1 \\ \hline
F335M & Mosaic Part 1 & BRIGHT2 & 2 & 1 & 171.8 & 3 \\
 & Mosaic Part 2 & BRIGHT2 & 7 & 1 & 601.3 & 1 \\ \hline
F444W & Mosaic Part 1 & BRIGHT2 & 2 & 1 & 171.8 & 3 \\
 & Mosaic Part 2 & BRIGHT2 & 7 & 1 & 601.3 & 1 \\
\enddata
\end{deluxetable}

\begin{deluxetable}{ c  c  c  c  c }
\tablewidth{0pt}
\tablecaption{MIRI Prime Observation Parameters. \label{tab:MIRI_obs_param}}
\tablehead{
\colhead{Filter} & \colhead{Read mode} & \colhead{Groups/Int} & \colhead{Integrations/Exp} & \colhead{Total Exposure Time (s)}
}
\startdata
F770W  & FASTR1 & 28 & 1 & 310.804 \\
F1000W  & FASTR1 & 36 & 1 & 399.606 \\
F1130W  & FASTR1 & 28 & 1 & 310.804 \\
F1500W  & FASTR1 & 32 & 1 & 355.205 \\
F2100W  & FASTR1 & 51 & 1 & 566.108 \\
\enddata
\end{deluxetable}

\begin{splitdeluxetable*}{l|ccccccBl|ccccc}
\tabletypesize{\scriptsize}
\tablewidth{0pt} 
\tablecaption{STARBUGII Parameters used for aperture and PSF photometry. \label{tab:source_detection_params}}
\tablehead{
\colhead{Parameter} & \colhead{F115W}& \colhead{F187N} & \colhead{F200W} &
\colhead{F277W} & \colhead{F335M} & \colhead{F444W}\\
\colhead{} & \colhead{}& \colhead{} & \colhead{} & \colhead{} & \colhead{} & \colhead{} & \colhead{Parameter} & \colhead{F770W} & \colhead{F1000W} &
\colhead{F1130W} & \colhead{F1500W} & \colhead{F2100W}
} 
\startdata 
SIGSKY  & 2.0 & 4.0 & 2.0 & 1.7 & 2.0 & 1.8 & SIGSKY & 1.5 & 1.5 & 1.5 & 1.2 & 1.0 \\
SIGSRC  & 4.0/5.0 & 4.5/10.0 & 4.0/5.0 & 4.0/5.0 & 4.0/5.0 & 4.0/5.0 & SIGSRC & 3.5 & 3.0 & 3.0 & 3.0 & 3.0 \\
SHARP\_LO  & 0.5 & 0.4 & 0.5 & 0.45 & 0.4 & 0.4 & SHARP\_LO & 0.30 & 0.30 & 0.30 & 0.20 & 0.20\\
SHARP\_HI  & 1.05 & 1.1 & 1.2 & 1.15 & 1.2 & 0.9 & SHARP\_HI & 1.2 & 0.90 & 1.2 & 1.1 & 1.1\\
ROUND1\_HI/LO  & $\pm$1.3 & $\pm$1.2 & $\pm$1.1 & $\pm$1.3 & $\pm$1.1 & $\pm$1.5 & ROUND1\_HI/LO & $\pm$1.5 & $\pm$1.6 & $\pm$1.4 & $\pm$1.6 & $\pm$1.3 \\
ROUND2\_HI/LO  & $\pm$1.3 & $\pm$1.0 & $\pm$1.1 & $\pm$1.3 & $\pm$1.1 & $\pm$1.5 & ROUND2\_HI/LO & $\pm$1.5 & $\pm$1.6 & $\pm$2.0 & $\pm$2.0 & $\pm$2.0 \\
SMOOTH\_LO  & 0 & 0.4 & 0.55 & 0.4 & 0 & 0 & SMOOTH\_LO & - & - & - & - & -\\
SMOOTH\_HI  & 1.6 & 1.0 & 1.05 & 1.4 & 1.5 & 1.1 & SMOOTH\_HI & 1.05 & 1.05 & 1.05 & 1.05 & 1.05\\
RICKER\_R  & 1.0 & 1.0 & 1.0 & 1.0 & 1.0 & 1.0 & RICKER\_R & 1.0/1.8 & 1.0/2.2 & 1.0/2.2 & 1.0/2.15 & 1.0/3.0\\ \hline
APPHOT\_R & 1.5 & 1.5 & 1.5 & 1.5 & 1.5 & 1.5 & APPHOT\_R & - & - & - & - & -\\
ENCENERGY & - & - & - & - & - & - & ENCENERGY & 0.7 & 0.7 & 0.7 & 0.7 & 0.7 \\
SKY\_RIN  & 3.0 & 3.0 & 3.0 & 3.0 & 3.0 & 3.0 & SKY\_RIN & 5.0 & 5.0 & 5.0 & 6.5 & 9.5\\
SKY\_ROUT  & 4.5 & 4.5 & 4.5 & 4.5 & 4.5 & 4.5 & SKY\_ROUT & 6.5 & 6.5 & 6.5 & 8.0 & 10.5\\
BGD\_R  & 0 & 0 & 0 & 0 & 0 & 0 & BGD\_R & 2 & 2.5 & 2.5 & 3 & 5\\
BOX\_SIZE  & 2 & 2 & 2 & 2 & 2 & 2 & BOX\_SIZE & 5 & 5 & 5 & 5 & 8\\
CRIT\_SEP  & 4 & 4 & 4 & 5 & 6 & 6 & CRIT\_SEP & 8 & 8 & 8 & 8 & 8\\ \hline
MATCH\_THRESH  & 0.06 & 0.08 & 0.1 & 0.1 & 0.1 & 0.1 & MATCH\_THRESH & 0.15 & 0.2 & 0.2 & 0.25 & 0.25\\
NEXP\_THRESH  & 3 & 3 & 3 & 3 & 3 & 3 & NEXP\_THRESH & 3 & 3 & 3 & 3 & 3 \\
\enddata
\end{splitdeluxetable*}

\begin{deluxetable}{ l c c c }
\tablewidth{0pt} 
\tablecaption{Calculated number of expected sources in the NIRCam filters using the deep exposure `background' areas shown in Fig. \ref{fig:density_map_2}. The last column in the table shows the percentage of the sources expected in the whole field based on the deep exposure `backgrounds' with respect to the number of sources detected in the field. \label{tab:background_NIRCam_2}}
\tablehead{
\colhead{Filter} & \colhead{Source density}& \colhead{\# of expected sources} & \colhead{\% Sources expected} \\
\colhead{} & \colhead{deep exposure} & \colhead{in whole field based on}& \colhead{over sources} \\
\colhead{} & \colhead{backgrounds [arcmin$^{-2}$]} & \colhead{deep exposure backgrounds} & \colhead{detected}
} 
\startdata 
F115W  & 5,131 & 154,994 & 76.9  \\ 
F187N  & 1145 & 34,585 & 123.6 \\ 
F200W  & 5,197 & 156,973 & 75.9  \\ 
F277W  & 4,071 & 124,511 & 71.6 \\ 
F335M  & 3,250 & 99,397 & 89.5  \\ 
F444W  & 2,525 & 77,230 & 73.1 \\ 
\enddata
\end{deluxetable}

\begin{deluxetable}{ l c c c }
\tablewidth{0pt} 
\tablecaption{Calculated number of expected sources in the NIRCam filters using the shallow exposure `background' areas shown in Fig. \ref{fig:density_map_1}. The last column in the table shows the percentage of the sources expected in the whole field based on the shallow exposure `backgrounds' with respect to the number of sources detected in the field. \label{tab:background_NIRCam_1}}
\tablehead{
\colhead{Filter} & \colhead{Source density}& \colhead{\# of expected sources} & \colhead{\% Sources expected} \\
\colhead{} & \colhead{shallow exposure} & \colhead{in whole field based on}& \colhead{over sources} \\
\colhead{} & \colhead{backgrounds [arcmin$^{-2}$]} & \colhead{shallow exposure backgrounds} & \colhead{detected}
} 
\startdata 
F115W  & 3,631 & 109,724 & 54.4  \\ 
F187N  & 280 & 8,459 & 30.2  \\ 
F200W  & 3,878 & 117,154 & 56.6  \\ 
F277W  & 3,629 & 110,995 & 63.8  \\ 
F335M  & 1,881 & 57,519 & 51.8  \\ 
F444W  & 1,973 & 60,332 & 57.1  \\
\enddata
\end{deluxetable}

\begin{deluxetable}{ l c c c }
\tablewidth{0pt} 
\tablecaption{Calculated number of contaminating sources in the MIRI filters using the `background' areas shown in Fig. \ref{fig:density_map_3}. The last column in the table shows the percentage of the sources expected in the whole field based on the `backgrounds' with respect to the number of sources detected in the field. \label{tab:background_MIRI}}
\tablehead{
\colhead{Filter} & \colhead{Source density}& \colhead{\# of expected sources} & \colhead{\% Sources expected} \\
\colhead{} & \colhead{[arcmin$^{-2}$]} & \colhead{in whole field}& \colhead{over sources detected} \\
} 
\startdata 
F770W  & 351 & 4,485 & 39.7 \\ 
F1000W  & 179 & 2,285 & 29.9 \\ 
F1130W  & 88 & 1,118 & 17.7 \\ 
F1500W  & 56 & 717 & 20.0 \\ 
F2100W  & 14 & 182 & 18.4 \\ 
\enddata
\end{deluxetable}

\begin{figure*}
    \centering
    \includegraphics[width=0.8\linewidth]{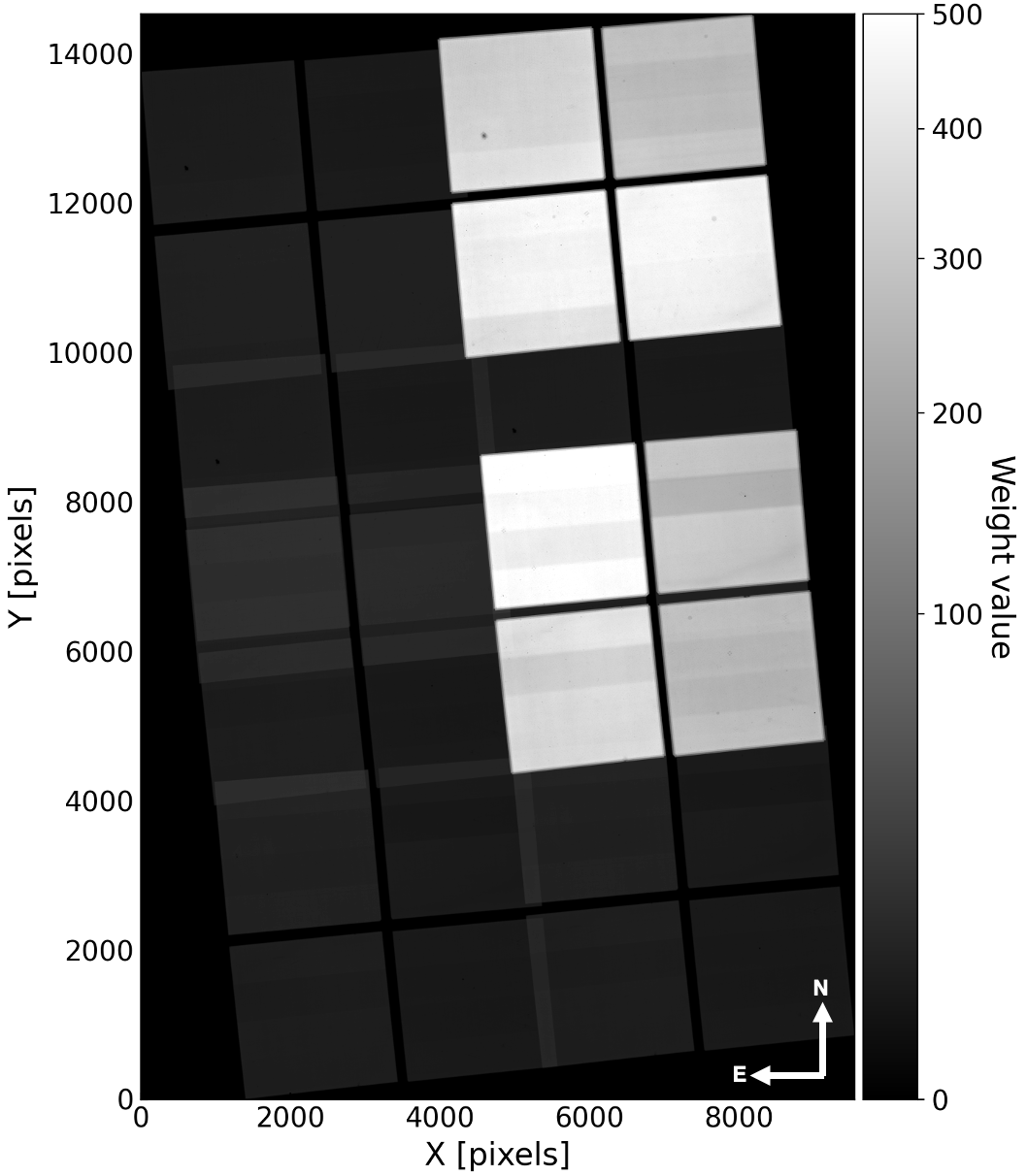}
    \caption{The F115W 2-D weight image showing the relative weights given to the pixels based on the exposure time. The x- and y-axis are in pixels, as this extension of the FITS file does not have a world coordinate system. The lighter the colour, the higher the weight. The two lighter quadrants show the deeper exposure areas.}
    \label{fig:weight_image}
\end{figure*}


\end{document}